\documentclass[9pt,shortpaper,twoside,web]{ieeecolor}
\usepackage{generic}
\usepackage{cite}
\usepackage{amsmath,amssymb,amsfonts}

\usepackage{algorithmic}
\usepackage{graphicx}
\usepackage{comment}
\usepackage[switch]{lineno} 
\usepackage{textcomp}
\usepackage{multicol}
\usepackage{caption}
\usepackage[table]{xcolor}
\usepackage{tabularx}
\usepackage{siunitx}
\usepackage{epstopdf}

\definecolor{DeepGreen}{RGB}{0,100,0}
\definecolor{headerblue}{RGB}{0, 102, 204}
\usepackage{array}
\usepackage{longtable}

\definecolor{headercolor}{RGB}{0, 102, 204}
\definecolor{rowgray}{gray}{0.95}
\definecolor{lightblue}{RGB}{220,230,241}

\definecolor{headerblue}{RGB}{0, 102, 204}
\definecolor{rowgray}{gray}{0.95}

\usepackage{hyperref}
\def\BibTeX{{\rm B\kern-.05em{\sc i\kern-.025em b}\kern-.08em
    T\kern-.1667em\lower.7ex\hbox{E}\kern-.125emX}}
\begin{document}
\title{Effect of nearby Metals on \\ Electro-Quasistatic Human Body Communication}
\author{Samyadip Sarkar, Arunashish Datta, David Yang, Mayukh Nath \\ {\em Student Member, IEEE}, Shovan Maity, {\em Member, IEEE} and Shreyas Sen, {\em Senior Member, IEEE}
\tiny{
\thanks{Samyadip Sarkar, Arunashish Datta, David Yang, Mayukh Nath, Shreyas Sen are with Elmore Family School of Electrical and Computer Engineering, Purdue University, West Lafayette, IN, USA (e-mail: \{sarkar46, datta30, yang996, shreyas\}@purdue.edu, nathm@alumni.purdue.edu).}
\thanks{Shovan Maity is with Quasistatics Inc. USA (e-mail: shovan@10xar.com).}}}

\maketitle

\begin{abstract}
In recent decades, Human Body Communication (HBC) has emerged as a promising alternative to traditional radio wave communication, utilizing the body's conductive properties for low-power connectivity among wearables. This method harnesses the human body as an energy-efficient channel for data transmission within the Electro-Quasistatic (EQS) frequency range, paving the way for advancements in Human-Machine Interaction (HMI). While previous research has noted the role of parasitic return paths in capacitive EQS-HBC, the influence of surrounding metallic objects on these paths—critical for EQS wireless signaling—has not been thoroughly investigated. This paper addresses this gap through a structured approach, analyzing how various conducting objects, ranging from non-grounded (floating) and grounded metals to enclosed metallic environments such as elevators and cars, affect the performance of the body-communication channel. We present a theoretical framework supported by Finite Element Method (FEM)-based simulations and experiments with wearable devices. Our findings reveal that metallic objects within $\sim$20 cm of the devices can reduce transmission loss by $\sim$10 dB. When the device's ground connects to a grounded metallic object, channel gain can increase by at least 20 dB. Additionally, the contact area during touch-based interactions with grounded metals depicts contact impedance-dependent
high-pass channel characteristics. The proximity to metallic objects enhances variability within a critical distance, with grounded metals
having an overall higher impact than floating ones. These insights are crucial for improving the reliability of body-centric communication links, thereby supporting applications in healthcare, consumer electronics, defense, and industrial sectors.
\end{abstract}


 \section{Introduction}
The trend of exponential miniaturization in semiconductor devices, combined with the emergence of cutting-edge sensors, is actively transforming the wearable technology industry. These innovations, with their wide-ranging applications, are improving healthcare, consumer electronics, and defense, while also revolutionizing human-machine interaction. Hence, the established wireless network among these devices, comprising in-body, on-body, and off-body communication—known as a Wireless Body Area Network (WBAN), is paving the way for a more connected and enhanced life \cite{seyedi2013survey, hasan2019comprehensive}. This study specifically focuses on the on-body communication links with wearable devices. Radiative radio frequency (RF)-based wireless communication methods, such as Bluetooth, WiFi, MedRadio, and Zigbee, have been conventional for wireless signaling among these Internet of Things (IoT) devices. However, 
 they face various challenges. Due to their inherent broadcasting nature, these technologies can create potential physical security vulnerabilities, highlighting the need for innovative approaches to protect user data. Furthermore, the higher power consumption ($\sim$10s of mW to 100s of mW) of these techniques results in a rapid depletion of battery life, underscoring the need for more energy-efficient solutions.

\begin{figure}[ht]
\centering
\includegraphics[width=0.48\textwidth]{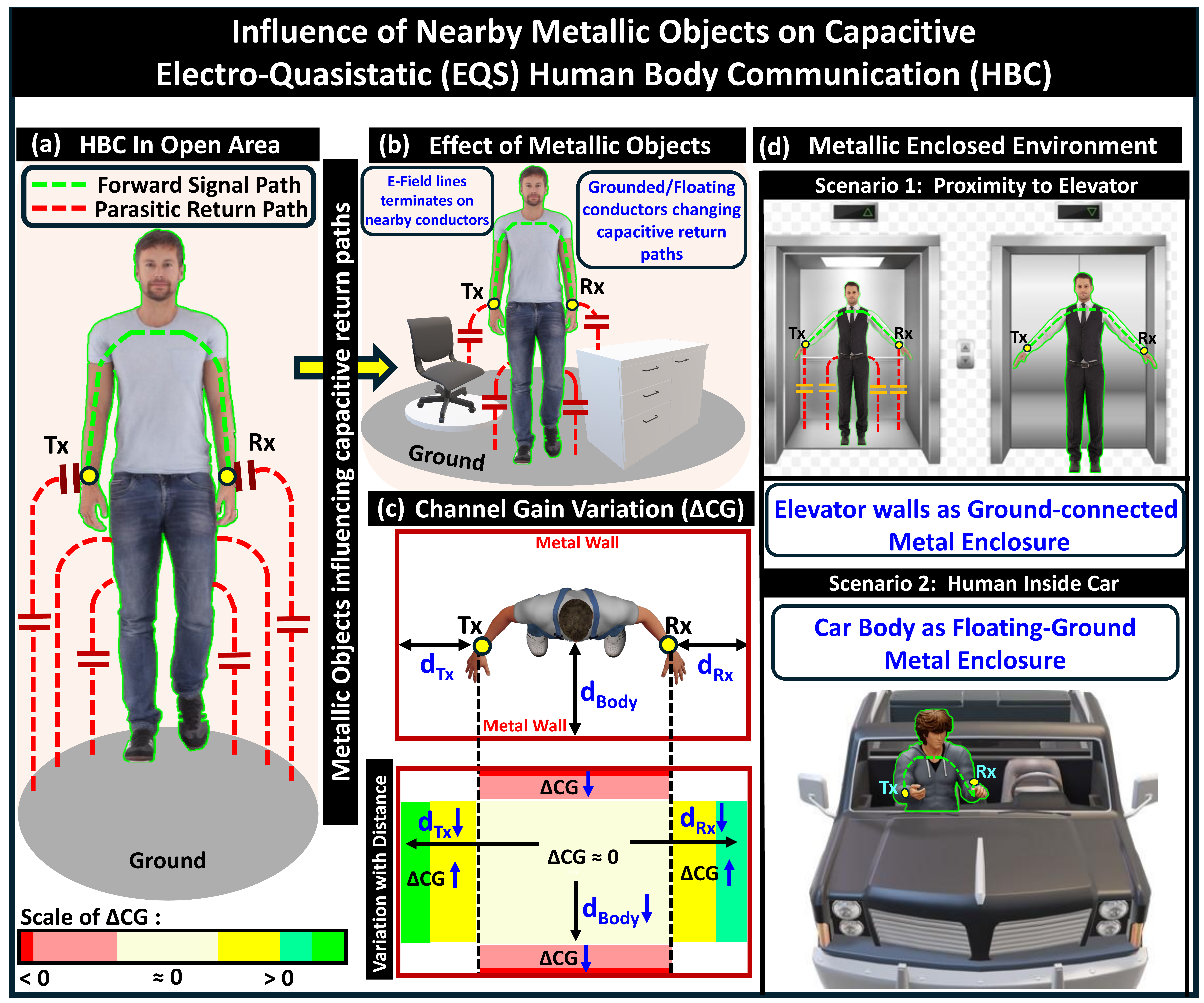}
\caption{This study examines how nearby metallic objects affect capacitive coupling-based Human Body Communication (HBC) in the Electro-Quasistatics (EQS) frequency range across different scenarios: (a) open space without metallic objects, (b) impact on capacitive parasitic return paths, (c) variations in channel gain near metallic objects, and (d) environments like being inside or near an elevator and (e) seated in a car.}
\label{fig:Introduction}
\end{figure}

Moreover, RF-based methods struggle in environments with numerous metal structures, such as buildings, hospitals, and vehicles. Metal can reflect or absorb radio waves, leading to multi-path propagation, signal loss, network failures, increased power consumption, and interference at higher frequencies. As a result, RF communication is not ideal for the energy constraints of ultra-low-power wearable devices. Besides RF-based techniques, other wireless communication technologies include inductive, acoustic, and optical methods, each presenting specific advantages and challenges. Typically operating below 30 MHz, quasistatic inductive links excel over short coverage areas, i.e., when transmitter and receiver coils are present in proximity, but their high sensitivity to precise alignment and rapid signal attenuation over long distances—making them suboptimal for long-range body communication \cite{park2019sub, wen2021channel,  hernandez2017magnetic, nath2022understanding, denisov2010ultrasonic, haerinia2020wireless}. Acoustic methods, though promising for their ability to penetrate 
deep tissues and device miniaturization, especially for implantable bio-electronic systems, struggle with reduced transduction efficiency, which limits their coverage and data rates (kb/s) \cite {santagati2017implantable, zhang2017bioacoustics, bos2018enabling, park2024recent, ghanbari2019sub, sonmezoglu2021monitoring}. Optical communication offers high bandwidth, flexibility with diverse modulation schemes, and immunity to electromagnetic interference, making it promising for biomedical settings. Moreover, its capacity to deliver both power and data over the same optical link enables secure, private, long-range, high-speed transmission. However, tissue-induced scattering, absorption, and reflection pose significant challenges, so selecting optimal wavelengths and tightly controlling transmitted power are essential for reliable biomedical operation \cite{yousif2019performance, hayal2023modeling, elsayed2024coding, elsayed2024performance, elsayed2025atmospheric, fuada2024study, fuada2025optical}.

Capacitive coupling-based Human Body Communication (HBC) in the Electro-Quasistatic (EQS) frequency regime ($\leq$30 MHz) offers an effective non-radiative communication method that significantly enhances physical security ($>$30$\times$ better than RF) and reduces power consumption ($\sim$100$\times$ lower than RF) \cite{das2019enabling, maity2019bodywire}. 
These factors make capacitive HBC advantageous, facilitating efficient data transmission that leverages the body's conductive properties.

We focus on capacitive coupling-based EQS-HBC, enabling low-loss, long-range ($\leq$2 m) communication with reduced power consumption, suitable for battery-powered wearables. In this system, a capacitive transmitter (Tx) couples an EQS signal through a skin-mounted electrode to the user's body, while an on-body capacitive receiver (Rx) captures it. The floating ground electrodes create parasitic return paths that are sensitive to environmental changes. With the rise of HBC-enabled devices—available in forms like wearables and implantables—alongside growing research in Human-Machine Interaction (HMI) \cite{sen2020body, chatterjee2023bioelectronic, maity2019bodywire, sarkar2025human}, a key question arises: \textbf{How does the presence of conducting objects affect the body-communication channel's link margin?} Hence, understanding these channel characteristics is essential for deploying this technology in various environments, such as elevators, vehicles, conceptualized in Fig. \ref{fig:Introduction}. 

Over the past couple of decades, the communication channel of capacitive EQS-HBC has been extensively studied. However, limited research addresses the impact of metallic surroundings on the HBC channel. Key findings from previous studies include: Lucev et al. \cite{lucev2012capacitive} examined channel loss variation in capacitive HBC using a benchtop ground-connected vector network analyzer (VNA), noting that improper ground isolation between Tx and Rx grounds led to optimistic loss estimates. Their findings showed high-pass characteristics with 50 $\Omega$ resistive termination leading to higher loss at lower frequencies. Xu et al. \cite{xu2012equation} investigated environmental couplings on the Electric-field (EF)-based intrabody communication (IBC) channel from 20 MHz to 100 MHz with battery-powered Tx and Rx, concluding that nearby conductive objects enhance channel gain. However, their study did not address touch-based interactions between users and metallic objects. Park et al. \cite{park2016channel} conducted channel gain measurements using various setups in 20 MHz to 150 MHz, comparing results from VNAs and spectrum analyzers. They reported that even with impedance matching, the optimistic loss estimates persisted. Maity et al. \cite{maity2018bio} evaluated the effects of source and termination impedance on the EQS-HBC characteristics of wearable prototypes, spanning from 10 kHz to 1 MHz, which promotes voltage mode signaling for broadband communication. Xu et al. \cite{xu2019modeling} studied channel loss in vehicles and found a 7 dB gain improvement with a wearable transmitter, a larger ground receiver serving as a spectrum analyzer, which resulted in more optimistic loss figures. Yang et al. \cite{yang2022physically} performed wearable-to-wearable measurements at 415 kHz, assessing variability due to posture and environmental conditions. Sarkar et al. \cite{sarkar2023electro} explored channel variability with positional changes relative to surrounding structures, finding substantial signal boosts with specific Rx setups. In Appendix IV \ref{appendix:d}, a summary is presented that highlights key comparisons drawn from previous studies. Overall, channel variability is significantly influenced by the presence of conductive structures, which affects the EQS-HBC signal characteristics. 

Addressing this issue, this paper presents a structured approach that delves into Circuit Model-based insights. The contributions of this paper are summarized below:

1. Developed a \textbf{theory and circuit model} to analyze the impact of grounded and non-grounded metallic objects on capacitive EQS-HBC channel behavior.

2. Conducted human body \textbf{channel loss measurements with wearable devices} and simulations to validate the theory across different environments.

3. Investigated the \textbf{effect of touch} on channel loss variation when in contact with metallic objects.

4. Examined \textbf{channel variability} and \textbf{deployment} of capacitive EQS-HBC in metallic enclosed spaces, such as elevators and cars.

The organization of the paper is as follows: Section II introduces the theoretical aspects of capacitive EQS-HBC, including the effect of both ground-connected and non-grounded metals. Section III discusses the impact of touch-based interactions with surrounding metals. Section IV describes the Finite Element Method (FEM)-based simulations in EQS, which capture channel variability in metallic enclosures. Section V covers the experimental setup and results for channel loss measurements using wearable devices. Section VI presents the correlation for the proposed model with the results from numerical simulation and experiments. Finally, Section VII discusses the key takeaways, and Section VIII concludes the paper.

\begin{figure}[ht]
\centering
\includegraphics[width=0.48\textwidth]{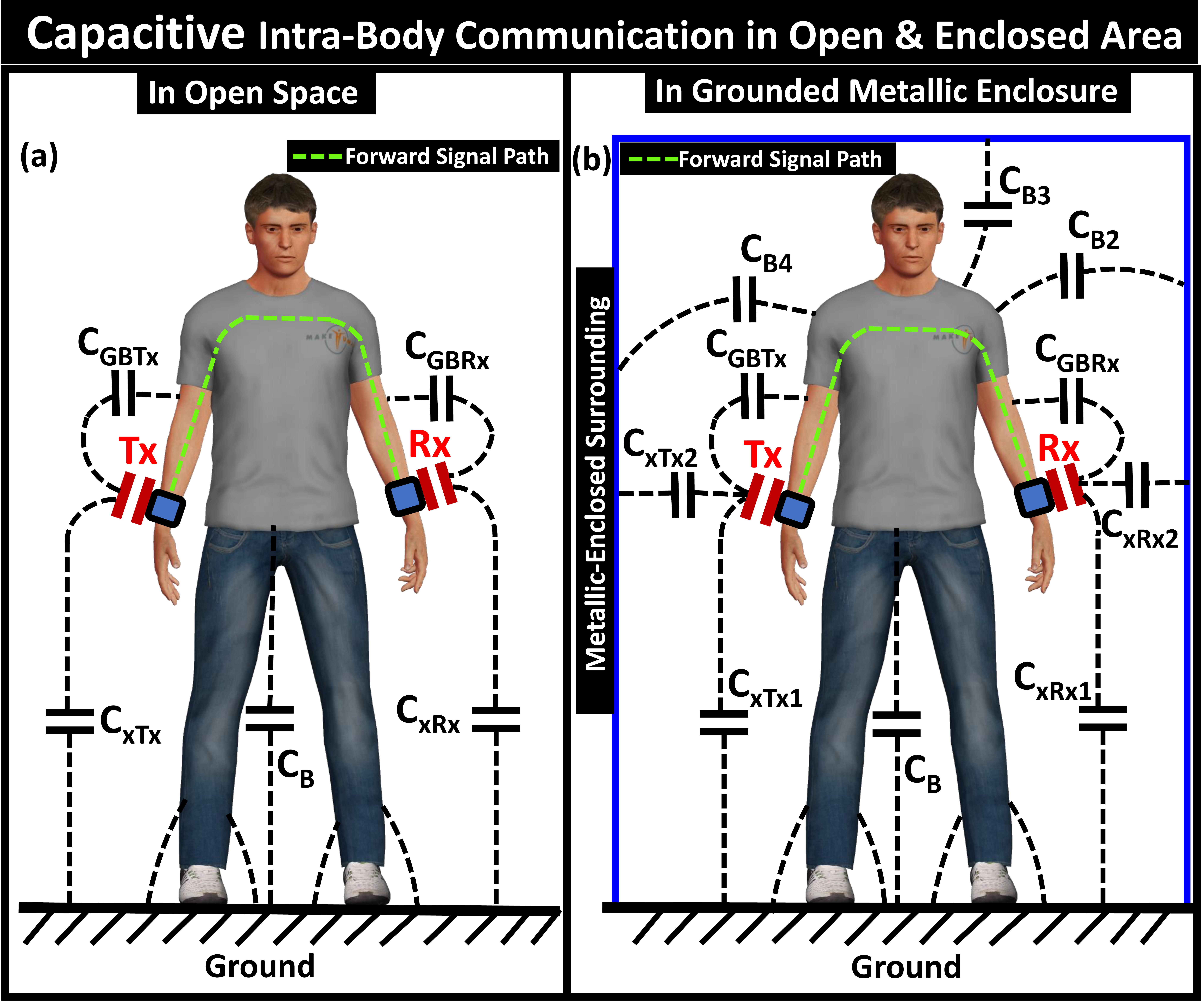}
\caption{Schematic of capacitive couplings in EQS-HBC: (a) Open Space, (b) Grounded metallic enclosure. Proximity to metal walls influence return path capacitances (C$_{xTx}$ $\&$ C$_{xRx}$) and body-to-earth coupling (C$_B$).}
\label{fig:Capacitive_Coupling_In_Metallic_Enclosure}
\end{figure}

\section{Influence of Surrounding Medium on Capacitive EQS-HBC}

The capacitive EQS-HBC utilizes single-ended excitation and voltage pick-up at the transmitting and receiving ends respectively. In the EQS range (f $\leq$30 MHz), the signal wavelength ($\lambda$ $\geq$10 m) greatly exceeds the dimensions of the human body ($\leq$2 m) and the transceiver devices ($\leq$0.03 m). This, length-scale concept when combined with a charge carrier relaxation time ($\tau$) that is much smaller than the operational time scale (T) (i.e., $\tau << T$ where T is in the order $\mu$s in EQS frequency regime), validates a lumped-element approximation for biophysical models of the body communication channel. Previous studies indicate that the transmission characteristics are strongly influenced by the parasitic return path, which varies with device size (such as for disk-shaped electrode radius: a) and their position relative to the human body),  but is nearly independent with device height (h) i.e., devices are well above the earth's ground for miniaturized wearables (i.e., $a << h$) \cite{maity2018bio, nath2019toward, datta2021advanced}. As schematically captured in Fig. \ref{fig:Capacitive_Coupling_In_Metallic_Enclosure} (a), for wearable Tx and Rx, return path capacitances (C$_\mathrm{xTx}$, C$_\mathrm{xRx}$ $\sim$ 100's of femto farads) are typically much lower than the combined effects of body-shadowing capacitance (C$_\mathrm{GBRx}$) and load capacitance (C$_\mathrm{L}$ $\sim$5 pF), both of which are lower than the body capacitance (C$_\mathrm{B}$ $\sim$ 100's of pF). Additionally, parasitic capacitances can change based on the proximity to metallic objects. It is worthwhile to note that better confinement makes the EQS signal around the subject's body \cite{das2019enabling} less vulnerable to the conducting objects, far away from the subject (i.e., beyond the leakage limit). However, the presence of metallic objects within the leakage limit of the EQS signal in the form of a grounded or a floating metal or sometimes even as a metallic enclosure, presented in Fig. \ref{fig:Capacitive_Coupling_In_Metallic_Enclosure} (b), alters the extent of the coupled electric field (E-Field) on the body at the transmitting end, i.e., influencing the quasistatic charge sharing between the Tx and subject and thus changing the induced body-potential ($V_B$) as shown in Eq. \ref{eq1}.
  \begin{equation}
   \centering 
   V_B \approx \left(\frac{C_{xTx}}{C_{B}} \right) V_{Tx} 
   \label{eq1}
\end{equation}
With the source resistance (typically $R_S = 50 \Omega$) of the voltage source being orders of magnitude lower than the impedance of $C_{GBTx}$ at EQS frequencies i.e., $R_S << Z_{C_{GBTx}}$) for $C_{GBTx} \sim3$ pF ($C_{GBTx} = C_{PP} + C_F$  where signal plate-ground plate capacitance ($C_{PP}$) = 2.2 pF for a device with disc-shaped electrode of radius = 2.5 cm and fringe capacitance ($C_F$) $\approx$ 0.65 pF to 0.85 pF, a function of device position relative to body), the voltage drop across $R_S$ can be fairly ignored. 
Subsequently, the induced electric field at the receiving end also varies with change in $V_B$ and the variation in the ratio of return path capacitance ($C_{xRx}$) to the effective load capacitance ($C_{L(eff.)} = C_{GBRx} + C_L$) leading to a variability in received voltage (V$_{Rx}$) level as shown in Eq. \ref{eq2}. 
\begin{equation}
   \centering 
   V_{Rx} \approx \left(\frac{C_{xRx}}{C_{GBRx} + C_L} \right) V_B 
   \label{eq2}
\end{equation}

Assuming the Tx and Rx are long distance apart, i.e., neglecting the effect of capacitive coupling between their ground electrodes (inter-device coupling ($C_C$))\cite{datta2021advanced}, the channel loss ($L$) through the body can be approximately formulated as presented in Eq. \ref{eq3}. 
\begin{equation}
   \centering 
   L (dB) = -20\log_{10} \left( \frac{V_{Rx}}{V_{Tx}} \right)  \approx -20\log_{10} \left(\frac{C_{xTx}}{C_B} \frac{C_{xRx}}{C_{GBRx} + C_L} \right)
   \label{eq3}
\end{equation}

\begin{figure}[ht]
\centering
\includegraphics[width=0.48\textwidth]{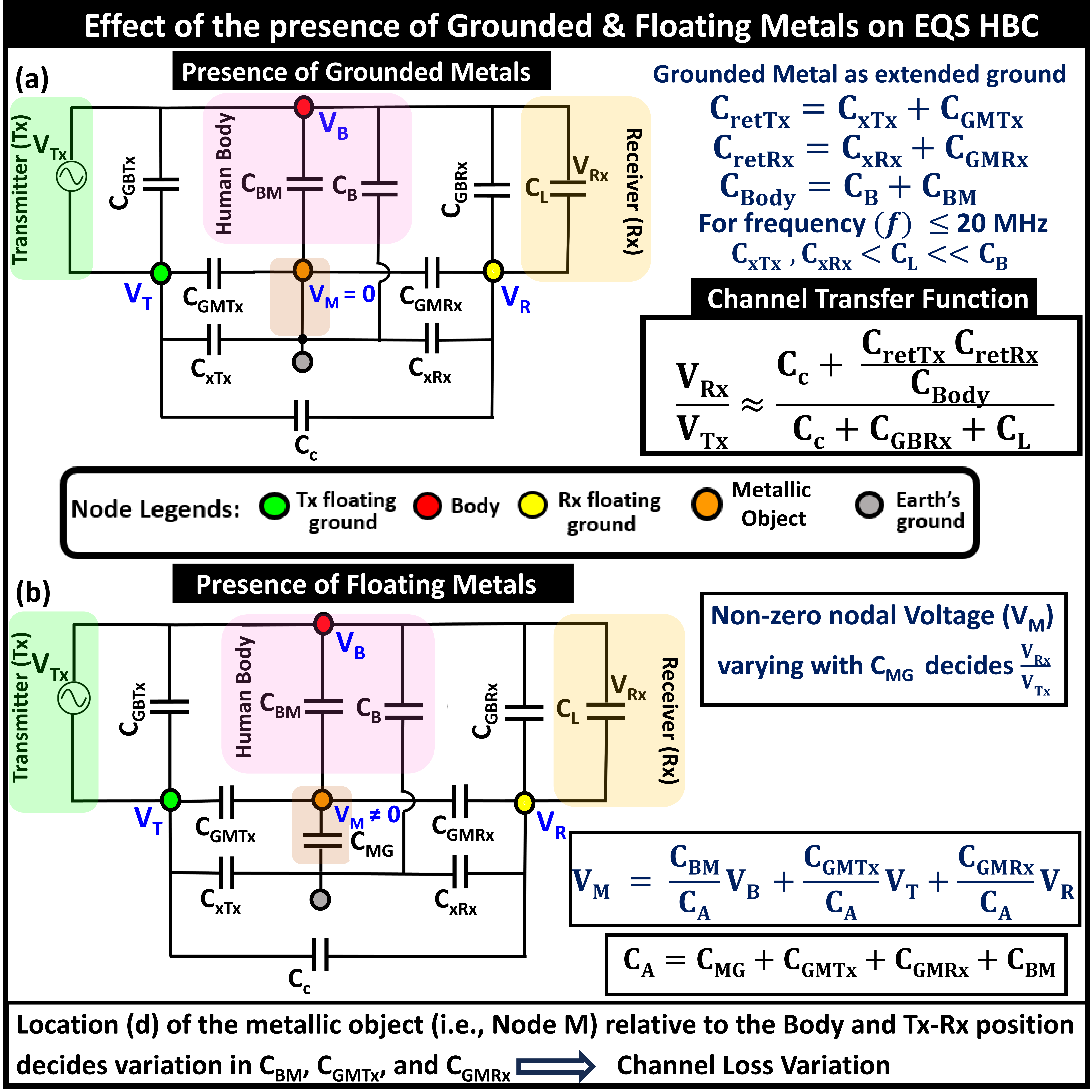}
\caption{Comparative analysis of the simplified equivalent circuit model: Effects of (a) grounded metals and (b) floating metals. The positioning of metallic objects relative to the subject's body and devices leads to variability in channel characteristics.}
\label{fig:Equivalent_Circuits_Grounded&Floating}
\end{figure}

\begin{figure}[ht]
\centering
\includegraphics[width=0.48\textwidth]{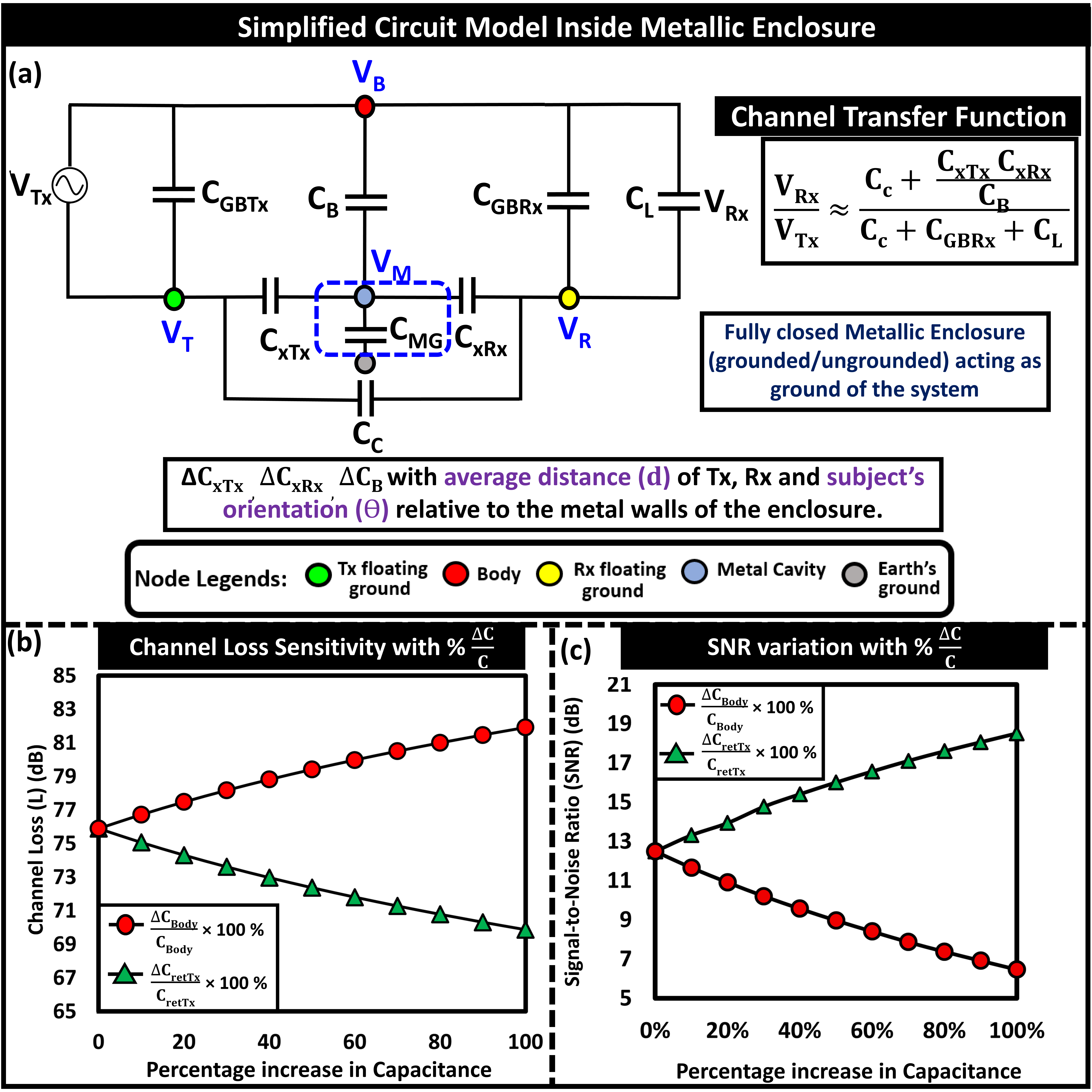}
\caption{(a) Simplified equivalent circuit model within a metallic enclosure, (b) Sensitivity of channel loss with fractional changes in $C_{retTx}$ and $C_{Body}$, (c) Variation in SNR. 
}
\label{fig:Equivalent_Circuits_Part1}
\end{figure}

In contrast to open spaces where electric field lines from devices and the human body terminate on the earth's ground, in metallic environments, these lines may end up on metallic objects, leading to significant variations in coupling capacitances. This affects the performance of communication channels, as illustrated in Fig. \ref{fig:Capacitive_Coupling_In_Metallic_Enclosure}. The bio-physical models, reflecting channel variability due to metals, are presented in Fig. \ref{fig:Equivalent_Circuits_Grounded&Floating} (a, b) for grounded, floating objects and Fig. \ref{fig:Equivalent_Circuits_Part1} (a) for metal enclosures. These are analyzed in the following subsections.

\subsection{Effect of nearby grounded metals}
Grounded metallic objects present in the vicinity of an EQS-HBC user can influence the channel loss of body-centric communication. The impact differs based on whether the objects are closer to the human body or the communicating devices, as with proximity, the metallic object starts to influence the induced body potential via its location-dependent effective loading. From the perspective of the E-field lines, it can also be interpreted that more E-field lines from the device grounds can now terminate on the earth's ground. This boosts the equivalent return path capacitances at Tx (i.e., C$_{retTx}$ = C$_{xTx}$ + C$_{GMTx}$) and at Rx (i.e., C$_{retRx}$ = C$_{xRx}$ + C$_{GMRx}$) and hence, increases $C_{ret}$ of the devices as ground comes closer which in-turn reduces the extent of body shadowing. Therefore, this results in a rise in the received signal level (since, $V_{Rx}$ $\propto$ C$_{retTx}$C$_{retRx}$), as shown in Fig. \ref{fig:Equivalent_Circuits_Grounded&Floating} (a). However, grounded metals, when present closer to the human body compared to the devices, enhance the equivalent capacitance of the subject's body (i.e., $C_{Body}$ = $C_{B}$ + $C_{BM}$). This results in an increased channel loss. Hence, with C$_{GMTx}$, C$_{GMRx}$, and C$_{BM}$ respectively, represent the couplings from the ground of Tx, ground of Rx, and user's body to the grounded metallic object, the overall channel loss (L$_{GM}$) can be approximated in Eq. \ref{eq4}. 
\begin{equation}
   \centering 
   L_{GM} (dB) \approx
   -20\log_{10}\left(\frac{C_c + \frac{C_{retTx} C_{retRx}}{C_{Body}}}{C_c + C_{GBRx}+ C_L}\right) 
   \label{eq4}
\end{equation}

\subsection{Effect of nearby floating metals} 
This section examines the impact of non-grounded metallic objects on EQS-HBC channel characteristics. Like grounded metals, the position of floating metals affects channel behavior based on their proximity to devices or the user's body. Instead of providing a direct ground, these objects influence channel loss through capacitive coupling (C$_{MG}$) with the earth's ground, which varies with the object's size and height. 
The equivalent circuit model is shown in Fig. \ref{fig:Equivalent_Circuits_Grounded&Floating} (b), and can be analyzed by assuming the nodal voltages at the ground of the Tx, ground of the Rx, body, and metal object as $V_T$, $V_R$, $V_B$, and $V_M$, respectively. The induced potential on the metallic object from the body and devices is expressed in Eq. \ref{eq5}
\begin{equation}
   \centering 
   V_{M} = \frac{C_{BM}}{C_A} V_B + \frac{C_{GMRx}}{C_A} V_R + \frac{C_{GMTx}}{C_A} V_T
   \label{eq5}
\end{equation}
where  C$_A$ = (C$_{MG}$ + C$_{GMRx}$ + C$_{GMTx}$ + C$_{BM}$). For a large object, the expression for C$_A$ can be approximately simplified as
C$_A$ = (C$_{MG}$ + C$_{BM}$) (Since, C$_{GMRx}$, C$_{GMTx}$ $<<$ C$_{MG}$, C$_{BM}$). The coupling between the body and a metallic object (C$_{BM}$) varies with changes in distance and orientation relative to the object. Additionally, the subject's body potential (V$_B$) is influenced by nearby metallic objects. Incorporating these effects, the nodal voltage at the ground of the Rx is given in Eq. \ref{eq6:main}.
\begin{subequations}\label{eq6:main}
    \begin{equation}
   \centering 
   V_{R} = \frac{B}{P} V_B + \frac{Q}{P} V_T
   \tag{\ref{eq6:main}}
\end{equation}

where the expression for $P$, $Q$, $B$, $C_{L(eff.)}$ and  are presented in Eq. \ref{eq6:main} (a, b, c, d)
\begin{equation}
P = \frac{1}{C_{L(eff.)}} \left [(C_{L(eff.)} + C_{xRx} + C_{GMRx}) + \frac{C_{GMRx}^2}{C_A}\right]
\label{eq6:main:a}
\end{equation}
\begin{equation}
    Q = -\frac{C_{GMRx}  C_{GMTx}}{C_{L(eff.)} C_A}
    \label{eq6:main:b}
\end{equation}
\begin{equation}
B = 1 - \frac{C_{GMRx}  C_{BM}}{C_{L(eff.)} C_A}
\label{eq6:main:c}
\end{equation}
\begin{equation}
C_{L(eff.)} = C_{GBRx} + C_L
\label{eq6:main:d}
\end{equation}
\end{subequations}

The grounded metal scenario can be treated as a special case of the floating metal scenario under the following approximations:
\\ With $Z_{C_{MG}} \approx 0$, $C_{BM} \approx C_B$, $C_{GMTx} \approx C_{xTx}$, $C_{GMRx} \approx C_{xRx}$, we get $V_M \approx 0$. The circuit model shown in Fig. \ref{fig:Equivalent_Circuits_Part1} (a) illustrates how transmission characteristics change within a metallic enclosure. Fig. \ref{fig:Equivalent_Circuits_Part1} (b) and Fig. \ref{fig:Equivalent_Circuits_Part1} (c) provide insight into the sensitivity of channel loss and the Signal-to-Noise Ratio (SNR) of the communication link, respectively, based on variations in coupling capacitances due to proximity to metallic objects.

\subsection{Sensitivity Analysis of EQS–HBC Transfer Function}
The relative sensitivity of channel transfer function \(T\) to a small change in the coupling capacitance \(C_i\) is expressed in Eq. \ref{eq7:main}
\begin{subequations}\label{eq7:main}
\begin{equation}
  S_{C_i}
  = \frac{C_i}{T}\,\frac{\partial T}{\partial C_i}
  \tag{\ref{eq7:main}}
\end{equation}
\subsection{Sensitivities of \(\frac{V_{B}}{V_{Tx}}\) $\&$ \(\frac{V_{Rx}}{V_{B}}\)}
Hence, the relative sensitivities are defined in Eqs. \ref{eq7:main} (a, b, c, d) .
\begin{equation}
    S_{C_{\mathrm{retTx}}}
    = \frac{C_{\mathrm{retTx}}}{\left(\frac{V_{B}}{V_{Tx}}\right)}\,
       \frac{\partial \left(\frac{V_{B}}{V_{Tx}}\right)}{\partial C_{\mathrm{retTx}}}
    = \frac{C_{\mathrm{Body}}}
           {C_{\mathrm{retTx}} + C_{\mathrm{Body}}}
           \label{eq7:main:a}
\end{equation}
\begin{equation}
    S_{C_{\mathrm{Body}}}
    = \frac{C_{\mathrm{Body}}}{\left(\frac{V_{B}}{V_{Tx}}\right)}\,
       \frac{\partial \left(\frac{V_{B}}{V_{Tx}}\right)}{\partial C_{\mathrm{Body}}}
    = -\,\frac{C_{\mathrm{Body}}}
             {C_{\mathrm{retTx}} + C_{\mathrm{Body}}}
             \label{eq7:main:b}
\end{equation}
\begin{equation}
  S_{C_{\mathrm{retRx}}}
    = \frac{C_{\mathrm{retRx}}}{\left(\frac{V_{Rx}}{V_{B}}\right)}\,
       \frac{\partial \left(\frac{V_{Rx}}{V_{B}}\right)}{\partial C_{\mathrm{retRx}}}
    = \frac{C_{\mathrm{L(eff.)}}}
           {C_{\mathrm{retRx}} + C_{\mathrm{L(eff.)}}}
           \label{eq7:main:c}
\end{equation}
\begin{equation}
  S_{C_{\mathrm{L(eff.)}}}
    = \frac{C_{\mathrm{L(eff.)}}}{\left(\frac{V_{Rx}}{V_{B}}\right)}\,
       \frac{\partial \left(\frac{V_{Rx}}{V_{B}}\right)}{\partial C_{\mathrm{L(eff.)}}}
    = - \frac{C_{\mathrm{L(eff.)}}}
             {C_{\mathrm{retRx}} + C_{\mathrm{L(eff.)}}}
             \label{eq7:main:d}
\end{equation}
\end{subequations}
The combined sensitivity can be obtained by combining the above sensitivities, as expressed in Eq. \ref{eq8:main} below.
\begin{subequations}\label{eq8:main}
    \begin{equation}
\frac{\partial T}{\partial C_i}
= \left(\frac{V_{Rx}}{V_{B}}\right) \;\frac{\partial \left(\frac{V_{B}}{V_{Tx}}\right)}{\partial C_i},
\quad C_i\in\{C_{\mathrm{retTx}},\,C_{\mathrm{Body}}\}
\tag{\ref{eq8:main}}
\end{equation}
\\
\begin{equation}
    \frac{\partial T}{\partial C_j}
 = \left(\frac{V_{B}}{V_{Tx}}\right) \;\frac{\partial \left(\frac{V_{Rx}}{V_{B}}\right)}{\partial C_j},
\quad C_j\in\{C_{\mathrm{retRx}},\,C_{\mathrm{L(eff.)}}\}
\label{eq8:main:a}
\end{equation}
\end{subequations}
Understanding this sensitivity helps to optimize link margins in transceiver design for better performance. Appendix II outlines how nearby metallic objects affect body channel performance, influencing Shannon capacity and bit error rates across various modulation schemes.

\section{Impact of Touch-based Interactions}
In capacitive intrabody communication, accidental contact with grounded or floating metal objects creates a low-impedance node that reduces the received signal. The potential of this node varies based on: \textbf{1.} whether the object is grounded or not—grounded metal with low driving impedance decreases the node's potential more effectively, \textbf{2.} differences in channel loss related to contact area ($A_{con}$), and \textbf{3.} the touch location relative to the communicating devices, as discussed in the following subsection. 
\begin{figure}[ht]
\centering
\includegraphics[width=0.48\textwidth]{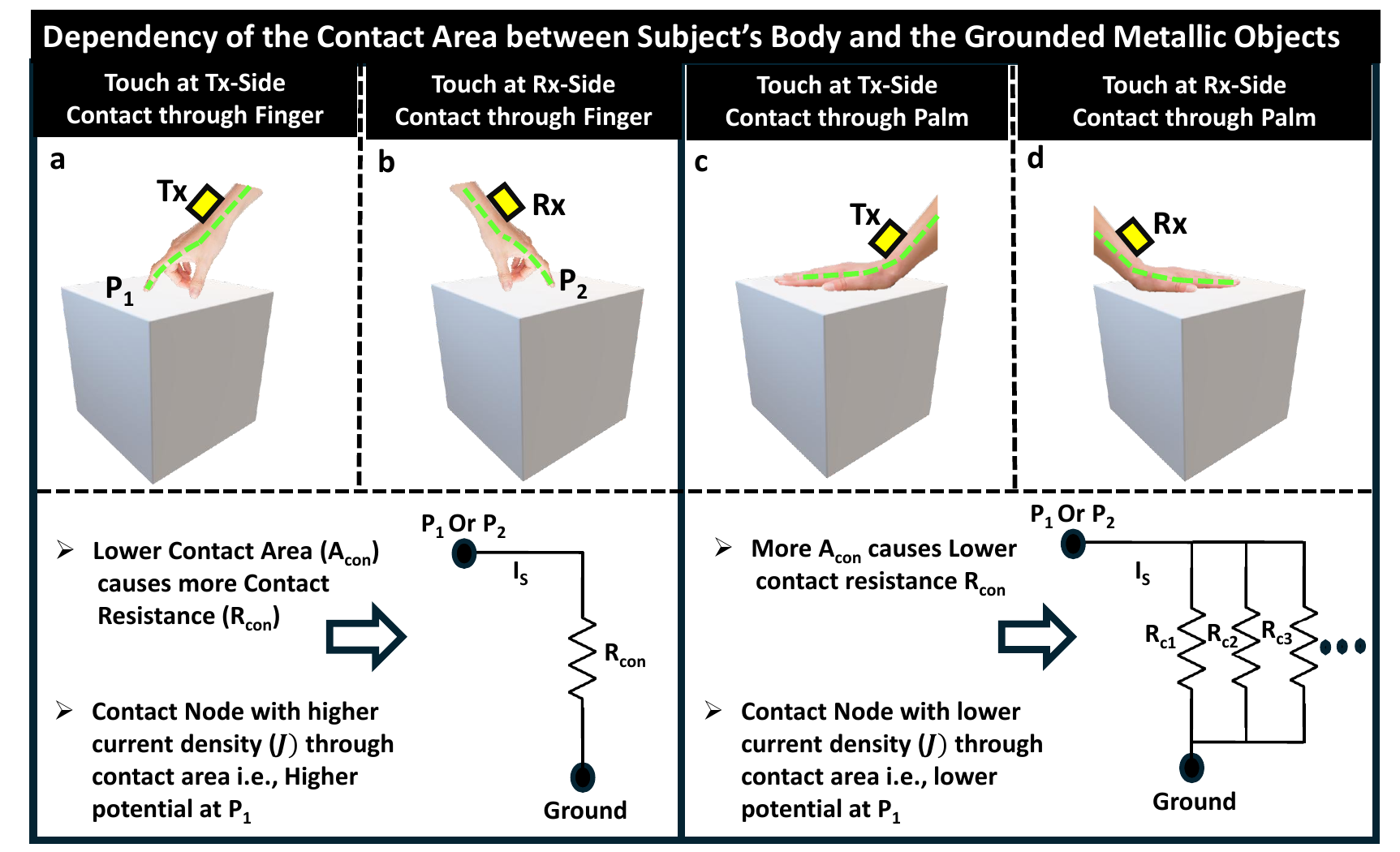}
\caption{During touch-based interactions between the subject and metallic objects, the area of contact and the location of touch is of critical importance in governing channel loss.}
\label{fig:Contact_Area_Schematic}
\end{figure}

\subsection{Influence of Contact Area} 

The contact area of the subject's body with the metallic objects can decisively impact the channel loss, as schematically illustrated in Fig. \ref{fig:Contact_Area_Schematic} (a, c) $\&$ (b, d). An increase in the effective area of contact ($A_{con}$) during touch, reduces the equivalent contact impedance ($Z_{con}$ = $R_{con} || C_{BM}$) at the point of contact (since, $R_{con}$ $\propto$ $\frac{1}{A_{con}}$ and $C_{BM} \propto {A_{con}}$). This makes the point of contact, a node offering area-dependent impedance while sinking current through it, thus affecting the channel characteristics. The touch event causes an RC-behavior through $Z_{con}$ at the contact location and thereby leads to high-pass channel behavior.

\begin{figure}[ht]
\centering
\includegraphics[width=0.48\textwidth]{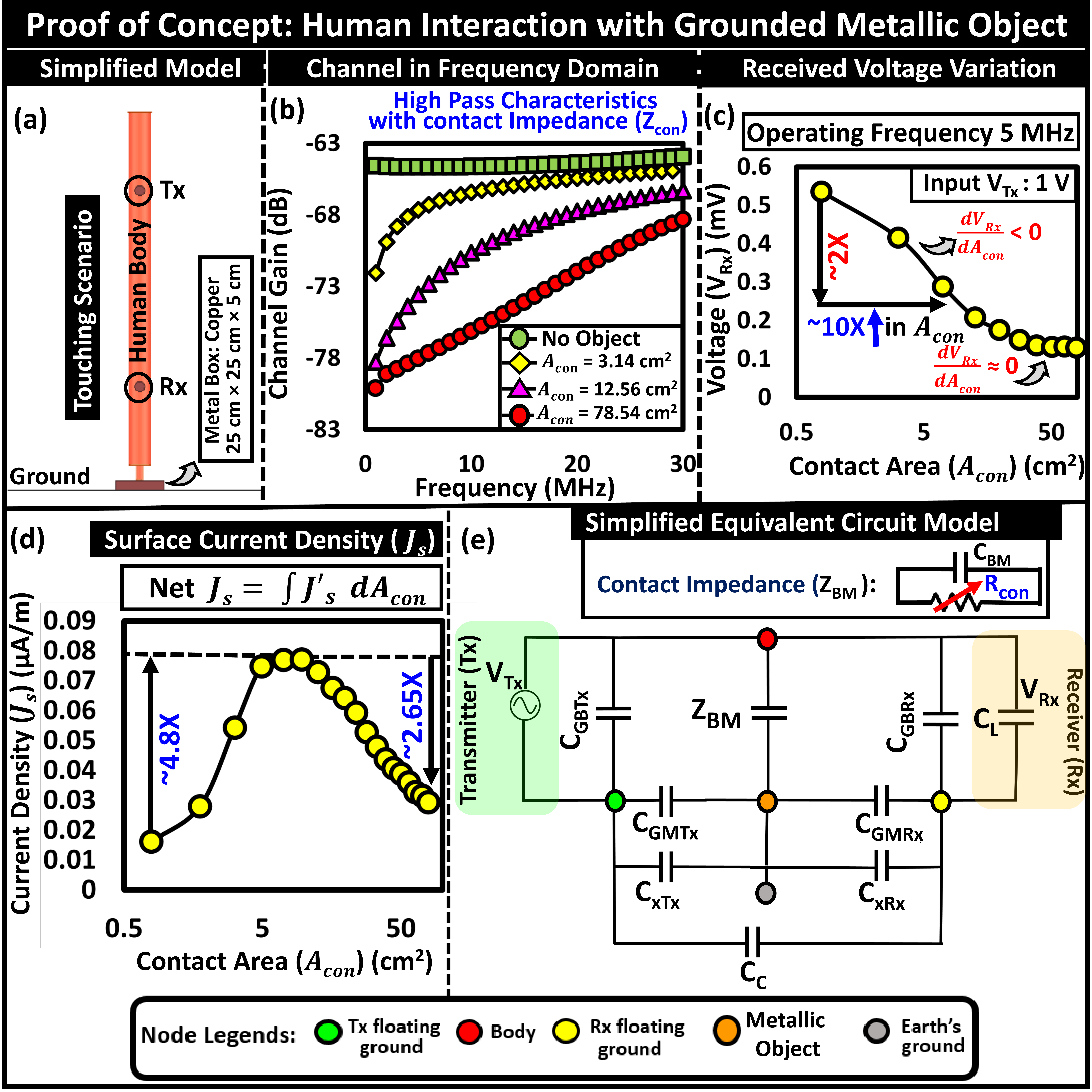}
\caption{Analyzing touch interactions between an EQS-HBC user and a metallic object: (a) A simplified human arm model with Tx and Rx contacting the metal. (b) Investigating how channel gain varies with frequency and contact area ($A_{con}$), factoring in contact impedance. (c) Examining the relationship between received voltage and $A_{con}$. (d) Observing current density variations with $A_{con}$. (e) Proposing a simplified circuit model for touching and non-touching scenarios.}
\label{fig:Simplified_Model}
\end{figure}

\subsection{Dependency on Contact Location}
The location of touch significantly affects channel loss, as shown in Fig. \ref{fig:Contact_Area_Schematic} (a, b) $\&$ (c, d).  To understand the effect of the surrounding medium on capacitive EQS-HBC (frequency $\leq$30 MHz), numerical electromagnetic simulations using ANSYS HFSS version  were conducted. The proof-of-concept model simulates a human arm with Tx and Rx, illustrated in Fig. \ref{fig:Simplified_Model} (a). It reveals frequency-dependent high-pass characteristics and the impact of contact resistance ($R_{con}$) during touch interactions, as shown in Fig. \ref{fig:Simplified_Model} (b). The analysis indicates how on-body voltage (V$_{out}$) varies with contact area ($A_{con}$) at 5 MHz, depicted in Fig. \ref{fig:Simplified_Model} (c). Results show that $V_{Rx}$ increases as $A_{con}$ decreases, significantly declining with an increase in $A_{con}$—approximately a 2$\times$ reduction with a 10$\times$ increase in $A_{con}$. This indicates a decreasing attenuation rate in $V_{Rx}$ after reaching a certain $A_{con}$, suggesting maxima within the $Z_{con}$ variation profile. Variations in $Z_{con}$ also affect current density ($J_s$) at the contact location, as illustrated in Fig. \ref{fig:Simplified_Model} (d). A simplified equivalent circuit model in Fig. \ref{fig:Simplified_Model} (e) shows that the impedance ($Z_{BM}$) between the body and metallic object is replaced by $Z_{con}$ during touch interactions. The voltage-mode transfer function ($T_{TGM}(s)$) during contact with grounded metal is approximated as shown in Eq. \ref{eq9}. 

\begin{equation}
   \centering 
   T_{TGM}(s) \approx \left(\frac{C_{retTx}C_{retRx}}{C_{L(eff.)} C_{BM} + \left(\frac{C_{L(eff.)}}{sR_{con}}\right)}\right)
   \label{eq9}
\end{equation}
where C$_{L(eff.)}$ includes the combined effect of body-shadowing (C$_{GBRx}$) and the plate-to-plate capacitance (C$_{L}$) at Rx i.e., C$_{L(eff.)}$ = (C$_{GBRx}$ + C$_{L}$). From the above channel gain expression, it can be stated that the body-channel exhibits frequency dependent variability with higher loss at lower frequencies i.e., presenting the characteristics of a first order high pass filer with a cut-off frequency (f$_c$) that is independent of C$_{L(eff.)}$ and is function of the contact impedance ($Z_{BM} (s) = \frac{R_{con}}{1+ sR_{con}C_{BM}}$), shown in Eq. \ref{eq10}. 
\begin{equation}
   \centering 
   f_c = \frac{1}{2 \pi R_{con} C_{BM} }
   \label{eq10}
\end{equation}

With $f_c$ being inversely proportional to $R_{con}$, the increase in $A_{con}$ results in a change in $f_c$ to a higher frequency as confirmed by the numerical simulation results in Fig. \ref{fig:Simplified_Model} (b). 
 In contrast, the impedance ($Z_{BM}$) becomes a purely capacitive ($C_{BM}$) during proximity-based non-touching interaction scenarios, i.e., replacing $Z_{con}$ with $\frac{1}{s C_{B}}$ where C$_{Body}$ = (C$_B$ + C$_{BM}$), the channel transfer function $T_{NTGM}(s)$ is expressed in Eq. \ref{eq11}
\begin{equation}
   \centering 
  T_{NTGM}(s) \approx \frac{C_{retTx} C_{retRx}}{C_{Body} C_{L(eff.)}}
  \label{eq11}
\end{equation}
resulting in frequency-independent, i.e., flat-band channel characteristics as presented previously in Eq. \ref{eq4} with C$_c$ = 0. 

Similarly, the subject's interaction with a floating metallic object can be mathematically formulated as follows:  
With a non-zero voltage at V$_M$, the voltage mode transfer gain $T_{NTFM} (s)$ in a simplified form is presented in Eq. \ref{eq12:main}
\begin{subequations}\label{eq12:main}
    \begin{equation}
\centering 
T_{NTFM}(s) = \left(\frac{1- \frac{\left(A-DS\right)}{DR+F}}{U +T \frac{\left(A-DS\right)}{DR+F}} \right)
\tag{\ref{eq12:main}}
\end{equation}
where the coefficients A, F, D, S, R, U, T take the following forms expressed in Eqs. \ref{eq12:main} (a-g).
\begin{equation}
A (s) =\left(1 - \frac{sC_{GMTx}Z_{BM}C_B}{C_{xTx}}\right)
\label{eq12:main:a}
\end{equation}
\begin{equation}
F(s) = (sC_{GMTx}\frac{C_{xRx}}{C_{xTx}} - sC_{GMRx})Z_{BM}
\label{eq12:main:b}
\end{equation}
\begin{equation}
   D = 1 + s (C_{GMRx} + C_{GMTx} + C_{MG})Z_{BM}
\label{eq12:main:c}
\end{equation}
\begin{equation}
    S = \frac{C_{L(eff.)}}{C_{GMRx}}
    \label{eq12:main:d}
\end{equation}
\begin{equation}
    R = -\frac{C_{xRx} + C_{L(eff.)} + C_{GMRx}}{C_{GMRx}}
    \label{eq12:main:e}
\end{equation}
\begin{equation}
    U = 1 + \frac{C_B}{C_{xTx}}
    \label{eq12:main:f}
\end{equation}
\begin{equation}
    T = \frac{C_{xRx}}{C_{xTx}}
    \label{eq12:main:g}
\end{equation}
\end{subequations}
Similar to grounded metal, during touch-based interaction with floating metal, the transfer characteristics ($T_{TFM}(s)$) can be obtained by replacing $Z_{BM}$ with $Z_{con}$ where $Z_{con}$ is defined in Eq. \ref{eq13} 
\begin{equation}
    Z_{con} = \frac{R_{con}}{1+sR_{con}C_{BM}}
    \label{eq13}
\end{equation}
The detailed derivations for the transfer functions are provided in the Appendix I.

\section{Simulation Setup $\&$ Results}
The following section depicts the descriptions of the FEM-based simulation setup for capacitive EQS-HBC in an environment surrounded by metallic objects.
\begin{figure}[ht]
\centering
\includegraphics[width=0.48\textwidth]{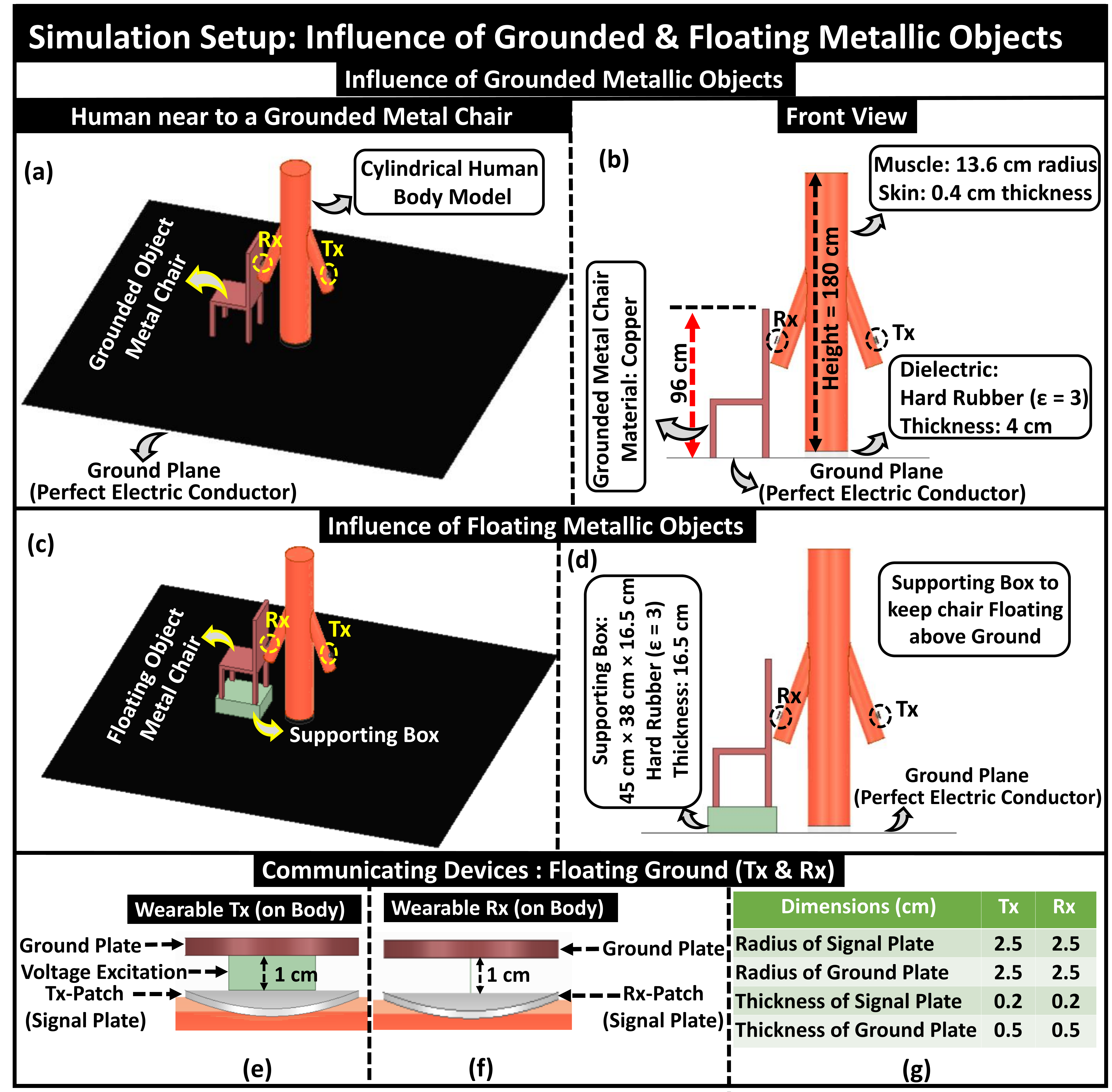}
\caption{Simplified cylindrical human body model for FEM simulations: (a) with a grounded metal chair, (b) front view and structural parameters, (c) with a floating metallic object, and (d) front view of this setup. Structure of devices: (e) wearable Tx, (f) wearable Rx, and (g) their dimensions.}
\label{fig:Sim_Setup_Grounded&Floating}
\end{figure}
\begin{figure}[ht]
\centering
\includegraphics[width=0.48\textwidth]{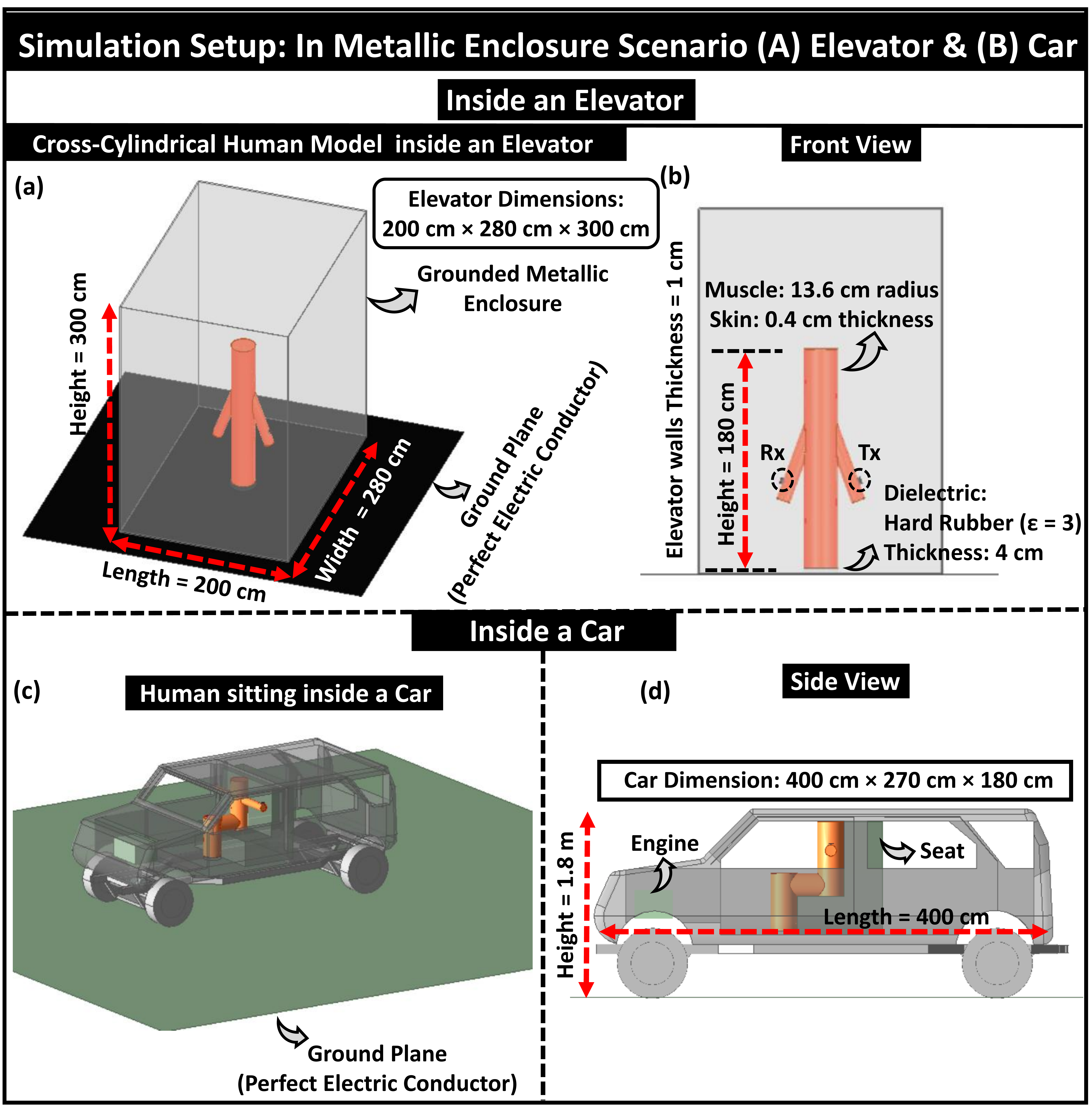}
\caption{Simulation Setup for channel variability in metallic enclosed spaces: Scenarios include (a) a human standing in an elevator, (b) front view of the setup, (c) a human sitting in a car, and (d) side view of the corresponding setup.}
\label{Sim_Setup_Elevator}
\end{figure}
\subsection{Simulation Setup}
A simplified cross-cylindrical human body model is utilized for faster FEM simulations, referencing tissue properties from Gabriel et al. \cite{Gabriel_1996}. Maity et al. \cite{maity2020safety} validated its accuracy by comparing electric and magnetic field distributions with those of the detailed VHP Female v2.2 model from Neva Electromagnetics \cite{neva_model}. The setup features a grounded copper chair on a plane with a perfect electric conductor boundary (intended to behave like an infinite ground plane or earth’s ground) assigned to it and a floating version supported by a rubber box, as shown in Fig. \ref{fig:Sim_Setup_Grounded&Floating} (a, b, c, d). Tx and Rx devices utilize disc-shaped electrodes that contact the skin, with floating ground electrodes, as detailed in Fig. \ref{fig:Sim_Setup_Grounded&Floating} (e, f, g). A copper cage surrounds the model, designed to study HBC in elevator scenarios, is illustrated in Fig. \ref{Sim_Setup_Elevator} (a, b). For variations in HBC channel characteristics in a car measuring 400 cm × 270 cm × 180 cm, the cabin is made of steel and the tires and seats are of rubber, as shown in Fig. \ref{Sim_Setup_Elevator} (c, d). The simulations utilize a Finite Element Boundary Integral (FEBI) method for accuracy, with High Frequency Structure Simulator (HFSS) from Ansys version 2023R2, under an academic license, for quasistatic simulations, and Ansys Maxwell to study coupling capacitance variability.

\subsection{Simulation Results and Discussion:}

\begin{figure}[h!]
\centering
\includegraphics[width=0.48\textwidth]{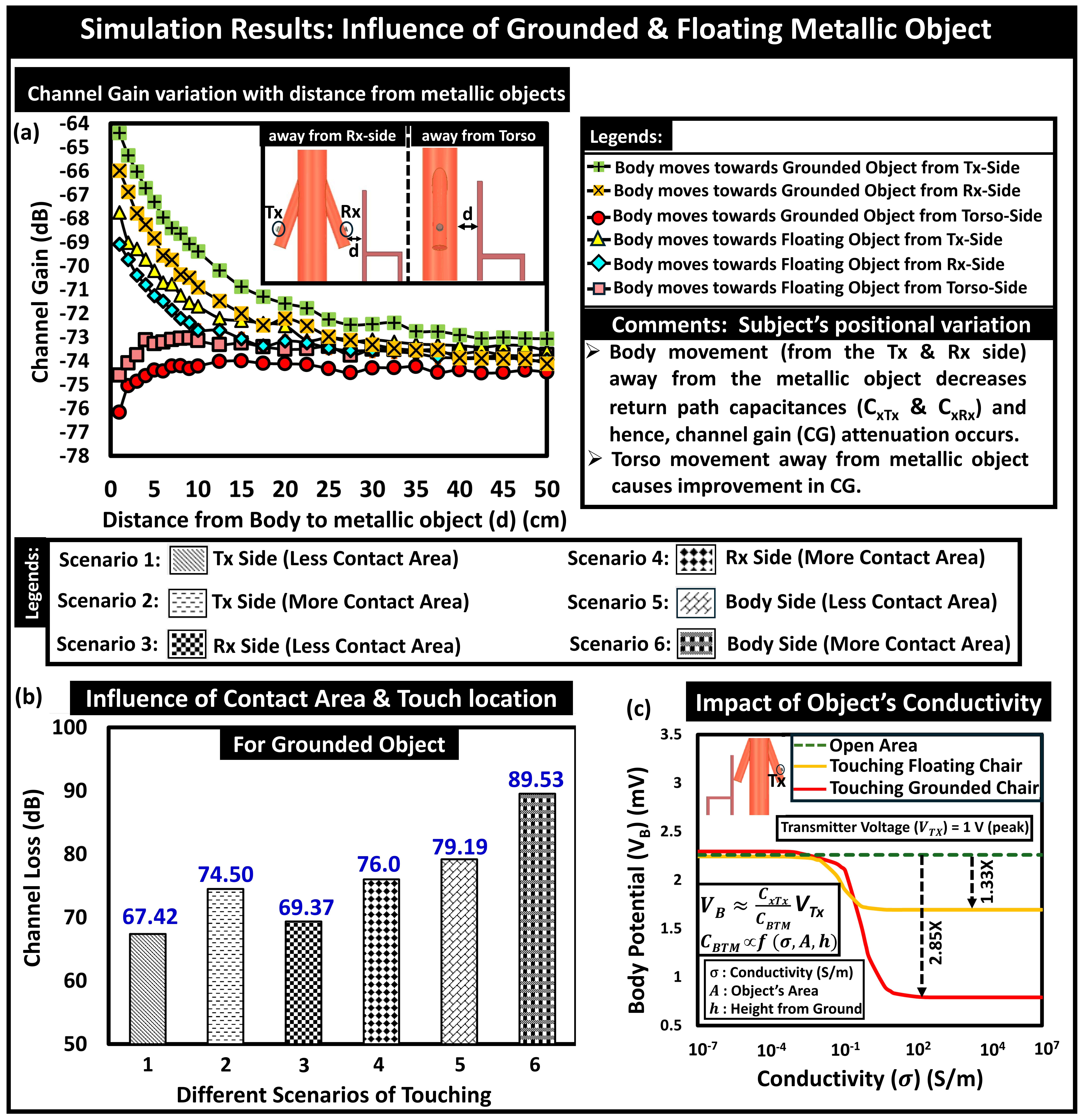}
\caption{Studying the impact of grounded and floating metallic objects on EQS-HBC channels includes key aspects such as: (a) how channel gain varies with distance from metallic objects, which significantly affects performance; (b) analysis of variability due to changes in contact area and touch location during interactions; and (c) the influence of the object's conductivity on body potential during touching scenarios.}
\label{fig:Contact_Area_Simulation_Results}
\end{figure}

The numerical simulation results on metallic objects near an EQS-HBC user's body are shown in Fig. \ref{fig:Contact_Area_Simulation_Results} (a). 
The orientation and distance from these objects significantly affect channel gain within the leakage limit ($\leq$20 cm) of EQS-HBC. Moving the Tx and Rx away from a metallic object reduces the received signal due to lowering in return path capacitance, while the proximity of metals to the user's body increases body capacitance and decreases the received signal. The contact area and its location also influence channel behavior (Fig. \ref{fig:Contact_Area_Simulation_Results} (b)). A larger contact area ($A_{con}$) with lower impedance ($Z_{con}$) leads to higher channel loss. The relationship between body potential ($V_{B}$) and the conductivity ($\sigma$) of the chair is portrayed in Fig. \ref{fig:Contact_Area_Simulation_Results} (c). As $\sigma$ exceeds a threshold ($\sigma_{c}$), $V_{B}$ decreases and stabilizes, depending on whether the object is floating or grounded. The mathematical expression relating $V_{B}$ with $V_{Tx}$ is presented in Eq. \ref{eq14} 
\begin{equation}
   \centering 
   V_{B} \approx \frac{C_{xTx}}{C_{BTM}} V_{Tx}
   \label{eq14}
\end{equation}
The capacitance ($C_{BTM}$) between body-metal and earth ground is influenced by the object's $\sigma$, area ($A$), and height ($h$). Higher contact impedance ($Z_{BM}$) is determined by the conductor's resistivity ($\rho\propto\frac{1}{\sigma}$). Touching a grounded metal chair reduces body potential more than touching a non-grounded chair.

\begin{figure}[h!]
\centering
\includegraphics[width=0.48\textwidth]{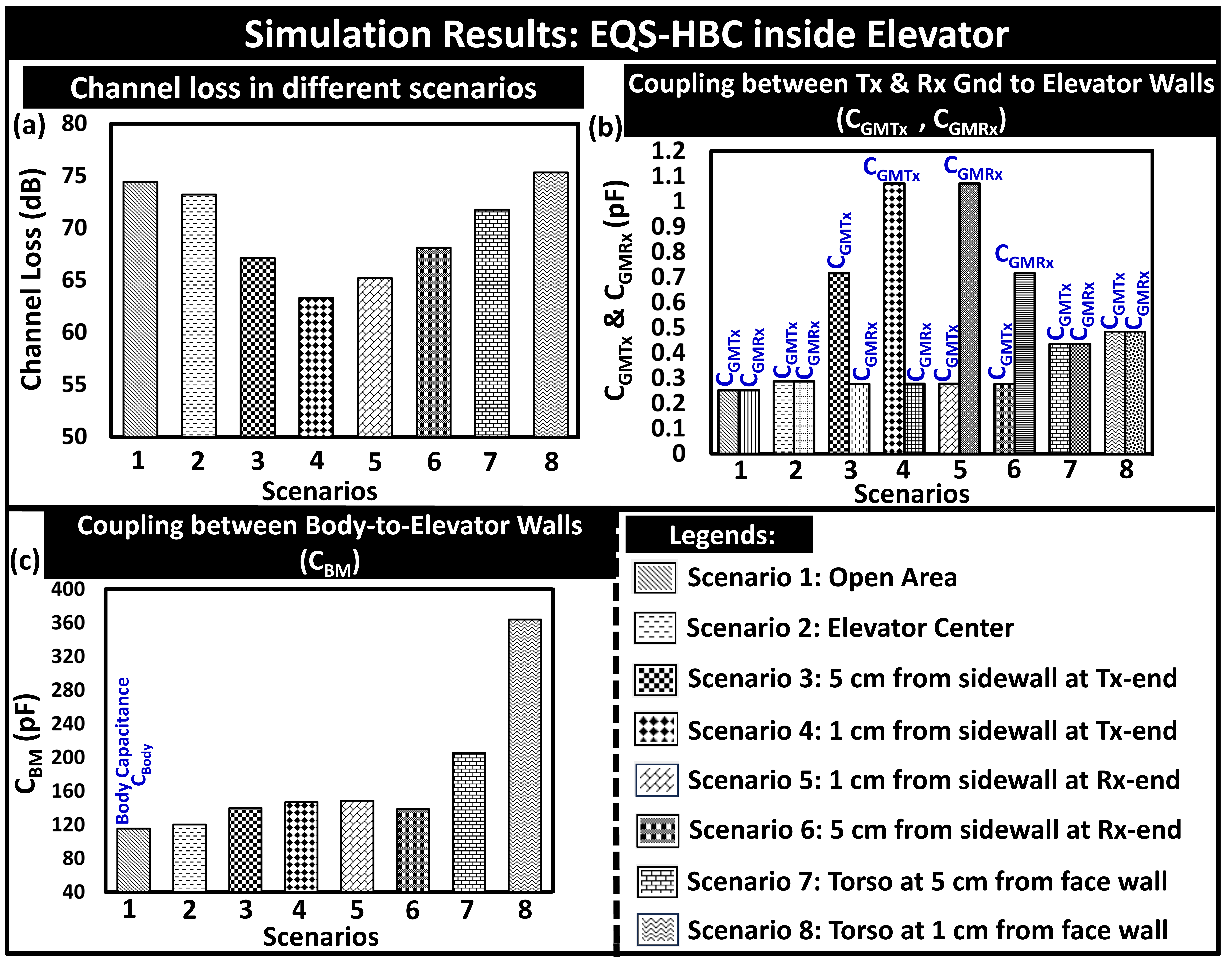}
\caption{EQS-HBC channel variability inside an elevator: (a) compares measured channel loss with the subject's body position under various scenarios, (b) illustrates body-to-metal cage coupling capacitance (C$_{BM}$) variations, and (c) presents the corresponding return path capacitances (C$_{GMTx}$ $\&$ C$_{GMRx}$).}
\label{fig:Sim_Results_Elevator}
\end{figure}

\begin{figure}[h!]
\centering
\includegraphics[width=0.48\textwidth]{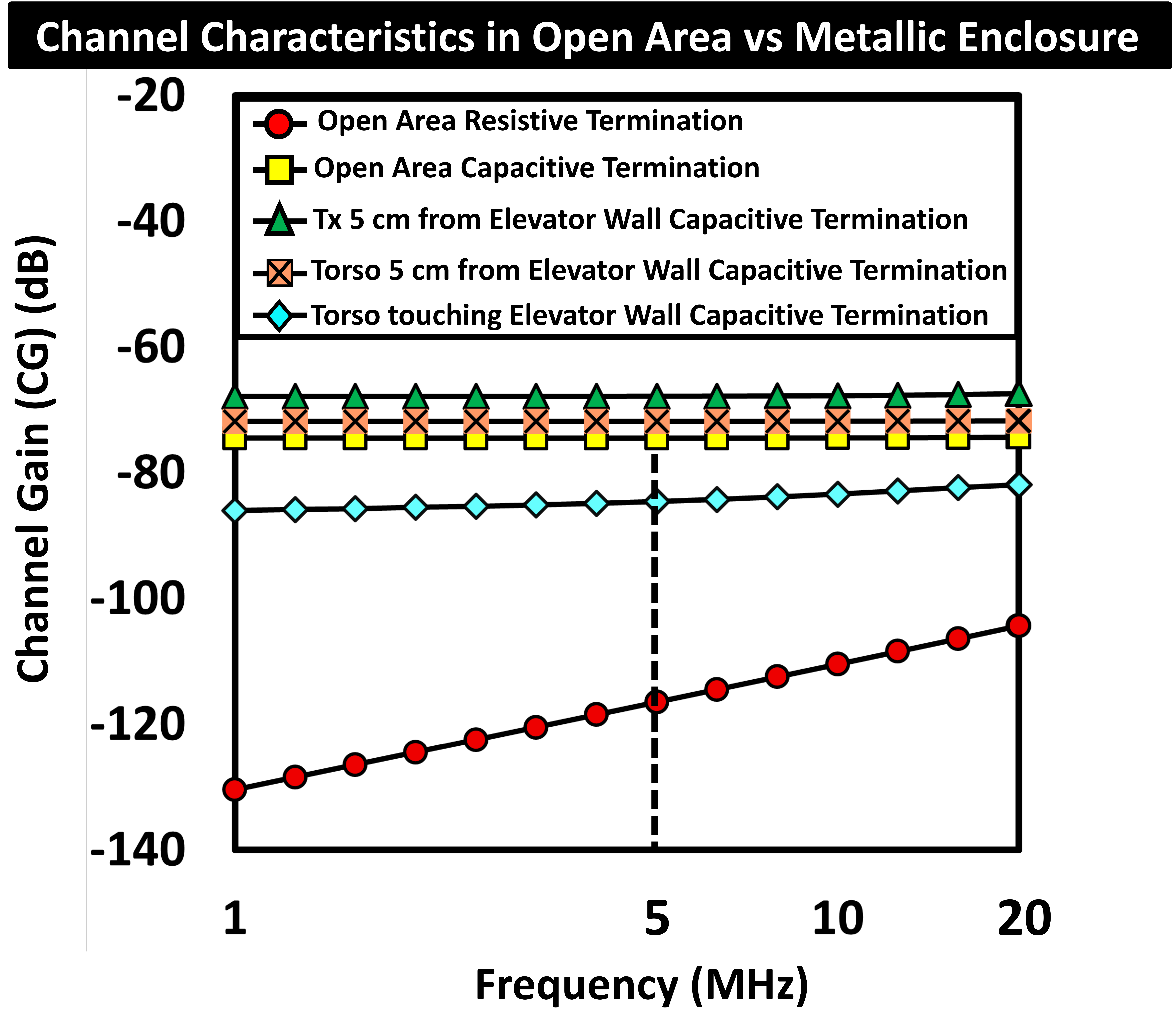}
\caption{Channel characteristics in the EQS frequency regime: highlight the benefits of high impedance capacitive termination at the Rx compared to 50 $\Omega$ resistive termination. Movement from the Tx $\&$ Rx side towards the elevator walls enhances channel gain, but when the torso contacts the wall, it attenuates the received signal while allowing high-pass characteristics through the contact impedance.}
\label{fig:Freq_Sweep_Results}
\end{figure}
\subsection{HBC inside Metallic Enclosure}
\subsubsection{Case Study A: HBC inside an Elevator}
The simulation results in Fig. \ref{fig:Freq_Sweep_Results} demonstrate the advantages of high impedance capacitive termination (Z$_{C_{L (eff.)}}$) in voltage mode communication compared to low impedance (R$_L$ = 50 $\Omega$) resistive termination in the EQS. Capacitive termination at the receiver end provides frequency-independent channel loss, whereas 50$\Omega$ termination exhibits a 20 dB/decade increase in transmission at lower frequencies. In the simulation, the presence of a grounded metallic enclosure around the human body enhances effective capacitive coupling, resulting in an approximately 5 dB increase in channel gain compared to an ideal open-space scenario. Changes in the body position within the elevator affect channel loss due to variations in coupling between the body and the surrounding environment. As the body moves toward the elevator walls, return path capacitance (C$_{GMTx}$ or C$_{GMRx}$) improves channel gain, but increased body-to-metal cage coupling (C$_{BM}$) can limit this gain. Variations in these capacitances are shown in Fig. \ref{fig:Sim_Results_Elevator} (a, b, c). When the body touches the elevator walls, the received signal diminishes significantly due to the low-impedance path to ground at the contact location. 

\subsubsection{Case Study B: HBC inside a Car}
This section examines the operational variability of capacitive EQS-HBC within a non-grounded metallic enclosure, such as a car. When a subject is seated in the car's center and communication devices are on their body, the surrounding metal frame increases transmission loss due to coupling effects. 
\begin{figure}[h!]
\centering
\includegraphics[width=0.48\textwidth]{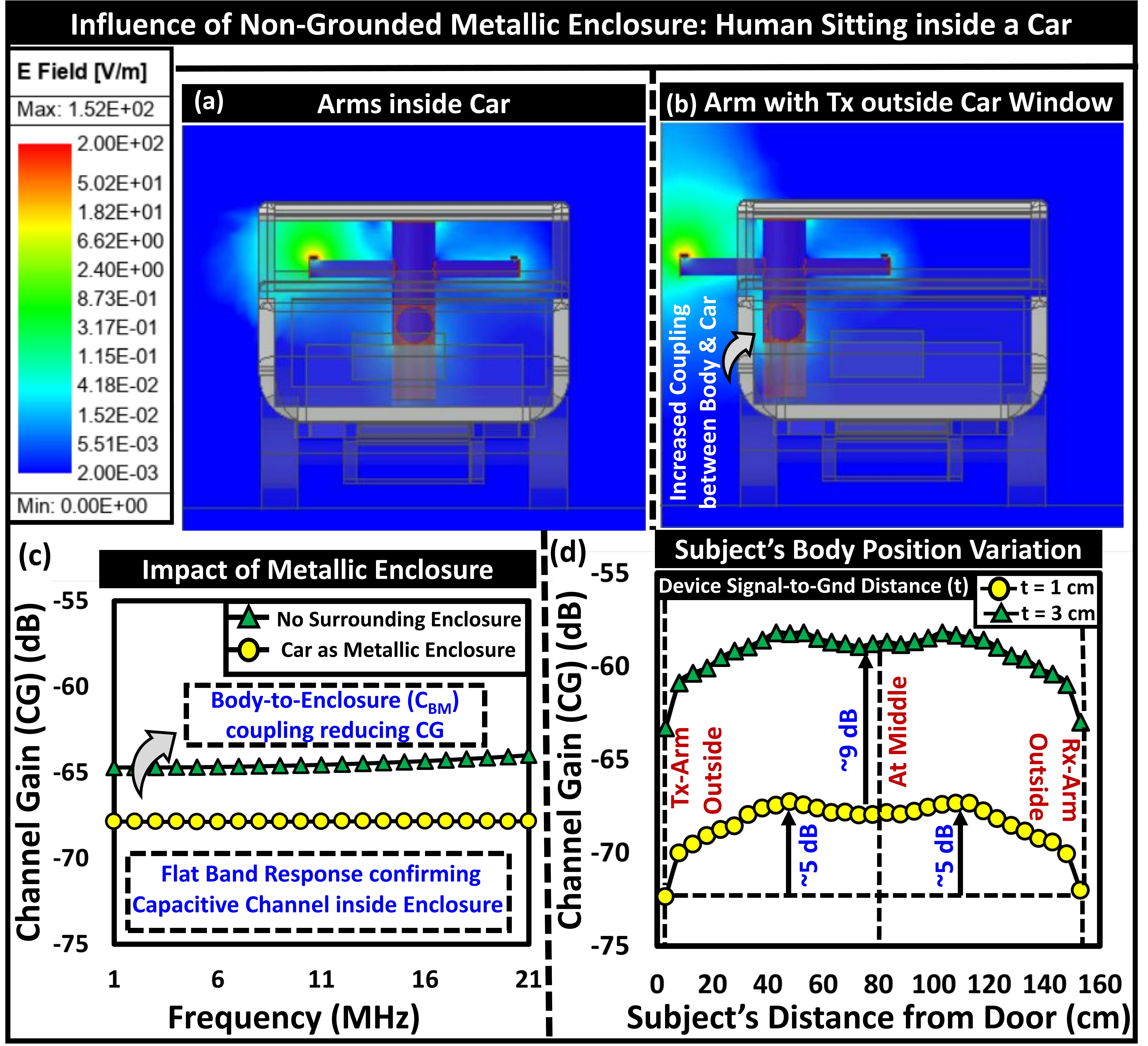}
\caption{Analyzing the impact of a non-grounded metallic enclosure: EQS-HBC user inside a car. It compares electric field (E-field) plots for a person sitting (a) at the center of the car and (b) by the door with their arm extended outside. Additionally, (c) shows the effective flat-band channel characteristic despite the metallic frame, while (d) illustrates how channel gain varies with the user's position.}
\label{fig:SimulationResult_Car}
\end{figure}

The flat-band nature of the channel response, shown in Fig. \ref{fig:SimulationResult_Car} (c), illustrates the capacitive behavior of the communication channel in a metal enclosure. Factors influence capacitive coupling: the subject's posture and the device's placement relative to the car's structure. For example, when a subject in a T-pose (posture with arm extended) adjusts their arms closer to the car doors, signal levels improve due to enhanced coupling in the return paths (C$_{GMTx}$ or C$_{GMRx}$). Transitioning from being fully inside the vehicle to extending an arm outside can shift from a capacitive to a galvanic mode. The transmission loss may increase despite a decrease in electric field lines reaching the ground, as proximity to the metal frame matters. Fig. \ref{fig:SimulationResult_Car} (b, d) shows about 5 dB channel gain attenuation when the torso is roughly 3 cm from the car door, with larger devices yielding about a 9 dB benefit from greater signal-to-ground separation.

\begin{figure}[ht]
\centering
\includegraphics[width=0.48\textwidth]{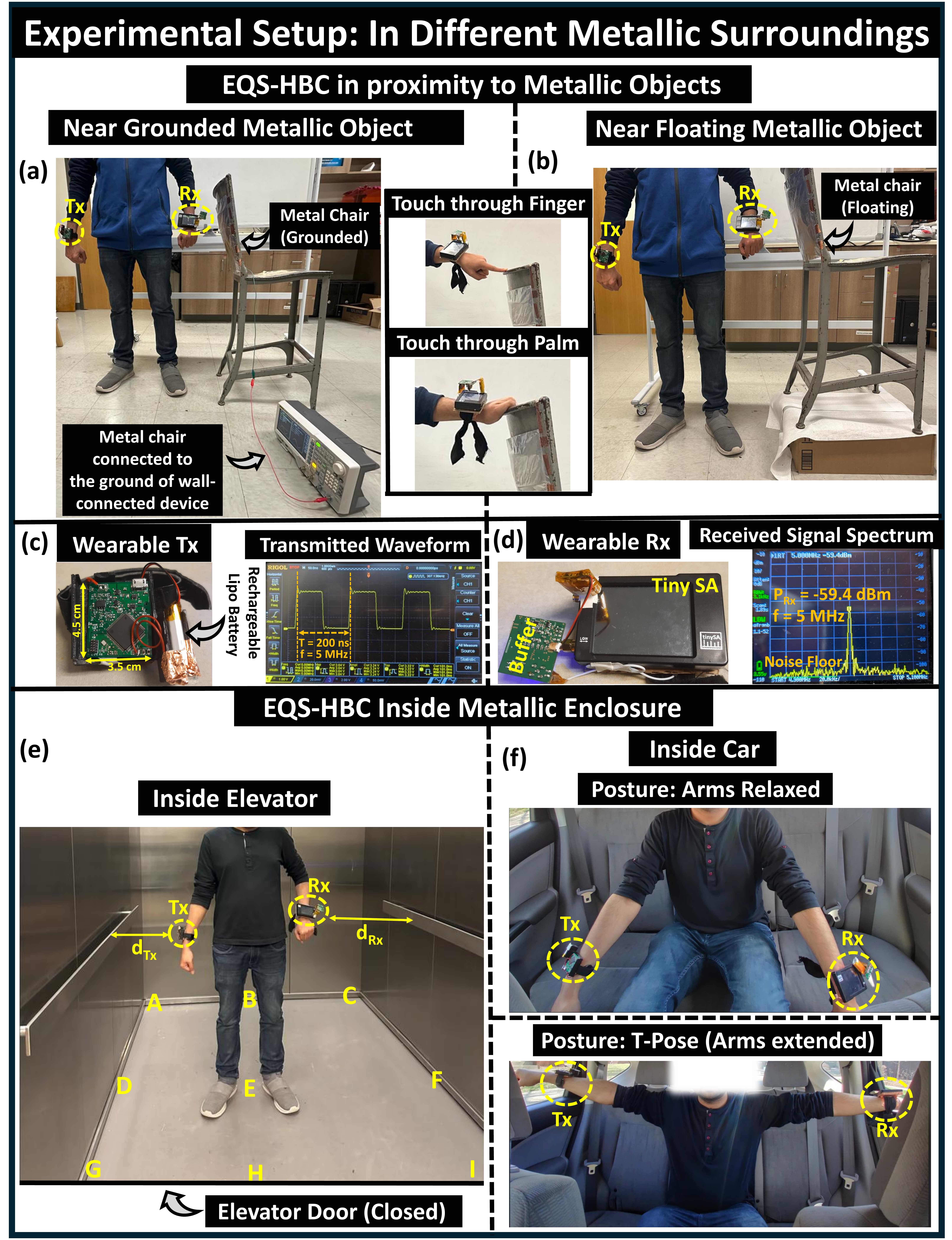}
\caption{Experimental Setup: Influence of nearby metallic objects: Subject with on-body communicating devices (Tx $\&$ Rx) stands near (a) a grounded metallic object, (b) a non-grounded or floating metallic object. Variation in contact area while touching through finger and palm. Wearable Communicating Devices: (c) Tx Setup, (d) Rx Setup (tinySA spectrum analyzer with buffer for high impedance capacitive termination), Channel variability inside metallic enclosed surroundings: (e) Inside an elevator, (f) Inside a Car. }
\label{fig:Experimental_Setup}
\end{figure}

\section{Experimental Setup $\&$ Results}
This section summarizes the experimental setup and results from battery-powered wearable devices, focusing on communication performance in metallic environments. It examines interactions with grounded and non-grounded metallic objects (Fig. \ref{fig:Experimental_Setup} (a, b)) and within metallic enclosures like elevators and cars (Fig. \ref{fig:Experimental_Setup} (e, f)). 
\subsubsection{Tx Setup} 
We designed a customized printed circuit board (PCB) measuring 3.5 cm × 4.5 cm, featuring an embedded NXP LPC55S6x ARM Cortex Microcontroller unit (MCU). This PCB serves as the wearable Tx in the EQS regime, as shown in Fig. \ref{fig:Experimental_Setup} (c). It is powered by a compact 3.7 V rechargeable battery housed in a 3D-printed enclosure. One GPIO pin of the MCU is programmed to generate a 3.3 V (peak-to-peak) signal at 5 MHz, which connects to the signal coupler.

\subsubsection{Rx Setup} 
\paragraph{Portable Spectrum Analyzer} For the wearable signal receiver in the EQS regime, we used a handheld spectrum analyzer (tinySA Basic), which operates from 100 kHz to 960 MHz. The signal electrode at the receiver employs copper tape for optimal skin contact. To accommodate the device's 50 $\Omega$ input impedance, we added a customized buffer at the input for high-impedance measurements. 
\paragraph{Buffer Setup} We designed a buffer using an operational amplifier (opamp: OPA2836) from Texas Instruments, with an input capacitance of $\sim$2 pF to capacitively load the receiver. The buffer output connects to the input of the spectrum analyzer, with the grounds of both devices forming an effective Rx ground area of 20 cm$^2$. The calibration details of the devices are provided in Appendix III \ref{appendix:c}.

\subsubsection{Experimental Procedure} 
 Channel loss measurements were conducted in various environmental settings, including a standard laboratory, an elevator, and a car. Informed consent was obtained from the participant, and the experiments involving human subjects complied with the guidelines approved by the Institutional Review Board (IRB Protocol 1610018370). Due to the dependence of return path capacitance on electrode orientation and the subject's body posture, we confirmed the trends in the experimental results through consistent measurements while maintaining steady body posture and device placement. To reduce over-the-air device-to-device coupling, the wearable devices were positioned on opposite arms of the subject. In the grounded metal scenario, we used a metal chair (approximately 96 cm tall with a 45 cm$\times$45 cm sitting area) connected to the ground of a wall-connected instrument to establish the setup shown in Fig. \ref{fig:Experimental_Setup} (a) i.e., for our experiments, the ground of the wall socket is considered to represent the earth’s ground. To emulate the floating metal scenario, while removing the ground connection, we elevated the chair by placing it on a cut-board box (45.72 cm $\times$ 35.56 cm $\times$ 16.51 cm) to create an approximate 16.51 cm gap from the ground, as depicted in Fig. \ref{fig:Experimental_Setup} (b). For the EQS-HBC measurements inside the Elevator, the subject with body-worn Tx and Rx changes their position relative to the elevator walls. Within the car, the subject remains seated with the on-body communicating devices, as illustrated in Fig. \ref{fig:Experimental_Setup} (e, f). 
To experimentally assess the impact of metallic objects on body channel performance in the EQS regime, the coupled voltage onto the human body from the wearable Tx is capacitively received by a wearable receiver setup. Voltage measurements are taken by converting the recorded power levels ($P_{Rx}$ (dBm)) from a TinySA spectrum analyzer with a buffer at its input into the peak received voltage $V_{Rx}$ (peak). After accounting for calibrations of the wearable transmitter, receiver, and buffer, the channel gain is calculated as follows: Channel Gain (G) = $20 \log_{10} \left(\frac{V_{Rx}(peak-peak)}{V_{Tx}(peak-peak)}\right)$.

We have ensured the statistical significance of our measurements by obtaining consistent data points while repeating each experiment over several weeks. The experimental datasets are plotted for the nominal and repeatable scenarios with statistical conformity provided in terms of plotting the mean value of the data points and standard deviation (shown with the error bars). A comparative analysis of the results was performed to identify the key factors crucial in establishing robust body-centric communication links.

\begin{figure}[h!]
\centering
\includegraphics[width=0.48\textwidth]{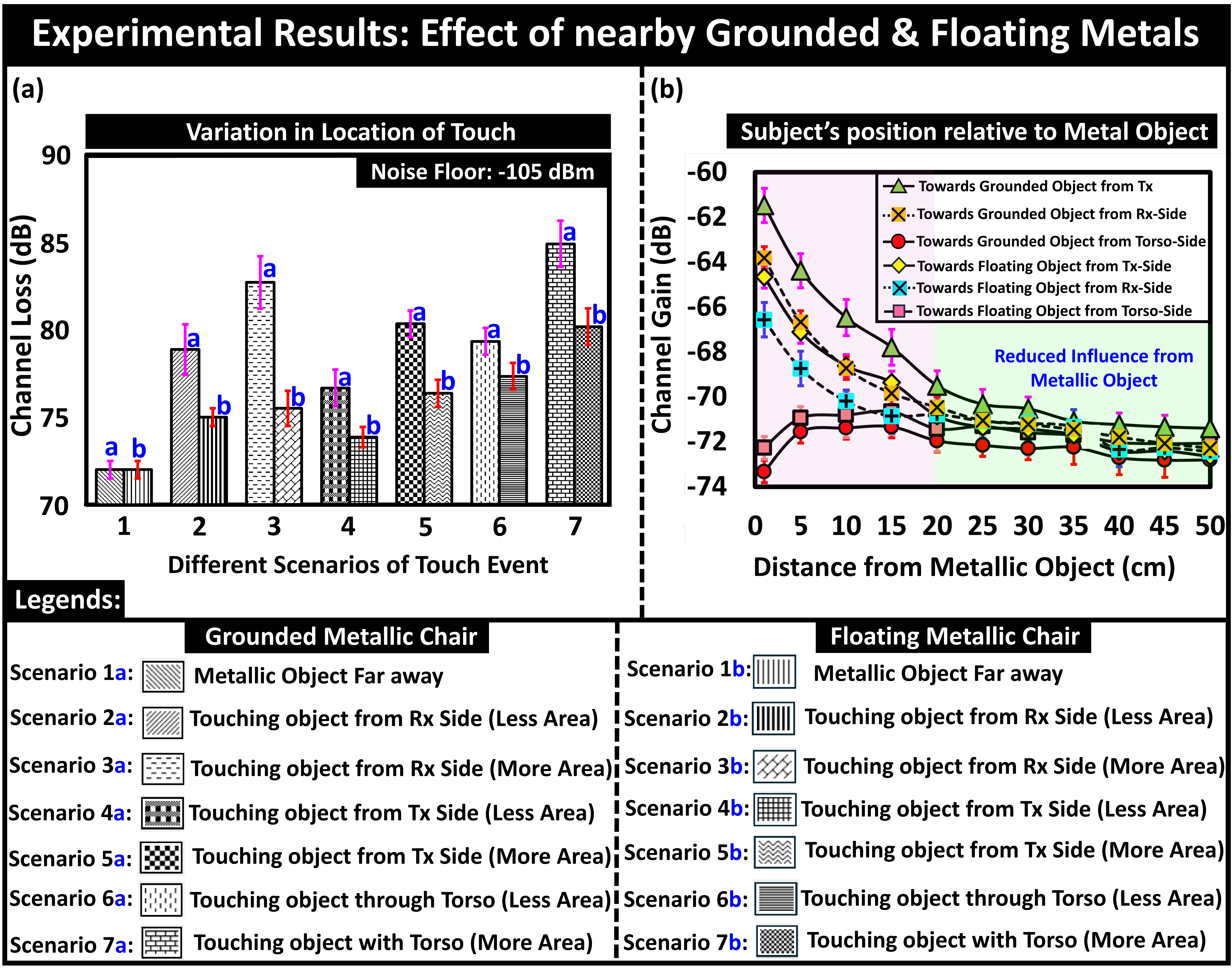}
\caption{Experimental results depict the influence of grounded and floating metallic objects on the EQS-HBC channel characteristics under different touching and non-touching scenarios.}
\label{fig:Experimental_Results-1}
\end{figure}

\subsection{Experimental Results}
During touch-based interactions with metallic objects, an increase in contact area leads to higher channel loss, as shown in Fig. \ref{fig:Experimental_Results-1} (a). This is due to reduced contact impedance, aligning with results from FEM simulations. Proximity of grounds of the Tx $\&$ Rx to grounded or floating metallic objects enhances the received signal by increasing return path capacitance, with greater improvements observed at shorter distances (within leakage limit $\leq$20 cm) to the metal chair, as illustrated in Fig. \ref{fig:Experimental_Results-1} (b). However, having metal objects close to the subject instead of the devices results in attenuated signal levels due to increased body capacitance.

\begin{figure}[h!]
\centering
\includegraphics[width=0.48\textwidth]{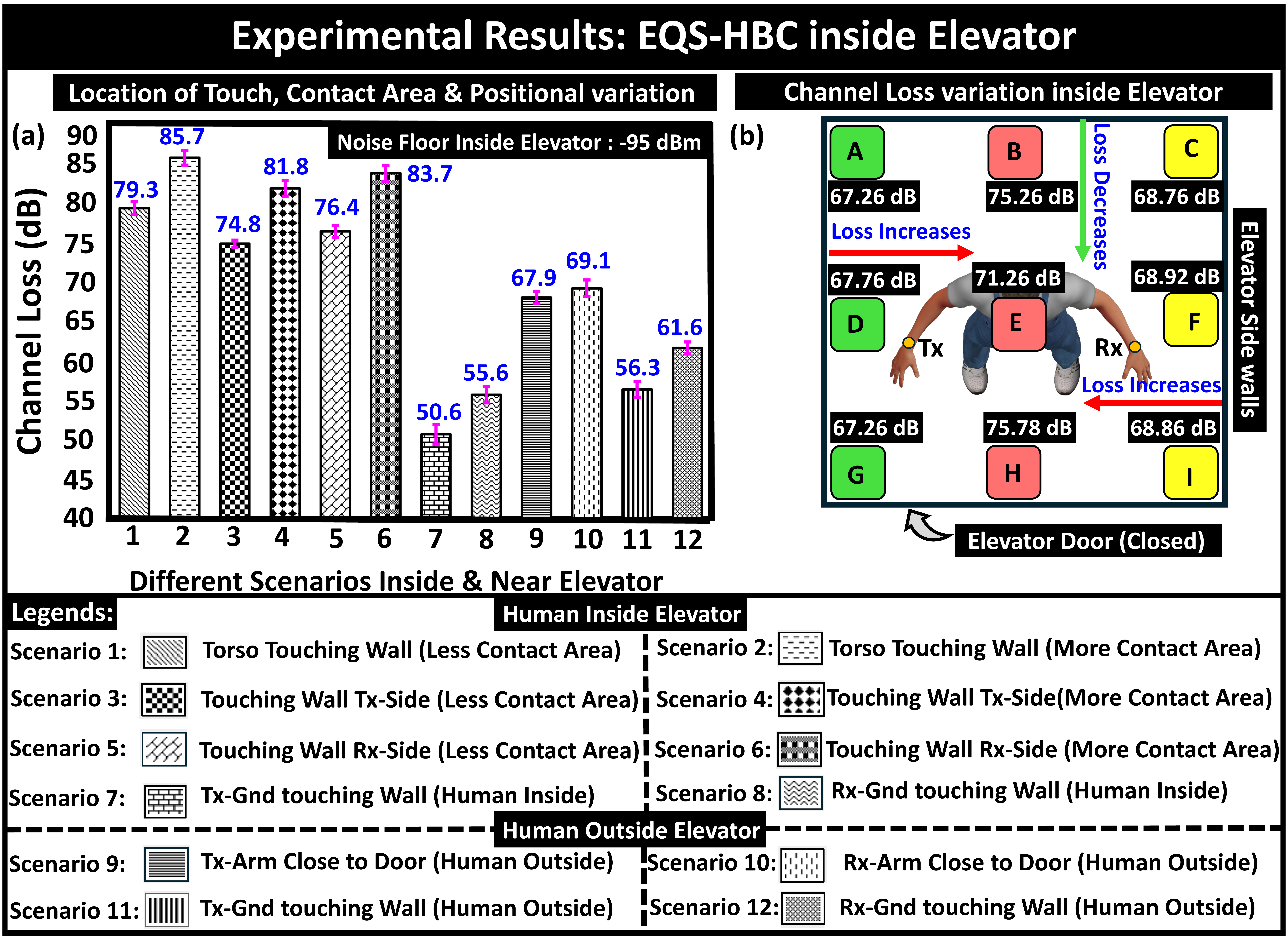}
\caption{Comparison of results from EQS-HBC channel loss measurements inside an elevator scenario. Studying the channel variability with positional variation of the subject's body relative to the elevator walls under touching and non-touching scenarios.}
\label{fig:Experimental_Results-2}
\end{figure}

Inside a metallic enclosure like an elevator, channel loss varies with touch location and contact area, as seen in Fig. \ref{fig:Experimental_Results-2} (a). Positions closer to the elevator walls (B and H) result in higher channel loss due to more increase in $C_{Body}$ in comparison to $C_{ret}$, while staying near the side walls (A, D, G, C, F, I) yields lower loss due to an increased return path, depicted in Fig. Fig. \ref{fig:Experimental_Results-2} (b). The channel losses in the elevator at positions A, G, C, and I are approximately similar to those at positions D and F but lower than at positions B and H. The fundamental reason is that as the subject's torso gets closer to the elevator walls, the capacitive coupling between the user's body and the elevator increases more than the return path capacitances at the Tx and Rx. However, at corner locations or near the elevator sidewalls, the increase in return path capacitances becomes more significant than the increase in body capacitance, resulting in reduced channel loss. Confirming this trend, the capacitance variations previously shown in Fig. \ref{fig:Sim_Results_Elevator} effectively illustrate how the coupling between Body-to-Elevator Walls and ground of the Tx/Rx to Elevator Walls changes as the subject changes its position inside the elevator.

\begin{figure}[h!]
\centering
\includegraphics[width=0.48\textwidth]{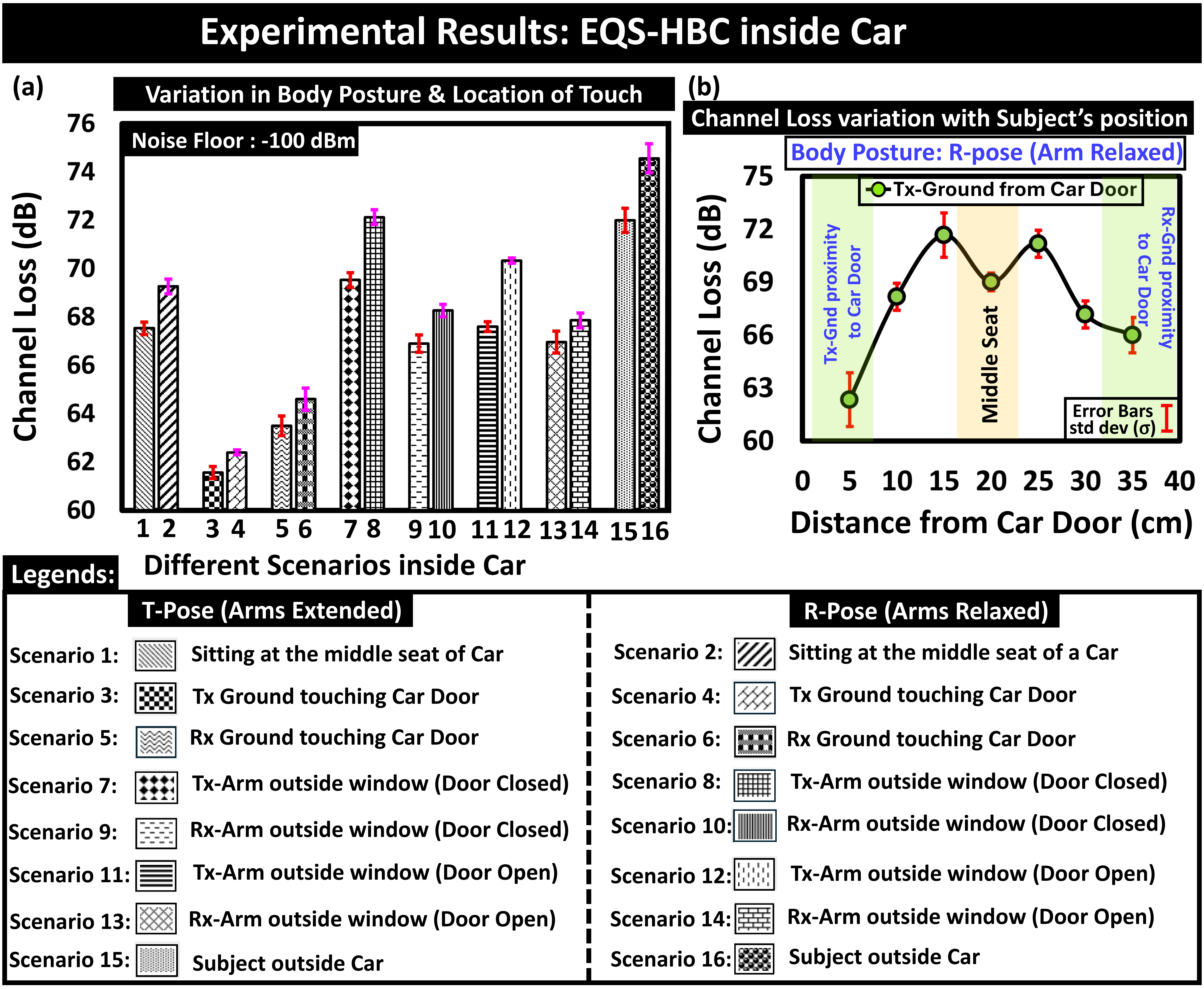}
\caption{Comparative analysis of the measured channel loss inside a car scenario with subject's posture variation and the variability resulting from the location of the on-body devices relative to the car body.}
\label{fig:Experimental_Results-3}
\end{figure}

In the car scenario, channel loss increases due to proximity to the car cabin. Body posture significantly affects path loss; a relaxed posture (R pose i.e., arms near the torso) shows higher loss than a T-pose (i.e., posture with arm extended, shown Fig. \ref{fig:Experimental_Results-3} (a)) due to higher body shadowing. Channel gains are greater when the device grounds touch the car door. However, the improvement is less pronounced than in elevators due to the presence of dielectric materials that prevent direct contact with the ground. Extending an arm outside the car window may seem to enable a capacitive transmission (Tx) and galvanic reception (Rx) mode of communication; however, unbalanced return path capacitances result in a dominant capacitive behavior \cite{modak2022bio}. Open car doors experience lower path loss due to reduced body capacitance. The variation in subject’s position inside the car
body is illustrated in Fig. \ref{fig:Experimental_Results-3} (b). Discrepancies between simulation and experimental results arise from the effective communication device area, with optimized designs likely to further improve channel gain. 

\begin{figure}[h!]
\centering
\includegraphics[width=0.48\textwidth]{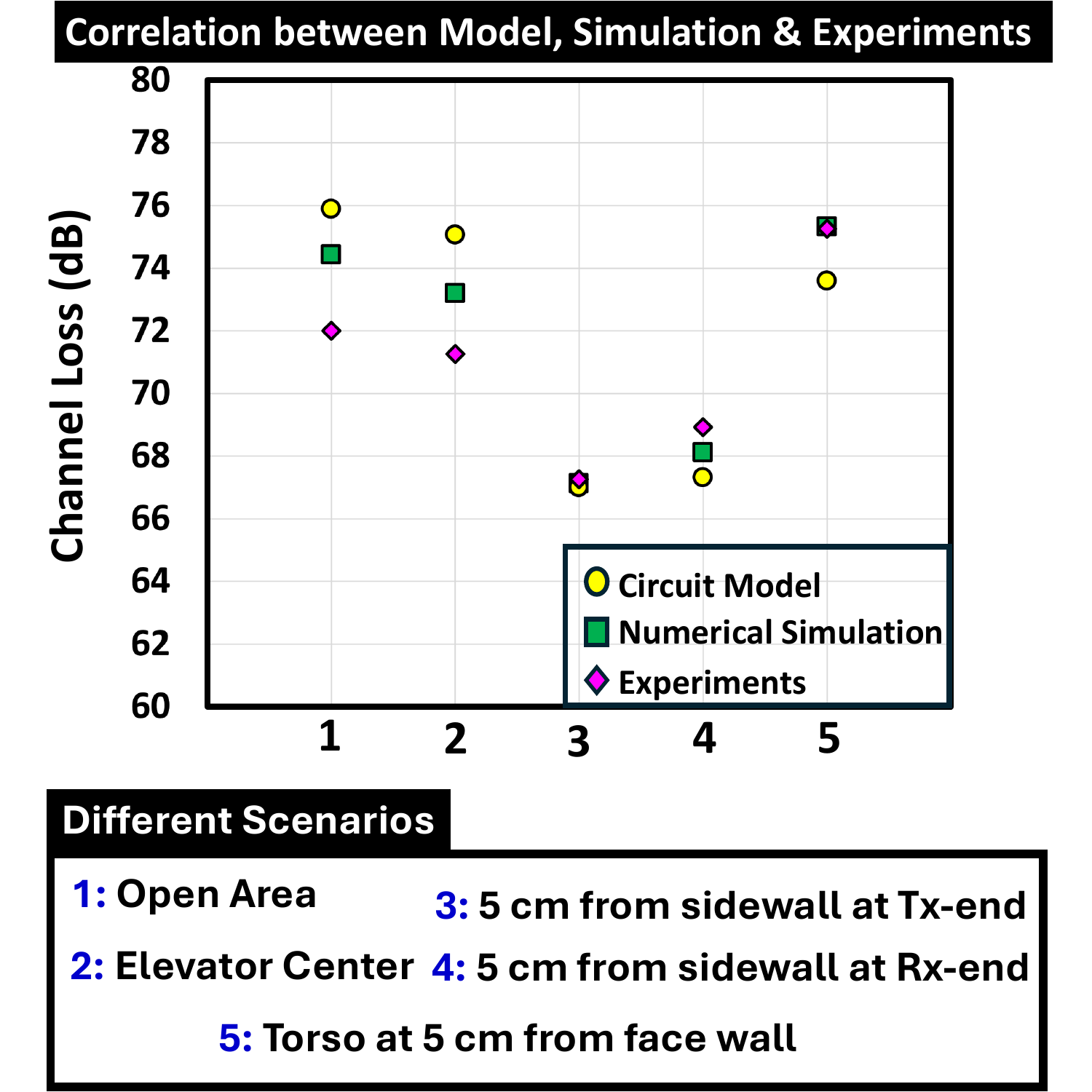}
\caption{Correlation between the Circuit Model, Numerical Simulation $\&$ Experiments for different scenarios inside elevator.}
\label{fig:Correlation}
\end{figure}

\section{Correlation among model, simulation, and experiments}
Here, we present a scatter plot that illustrates the correlation among the proposed circuit model, numerical simulations, and experimental results across various scenarios for the EQS-HBC user inside an elevator, as shown in Fig. \ref{fig:Correlation}. This analysis encompasses a variety of scenarios: Scenario 1 corresponds to measurements conducted in an open area, providing a baseline for comparison. In contrast, Scenarios 2, 3, 4, and 5 explore the effects of human presence within an elevator. 
Based on the scatter plot, we can interpret that the trend of variation among the circuit model, numerical simulations, and experiments is consistent. Specifically, when the subjects are positioned closer to the elevator walls at the transmitter (Tx) and receiver (Rx) sides, there is an increase in the return path capacitance, resulting in reduced channel loss, as observed in scenarios 3 and 4. However, it is important to note that the numerical values of channel loss vary slightly. This discrepancy arises from differences in the concept of the Earth's ground used in the simulation model compared to the experiments, as well as variations in the size of the communication devices that affect the return path capacitance.

An EQS body communication channel, which can be modeled as a purely passive, linear, time-invariant (LTI) network of resistors and capacitors, is reciprocal in the transimpedance sense ($Z_{21} = Z_{12}$). This is due to the symmetry of its admittance matrix, regardless of various subject-metal interactions (such as grounded or floating metal or events based on proximity or touch). However, apparent non-reciprocity may arise from mismatched terminations or time variance due to touch. With different Tx and Rx terminations, this can occur in voltage mode EQS signaling or if the network configuration changes between forward and reverse tests. Consequently, the channel gain may differ between forward and reverse paths, leading to apparent non-reciprocity in voltage transfer ($\frac{V_{Rx}}  {V_{Tx}}  \neq \frac{V_{Tx}}  {V_{Rx}}$)  This discrepancy arises from unequal source or load terminations or time-varying contact. Additionally, if the electrode-skin interface enters a nonlinear regime (such as saturation, DC offset, or polarization), strict reciprocity may no longer hold. This issue can be mitigated by ensuring operation within the linear range of the electrode-skin impedance.
\section{Discussion}
In this proof-of-concept study, we investigated how surrounding conductive objects affect capacitive EQS-HBC by influencing parasitic return paths between devices, the user’s body, and the environment. A metal object's high conductivity ($\sigma$ $\sim$ $10^7$ S/m) introduces a low-impedance node in the return path, increasing return-path capacitance ($C_{ret}$) and reducing impedance ($Z_{ret}$ $\approx$ $\frac{1}{j \omega C_{ret}}$), ultimately lowering transmission loss. Key findings include: (a) Conductive objects near the transmitter (Tx) or receiver (Rx) enhance parasitic return-path coupling ($C_{retTx}$ and $C_{retRx}$), improving channel gain. (b) Objects near the body raise body capacitance, which can attenuate on-body received signal levels. (c) Grounded metallic objects with higher $C_{ret}$ impact channel performance more than floating ones. (d) The contact area during user interactions with grounded objects significantly affects channel characteristics. Overall, ambient metals can either aid or hinder EQS-HBC performance based on their location, distance, and size.
To develop robust, adaptive HBC transceivers, the following studies can be pursued: To address the challenges posed by the presence of multiple surrounding objects, our future work will focus on:
\textbf{a.} Performance Analysis: Investigating the performance of body-centric HBC in rooms with multiple metallic fixtures, particularly how overlapping return-path capacitances from various conductive objects affect performance.
\textbf{b.} Dynamic Body-Metal Interactions: Utilizing subject motion-capture data and time-domain channel measurements to examine transient coupling effects from conducting objects during regular activities.
\textbf{c.} Broadband Characterization: Extending the operational range of the body communication link beyond the EQS regime, especially when the structural resonances of the human body and metallic objects become significant. Large metal structures may exhibit self-resonance in the EQS context, and their distributed capacitance and inductance can form LC resonances that introduce narrow-band notches, potentially limiting the channel capacity of EQS-HBC.
These studies aim to establish a more reliable and energy-efficient link between the human body and its environment.

\section{Conclusion}
This paper explores the variability in body channel characteristics associated with capacitive Human Body Communication (HBC) in the presence of surrounding metallic objects within the Electro-Quasistatics (EQS) frequency regime. It addresses how these objects impact channel loss due to parasitic capacitances in the return path, analyzing both grounded and non-grounded metals across various scenarios, including environments such as elevators and vehicles. The study utilizes biophysical models to assess these influences. Our findings indicate that when subjects stay close to metallic objects ($\leq$ 20 cm from devices and $\leq$5 cm from the user), transmission loss can vary by $\sim$10 dB. This variation can exceed $\sim$20 dB when the ground of the devices contacts the metallic object. Grounded metals have a greater impact on body-channel performance compared to non-grounded ones due to stronger coupling with the earth's ground. Improvement in channel gain inside the car ($\leq$7 dB) remains lower than the improvement in channel gains observed in elevators ($\leq$21 dB). The variability in channel performance is influenced by factors such as device size, subject orientation, and proximity to metallic objects. This research highlights the potential of capacitive EQS-HBC in various settings and may contribute to the design of energy-efficient wearable devices that enhance Human-Machine Interaction.

\section*{Acknowledgement}
This work was supported by Quasistatics Inc. under Grant 40003567. The authors thank Meghna Roy Chowdhury, Ph.D. student at SparcLab, for her help during the experiments.

\bstctlcite{BSTcontrol}
\bibliographystyle{IEEEtran}
\bibliography{references}

\begin{thebibliography}{10}
\providecommand{\url}[1]{#1}
\csname url@samestyle\endcsname
\providecommand{\newblock}{\relax}
\providecommand{\bibinfo}[2]{#2}
\providecommand{\BIBentrySTDinterwordspacing}{\spaceskip=0pt\relax}
\providecommand{\BIBentryALTinterwordstretchfactor}{4}
\providecommand{\BIBentryALTinterwordspacing}{\spaceskip=\fontdimen2\font plus
\BIBentryALTinterwordstretchfactor\fontdimen3\font minus \fontdimen4\font\relax}
\providecommand{\BIBforeignlanguage}[2]{{%
\expandafter\ifx\csname l@#1\endcsname\relax
\typeout{** WARNING: IEEEtran.bst: No hyphenation pattern has been}%
\typeout{** loaded for the language `#1'. Using the pattern for}%
\typeout{** the default language instead.}%
\else
\language=\csname l@#1\endcsname
\fi
#2}}
\providecommand{\BIBdecl}{\relax}
\BIBdecl

\bibitem{seyedi2013survey}
M.~Seyedi, B.~Kibret, D.~T. Lai, and M.~Faulkner, ``A survey on intrabody communications for body area network applications,'' \emph{IEEE Transactions on Biomedical Engineering}, vol.~60, no.~8, pp. 2067--2079, 2013.

\bibitem{hasan2019comprehensive}
K.~Hasan, K.~Biswas, K.~Ahmed, N.~S. Nafi, and M.~S. Islam, ``A comprehensive review of wireless body area network,'' \emph{Journal of Network and Computer Applications}, vol. 143, pp. 178--198, 2019.

\bibitem{park2019sub}
J.~Park and P.~P. Mercier, ``A sub-10-pj/bit 5-mb/s magnetic human body communication transceiver,'' \emph{IEEE Journal of Solid-State Circuits}, vol.~54, no.~11, pp. 3031--3042, 2019.

\bibitem{wen2021channel}
E.~Wen, D.~F. Sievenpiper, and P.~P. Mercier, ``Channel characterization of magnetic human body communication,'' \emph{IEEE Transactions on Biomedical Engineering}, vol.~69, no.~2, pp. 569--579, 2021.

\bibitem{hernandez2017magnetic}
Y.~K. Hern{\'a}ndez-G{\'o}mez, G.~{\'A}lvarez-Botero, J.~Rodr{\'\i}guez, and F.~R. de~Sousa, ``Magnetic human body communication based on double-inductor coupling,'' in \emph{2017 IEEE International Instrumentation and Measurement Technology Conference (I2MTC)}.\hskip 1em plus 0.5em minus 0.4em\relax IEEE, 2017, pp. 1--6.

\bibitem{nath2022understanding}
M.~Nath, A.~K. Ulvog, S.~Weigand, and S.~Sen, ``Understanding the role of magnetic and magneto-quasistatic fields in human body communication,'' \emph{IEEE Transactions on Biomedical Engineering}, vol.~69, no.~12, pp. 3635--3644, 2022.

\bibitem{denisov2010ultrasonic}
A.~Denisov and E.~Yeatman, ``Ultrasonic vs. inductive power delivery for miniature biomedical implants,'' in \emph{2010 International Conference on Body Sensor Networks}.\hskip 1em plus 0.5em minus 0.4em\relax IEEE, 2010, pp. 84--89.

\bibitem{haerinia2020wireless}
M.~Haerinia and R.~Shadid, ``Wireless power transfer approaches for medical implants: A review,'' \emph{Signals}, vol.~1, no.~2, pp. 209--229, 2020.

\bibitem{santagati2017implantable}
G.~E. Santagati and T.~Melodia, ``An implantable low-power ultrasonic platform for the internet of medical things,'' in \emph{IEEE INFOCOM 2017-IEEE Conference on Computer Communications}.\hskip 1em plus 0.5em minus 0.4em\relax IEEE, 2017, pp. 1--9.

\bibitem{zhang2017bioacoustics}
C.~Zhang, S.~Hersek, Y.~Pu, D.~Sun, Q.~Xue, T.~E. Starner, G.~D. Abowd, and O.~T. Inan, ``Bioacoustics-based human-body-mediated communication,'' \emph{Computer}, vol.~50, no.~2, pp. 36--46, 2017.

\bibitem{bos2018enabling}
T.~Bos, W.~Jiang, J.~D’hooge, M.~Verhelst, and W.~Dehaene, ``Enabling ultrasound in-body communication: Fir channel models and qam experiments,'' \emph{IEEE transactions on biomedical circuits and systems}, vol.~13, no.~1, pp. 135--144, 2018.

\bibitem{park2024recent}
W.~Park, J.~Lee, W.~G. Chung, I.~Jeong, E.~Kim, Y.~W. Kwon, H.~Seo, K.~Lim, E.~Kim, and J.-U. Park, ``Recent advances in wireless energy transfer technologies for body-interfaced electronics,'' \emph{Nano Energy}, vol. 124, p. 109496, 2024.

\bibitem{ghanbari2019sub}
M.~M. Ghanbari, D.~K. Piech, K.~Shen, S.~F. Alamouti, C.~Yalcin, B.~C. Johnson, J.~M. Carmena, M.~M. Maharbiz, and R.~Muller, ``A sub-mm 3 ultrasonic free-floating implant for multi-mote neural recording,'' \emph{IEEE Journal of Solid-State Circuits}, vol.~54, no.~11, pp. 3017--3030, 2019.

\bibitem{sonmezoglu2021monitoring}
S.~Sonmezoglu, J.~R. Fineman, E.~Maltepe, and M.~M. Maharbiz, ``Monitoring deep-tissue oxygenation with a millimeter-scale ultrasonic implant,'' \emph{Nature Biotechnology}, vol.~39, no.~7, pp. 855--864, 2021.

\bibitem{yousif2019performance}
B.~B. Yousif and E.~E. Elsayed, ``Performance enhancement of an orbital-angular-momentum-multiplexed free-space optical link under atmospheric turbulence effects using spatial-mode multiplexing and hybrid diversity based on adaptive mimo equalization,'' \emph{IEEE access}, vol.~7, pp. 84\,401--84\,412, 2019.

\bibitem{hayal2023modeling}
M.~R. Hayal, E.~E. Elsayed, D.~Kakati, M.~Singh, A.~Elfikky, A.~I. Boghdady, A.~Grover, S.~Mehta, S.~A.~H. Mohsan, and I.~Nurhidayat, ``Modeling and investigation on the performance enhancement of hovering uav-based fso relay optical wireless communication systems under pointing errors and atmospheric turbulence effects,'' \emph{Optical and Quantum Electronics}, vol.~55, no.~7, p. 625, 2023.

\bibitem{elsayed2024coding}
E.~E. Elsayed, M.~R. Hayal, I.~Nurhidayat, M.~A. Shah, A.~Elfikky, A.~I. Boghdady, D.~A. Juraev, and M.~Morsy, ``Coding techniques for diversity enhancement of dense wavelength division multiplexing mimo-fso fault protection protocols systems over atmospheric turbulence channels,'' \emph{IET Optoelectronics}, vol.~18, no. 1-2, pp. 11--31, 2024.

\bibitem{elsayed2024performance}
E.~E. Elsayed, ``Performance enhancement of atmospheric turbulence channels in dwdm-fso pon communication systems using m-ary hybrid dppm-m-papm modulation schemes under pointing errors, ase noise and interchannel crosstalk,'' \emph{Journal of Optics}, pp. 1--17, 2024.

\bibitem{elsayed2025atmospheric}
E.~E. Elsayed, M.~A. Yakout, and A.~S. Samra, ``Atmospheric turbulence mitigation in adaptive mimo hybrid rf/fso links using mppm/m-qam modulation and tas-mrc diversity under pointing errors,'' \emph{Journal of Optical Communications}, 2025.

\bibitem{fuada2024study}
S.~Fuada, M.~S{\"a}rest{\"o}niemi, M.~Katz, S.~Soderi, and M.~H{\"a}m{\"a}l{\"a}inen, ``Study on fat as the propagation medium in optical-based in-body communications,'' in \emph{Nordic Conference on Digital Health and Wireless Solutions}.\hskip 1em plus 0.5em minus 0.4em\relax Springer, 2024, pp. 467--479.

\bibitem{fuada2025optical}
S.~Fuada, M.~S{\"a}rest{\"o}niemi, and M.~Katz, ``Optical wireless data and power transfer for in-body electronic devices: A proof-of-concept study using ex vivo porcine samples,'' \emph{IET Optoelectronics}, vol.~19, no.~1, p. e70012, 2025.

\bibitem{das2019enabling}
D.~Das, S.~Maity, B.~Chatterjee, and S.~Sen, ``Enabling covert body area network using electro-quasistatic human body communication,'' \emph{Scientific reports}, vol.~9, no.~1, pp. 1--14, 2019.

\bibitem{maity2019bodywire}
S.~Maity, B.~Chatterjee, G.~Chang, and S.~Sen, ``Bodywire: A 6.3-pj/b 30-mb/s- 30-db sir-tolerant broadband interference-robust human body communication transceiver using time domain interference rejection,'' \emph{IEEE Journal of Solid-State Circuits}, vol.~54, no.~10, pp. 2892--2906, 2019.

\bibitem{sen2020body}
S.~Sen, S.~Maity, and D.~Das, ``The body is the network: To safeguard sensitive data, turn flesh and tissue into a secure wireless channel,'' \emph{IEEE Spectrum}, vol.~57, no.~12, pp. 44--49, 2020.

\bibitem{chatterjee2023bioelectronic}
B.~Chatterjee, P.~Mohseni, and S.~Sen, ``Bioelectronic sensor nodes for the internet of bodies,'' \emph{Annual Review of Biomedical Engineering}, vol.~25, pp. 101--129, 2023.

\bibitem{sarkar2025human}
S.~Sarkar, D.~Yang, M.~Nath, A.~Datta, S.~Maity, and S.~Sen, ``Human-structure and human-structure-human interaction in electro-quasistatic regime,'' \emph{Communications Engineering}, vol.~4, no.~1, p.~26, 2025.

\bibitem{lucev2012capacitive}
{\v{Z}}.~Lucev, I.~Krois, and M.~Cifrek, ``A capacitive intrabody communication channel from 100 khz to 100 mhz,'' \emph{IEEE Transactions on Instrumentation and Measurement}, vol.~61, no.~12, pp. 3280--3289, 2012.

\bibitem{xu2012equation}
R.~Xu, W.~C. Ng, H.~Zhu, H.~Shan, and J.~Yuan, ``Equation environment coupling and interference on the electric-field intrabody communication channel,'' \emph{IEEE Transactions on biomedical engineering}, vol.~59, no.~7, pp. 2051--2059, 2012.

\bibitem{park2016channel}
J.~Park, H.~Garudadri, and P.~P. Mercier, ``Channel modeling of miniaturized battery-powered capacitive human body communication systems,'' \emph{IEEE Transactions on Biomedical Engineering}, vol.~64, no.~2, pp. 452--462, 2016.

\bibitem{maity2018bio}
S.~Maity, M.~He, M.~Nath, D.~Das, B.~Chatterjee, and S.~Sen, ``Bio-physical modeling, characterization, and optimization of electro-quasistatic human body communication,'' \emph{IEEE Transactions on Biomedical Engineering}, vol.~66, no.~6, pp. 1791--1802, 2018.

\bibitem{xu2019modeling}
Y.~Xu, Z.~Huang, S.~Yang, Z.~Wang, B.~Yang, and Y.~Li, ``Modeling and characterization of capacitive coupling intrabody communication in an in-vehicle scenario,'' \emph{Sensors}, vol.~19, no.~19, p. 4305, 2019.

\bibitem{yang2022physically}
D.~Yang, S.~Maity, and S.~Sen, ``Physically secure wearable--wearable through-body interhuman body communication,'' \emph{Frontiers in Electronics}, vol.~2, p. 807051, 2022.

\bibitem{sarkar2023electro}
S.~Sarkar, A.~Datta, M.~Nath, D.~Yang, S.~Maity, and S.~Sen, ``Electro-quasistatic human-structure coupling for human presence detection and secure data offloading,'' in \emph{2023 45th Annual International Conference of the IEEE Engineering in Medicine \& Biology Society (EMBC)}.\hskip 1em plus 0.5em minus 0.4em\relax IEEE, 2023, pp. 1--4.

\bibitem{nath2019toward}
M.~Nath, S.~Maity, and S.~Sen, ``Toward understanding the return path capacitance in capacitive human body communication,'' \emph{IEEE Transactions on Circuits and Systems II: Express Briefs}, vol.~67, no.~10, pp. 1879--1883, 2019.

\bibitem{datta2021advanced}
A.~Datta, M.~Nath, D.~Yang, and S.~Sen, ``Advanced biophysical model to capture channel variability for eqs capacitive hbc,'' \emph{IEEE Transactions on Biomedical Engineering}, 2021.

\bibitem{Gabriel_1996}
S.~{Gabriel et al.}, ``The dielectric properties of biological tissues: {II}. measurements in the frequency range 10 hz to 20 {GHz},'' \emph{Physics in Medicine and Biology}, vol.~41, no.~11, pp. 2251--2269, nov 1996.

\bibitem{maity2020safety}
S.~Maity, M.~Nath, G.~Bhattacharya, B.~Chatterjee, and S.~Sen, ``On the safety of human body communication,'' \emph{IEEE Transactions on Biomedical Engineering}, vol.~67, no.~12, pp. 3392--3402, 2020.

\bibitem{neva_model}
``{NEVA Electromagnetics LLC}{ | }{VHP-Female model v2.2 - VHP-Female College},'' \url{https://www.nevaelectromagnetics.com/vhp-female-2-2}, [accessed August 27, 2020].

\bibitem{modak2022bio}
N.~Modak, M.~Nath, B.~Chatterjee, S.~Maity, and S.~Sen, ``Bio-physical modeling of galvanic human body communication in electro-quasistatic regime,'' \emph{IEEE Transactions on Biomedical Engineering}, vol.~69, no.~12, pp. 3717--3727, 2022.

\end{thebibliography}

\pagebreak
\appendices

\section*{APPENDIX I: Transfer Function Derivations}
\label{appendix:a}
\subsection*{Proximity-based Interaction with Ground-connected Metallic Object}
\begin{figure}[ht]
\centering
\includegraphics[width=0.48\textwidth]{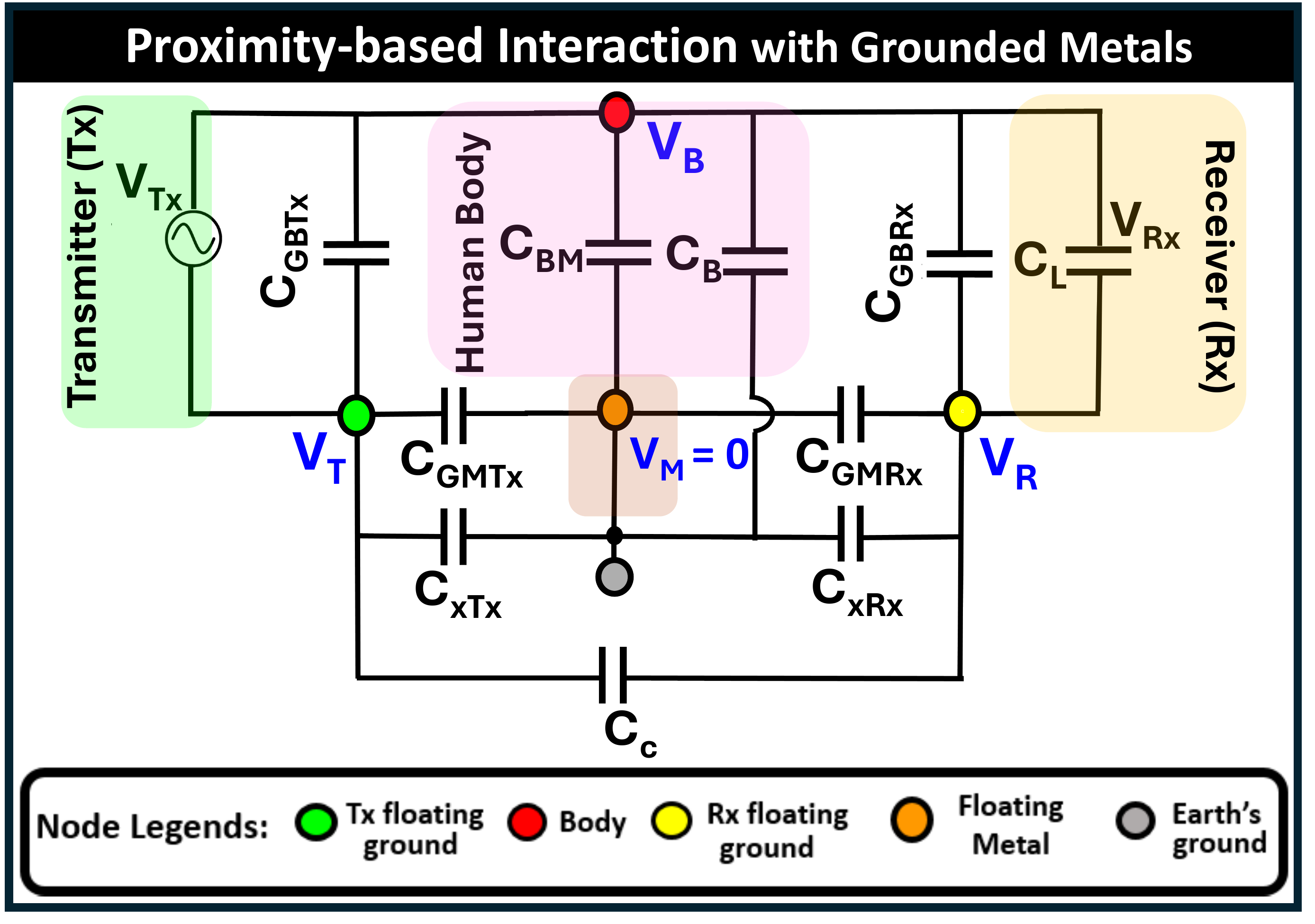}
\caption{Simplified equivalent circuit model for proximity-based interaction between subject and ground-connected metals.}
\label{Supp_fig1:Circuit_Gnd_Non-Touch}
\end{figure}

The simplified circuit model for the proximity-based interaction between EQS HBC user and a ground-connected metallic object is presented in Fig. \ref{Supp_fig1:Circuit_Gnd_Non-Touch}. For the return path capacitances at the Tx and Rx, we define the return path capacitances in the absence of any surrounding conductive objects such as $C_{xTx}$ and $C_{xRx}$. These values depend on the location of the devices (x) on the subject's body and the self capacitance of the devices, a function of geometry of the devices, represented as $C_{self}$. Specifically, the return path capacitances are expressed in Eqs. \ref{Supp_eq1a:main}, \ref{Supp_eq1a:main:a}.  
\begin{subequations}\label{Supp_eq1a:main}
    \begin{equation}
        \centering 
         C_{xTx} = x C_{selfTx} 
         \tag{\ref{Supp_eq1a:main}}
        \end{equation}
     \begin{equation}
        C_{xRx} = xC_{selfRx}
        \label{Supp_eq1a:main:a}
     \end{equation}
\end{subequations}                     
To distinguish between scenarios where ground-connected metallic objects are present and those where they are absent, we denote the net return path capacitance as $C_{retTx}$ and $C_{retRx}$. 
These are mathematically expressed in Eqs. \ref{Supp_eq2a:main}, \ref{Supp_eq2a:main:a}.
\begin{subequations}\label{Supp_eq2a:main}
    \begin{equation}
        \centering 
         C_{retTx} = C_{xTx} + C_{GMTx} 
         \tag{\ref{Supp_eq2a:main}}
        \end{equation}
     \begin{equation}
        C_{retRx} = C_{xRx} + C_{GMRx}
        \label{Supp_eq2a:main:a}
     \end{equation}
\end{subequations}     
Here, $C_{GMTx}$ and $C_{GMRx}$ represent the capacitances associated with the ground of the transmitter to the metallic object and from the ground of the receiver to the metallic object, respectively.
Assuming $C_B$ as the capacitance between the subject’s body and Earth’s ground when no other conductive objects are present in the surroundings. It is important to note that the presence of ground-connected metallic objects influences the subject’s body capacitance. To differentiate between scenarios in which ground-connected metallic objects are present and those in which they are absent, we denote the net body capacitance (including ground-connected metals) as  $C_{Body}$. This is expressed mathematically in Eq. \ref{Supp_eq3a}.
\begin{equation}
        \centering 
         C_{Body} = C_B + C_{BM}
         \label{Supp_eq3a}
        \end{equation}
where $C_{BM}$ represents the capacitance between the subject’s body and the metallic object.
Assuming the effect of inter-device coupling capacitance ($C_C$ $<$ 1 fF) to be negligible for Tx and Rx at more than 50 cm apart.
The return path impedance being couple of orders of magnitude higher than the impedance contribution of $Z_{Body}$ i.e., $Z_{Body} << Z_{retTx}$,
the induced potential on subject's body ($V_B$) is formulated in Eq. \ref{Supp_eq1}.
  \begin{equation}
   \centering 
   V_B = \frac{Z_{Body}}{Z_{Body} + Z_{retTx}} V_{Tx} \approx \frac{Z_{Body}}{Z_{retTx}}V_{Tx}= \left(\frac{C_{retTx}}{C_{Body}} \right) V_{Tx} 
   \label{Supp_eq1}
\end{equation}
Now, a fraction of $V_B$ gets picked-up at the receiver and the return path impedance being an order of magnitude higher than the impedance contribution of $Z_{L(eff.)}$ i.e., $Z_{L(eff.)} << Z_{retRx}$, the received voltage ($V_{Rx}$) as a function of $V_B$ is presented in Eq. \ref{Supp_eq2}.
 \begin{equation}
   \centering 
   V_{Rx} = \frac{Z_{L(eff.)}}{Z_{L(eff.)} + Z_{retRx}} V_{B} \approx \frac{Z_{L(eff.)}}{Z_{retRx}} V_{B}= \left(\frac{C_{retRx}}{C_{L(eff.)}} \right) V_{B} 
   \label{Supp_eq2}
\end{equation}
Hence, the transfer function is formulated by combining Eq. \ref{Supp_eq1} and Eq. \ref{Supp_eq2} and is presented in Eq. \ref{Supp_eq3}. 
\begin{equation}
   \begin{split}
   T_{NTGM}(s) & = \frac{V_{Rx}}{V_{Tx}}(s) \approx \frac{Z_{Body}}{Z_{retTx}} \times \frac{Z_{L(eff.)}}{Z_{retRx}} \\
   & \approx \frac{C_{retTx}}{C_{Body}} \times \frac{C_{retRx}}{C_{L(eff.)}}
   \label{Supp_eq3}
   \end{split}
\end{equation}
\begin{figure}[ht]
\centering
\includegraphics[width=0.48\textwidth]{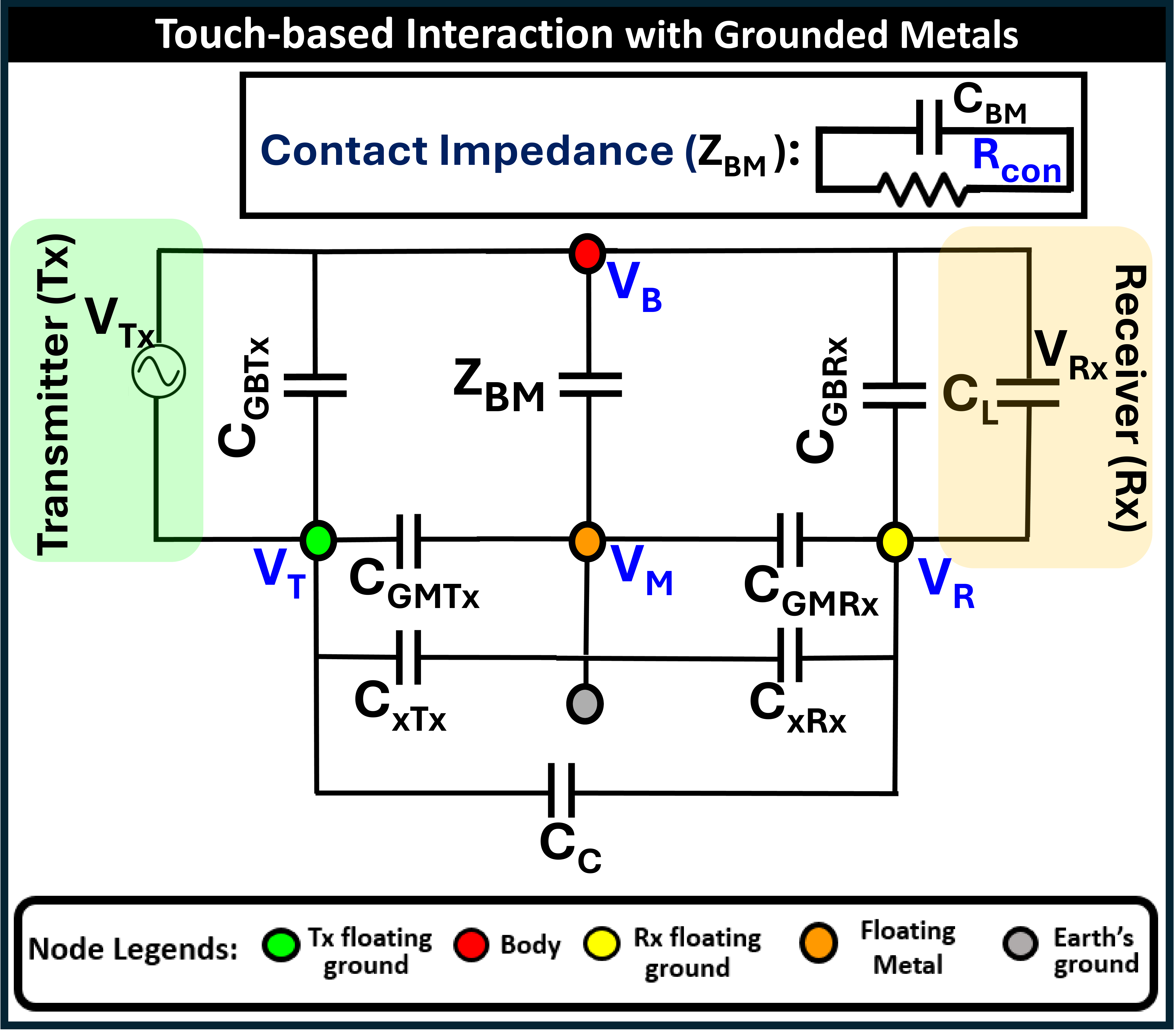}
\caption{Simplified equivalent circuit model for touch-based interaction between subject and ground-connected metallic obejcts.}
\label{Supp_fig1:Circuit_Gnd_Touch}
\end{figure}
\subsection*{Touch-based Interaction with Ground-Connected Metallic Object}
For the touch-based interaction with ground-connected metallic object, the impedance ($Z_{BM}$) between the body and the grounded metallic object considered to be the parallel combination of the contact resistance ($R_{con}$) and the contact capacitance ($C_{BM}$), shown in the equivalent circuit model in Fig. \ref{Supp_fig1:Circuit_Gnd_Touch}. Hence, the expression for $Z_{BM}$ is formulated in Eq. \ref{Supp_eq4}.
\begin{equation}
   \centering 
   Z_{BM}(s) = \frac{R_{con}}{1+sR_{con}C_{BM}}
   \label{Supp_eq4}
\end{equation}
 Now, in the impedance based transfer function derived in Eq. \ref{Supp_eq3}, substituting the value of Z$_{BM}$, we get the transfer function for touch-based interactions (T$_{TGM}$), derived in Eq. \ref{Supp_eq5}. 
\begin{equation}
   \begin{split}
   \centering 
   T_{TGM}(s) & = \frac{V_{Rx}}{V_{Tx}}(s) \approx \frac{\frac{R_{con}}{1 + s R_{con} C_{BM}}}{\frac{1}{s C_{retTx}}} \times \frac{\frac{1}{s C_{L(eff.)}}}{\frac{1}{s C_{retRx}}} \\
   &\approx \left(\frac{C_{retTx}C_{retRx}}{C_{L(eff.)} C_{BM} + \left(\frac{C_{L(eff.)}}{sR_{con}}\right)}\right)
   \label{Supp_eq5}
    \end{split}
\end{equation}
\begin{figure}[ht]
\centering
\includegraphics[width=0.48\textwidth]{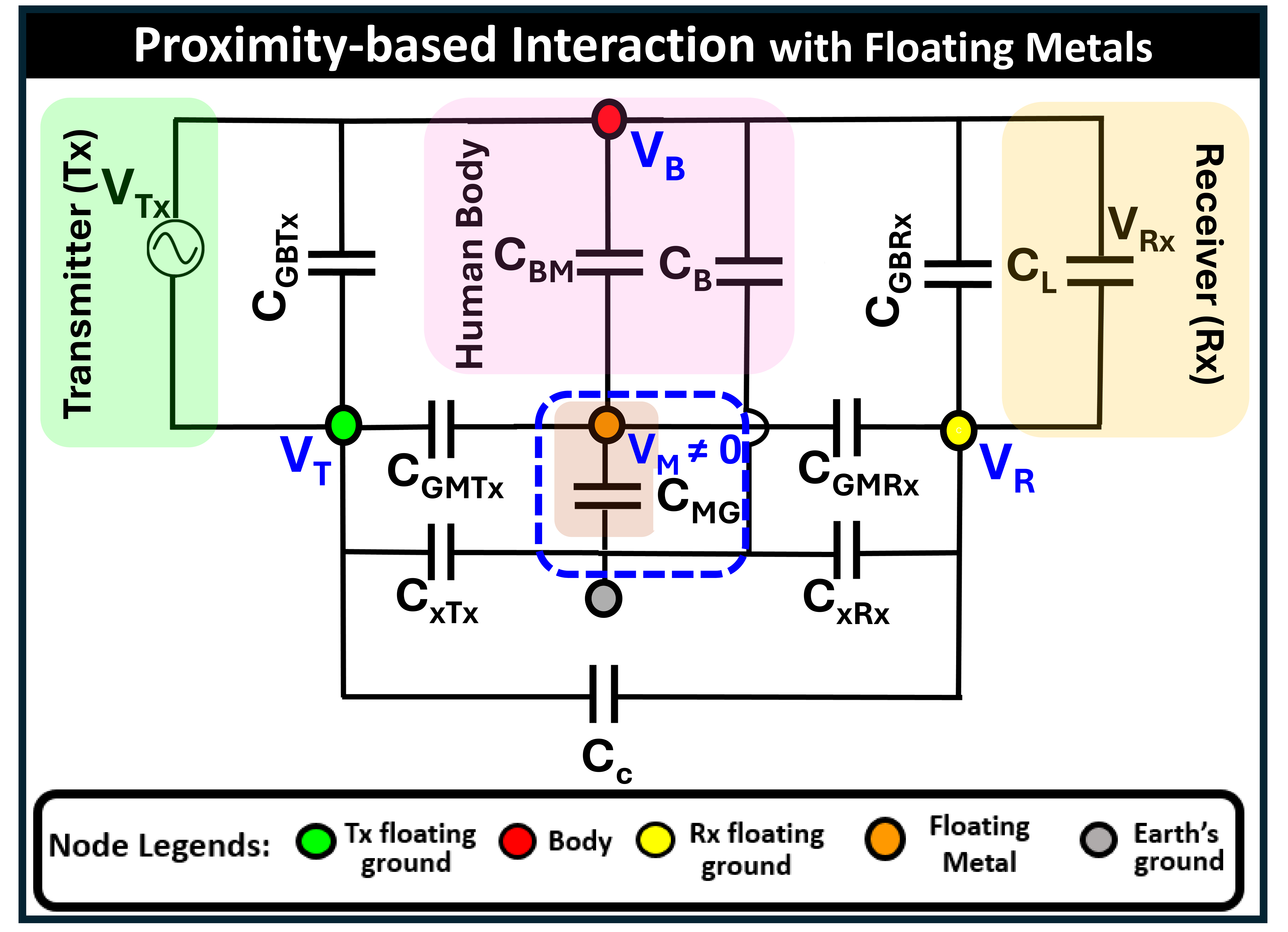}
\caption{Simplified equivalent circuit model for touch-based interaction between subject and Floating metallic obejcts.}
\label{Supp_fig1:Circuit_Floating_Non_Touch}
\end{figure}
\subsection*{Proximity $\&$ Touch-based Interaction with Floating Ground Metallic Object}
The circuit model illustrating the proximity-based interaction between a user of EQS HBC and a floating metallic object is shown in Fig. \ref{Supp_fig1:Circuit_Floating_Non_Touch}. Assuming the voltages at the following nodes: ground of the Tx, ground of the Rx, subject's body, and metal object are at potential $V_T$, $V_R$, $V_B$, and $V_M$ respectively. By KCL at the earth's ground node, we get the relation expressed in Eq. \ref{Supp_eq7}. 
\begin{equation}
    C_B V_{B} + C_{xTx} V_{T} + C_{xRx} V_{R} = 0
\label{Supp_eq7}
\end{equation}
Then, by KCL at the node with potential $V_M$, we get the relation expression in Eq. \ref{Supp_eq8}.
\begin{equation}
\begin{split}
 C_{BM} V_{B} + C_{GMTx} V_{T} + C_{GMRx} V_{R} - \\ (C_{MG} + C_{GMRx} + C_{GMTx} + C_{BM})V_M= 0
 \label{Supp_eq8}
\end{split}
\end{equation}
Hence, the induced potential on the metallic object ($V_M$) is presented in Eq. \ref{Supp_eq9:main}.
\begin{subequations}\label{Supp_eq9:main}
\begin{equation}
\centering
V_{M} = \frac{C_{BM}}{C_A} V_B + \frac{C_{GMRx}}{C_A} V_R + \frac{C_{GMTx}}{C_A} V_T
\tag{\ref{Supp_eq9:main}}
\end{equation}
where, 
\begin{equation}
C_A = (C_{MG} + C_{GMRx} + C_{GMTx} + C_{BM})\label{Supp_eq9:main:a}
\end{equation}
\end{subequations}
For a large object, the expression for C$_A$ can be approximately simplified as
C$_A$ $\approx$ (C$_{MG}$ + C$_{BM}$) (Since, C$_{GMRx}$, C$_{GMTx}$ $<<$ C$_{MG}$, C$_{BM}$). The coupling between the body-to-metallic object (C$_{BM}$) changes with the change in the average distance and orientation of subject relative to the metallic object. 
Now, by KCL at the node with potential $V_R$, we get the relation expression in Eq. \ref{Supp_eq10}.
\begin{equation}
    C_{L(eff.)} V_{B} - (C_{L(eff.)} + C_{xRx} + C_{GMRx}) V_{R} - C_{GMRx} V_{M} = 0
    \label{Supp_eq10}
\end{equation}
Since, the subject's body potential (V$_B$) also changes under the influence of metallic objects in the surroundings. Incorporating the effect from the surrounding metals, the nodal voltage at the ground of the Rx is expressed in Eq. \ref{Supp_eq11:main}.
\begin{subequations} \label{Supp_eq11:main}
\begin{equation}
   \centering 
   V_{R} = \frac{B}{P} V_B + \frac{Q}{P} V_T
   \tag{\ref{Supp_eq11:main}}
\end{equation}
where $P$, $Q$, and $B$ take the following form: 
\begin{equation}
P = \frac{1}{C_{L(eff.)}} \left [(C_{L(eff.)} + C_{xRx} + C_{GMRx}) + \frac{C_{GMRx}^2}{C_A}\right] 
\label{Supp_eq11:main:a}
\end{equation}
\begin{equation}
    Q = -\frac{C_{GMRx}  C_{GMTx}}{C_{L(eff.)} C_A}
    \label{Supp_eq11:main:b}
\end{equation}
\begin{equation}
B = 1 - \frac{C_{GMRx}  C_{BM}}{C_{L(eff.)} C_A}
\label{Supp_eq11:main:c}
\end{equation}
\begin{equation}
C_{L(eff.)} = C_{GBRx} + C_L
\label{Supp_eq11:main:d}
\end{equation}
\end{subequations}
\begin{figure}[ht]
\centering
\includegraphics[width=0.48\textwidth]{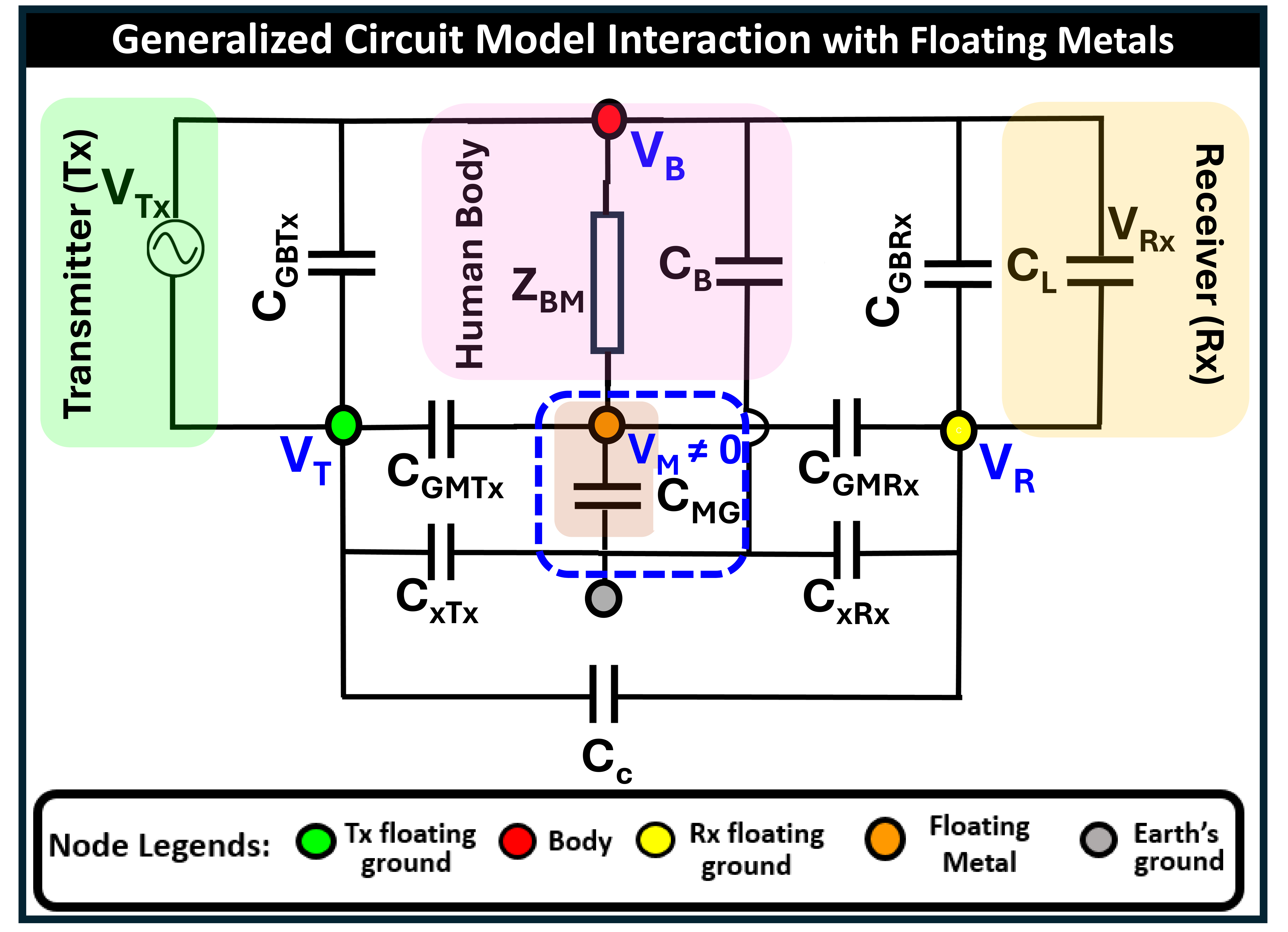}
\caption{Generalized circuit model for interaction between subject and Floating metallic obejcts.}
\label{Supp_fig1:Circuit_Floating_Touch}
\end{figure}
A generalized circuit model for the interaction between EQS HBC user and a floating metallic object is presented in Fig. \ref{Supp_fig1:Circuit_Floating_Touch}. By KCL at node M, we obtain the expression for $V_B$ in relation to $V_M$, $V_T$, and $V_R$, presented in Eq. \ref{Supp_eq12}. 
\begin{equation}
\begin{split}
V_B ={}& \bigl(1 + s\,(C_{GMRx} + C_{GMTx} + C_{MG})\,Z_{BM}\bigr)\,V_M \\ 
       &- s\,C_{GMRx}\,Z_{BM}\,V_R 
        - s\,C_{GMTx}\,Z_{BM}\,V_T
        \label{Supp_eq12}
\end{split}
\end{equation}
By KCL at the ground node of the Rx, we obtain another equation relating $V_R$ with $V_B$ and $V_M$, presented in Eq. \ref{Supp_eq13}.
\begin{equation}
V_R\:\bigl(C_{xRx} + C_{L(\mathrm{eff.})} + C_{GMRx}\bigr)
= C_{L(\mathrm{eff.})}\,V_B + C_{GMRx}\,V_M
\label{Supp_eq13}
\end{equation}
Again, by KCL at the ground node, we obtain another expression relating $V_{B}$, $V_R$ and $V_T$, presented in Eq. \ref{Supp_eq14:main}.
\begin{subequations}\label{Supp_eq14:main}
\begin{equation}
\centering
C_B V_B + C_{xRx} V_R + C_{xTx}V_T = 0
\tag{\ref{Supp_eq14:main}}
\end{equation}
where, $V_T$ can be expressed as  
\begin{equation}
V_T = -(\frac{C_{B}}{C_{xTx}} V_B + \frac{C_{xRx}}{C_{xTx}} V_R )\label{Supp_eq14:main:a}
\end{equation}
\end{subequations}
By substituting $V_T$ from Eq. \ref{Supp_eq14:main:a} in the expression for $V_B$ in Eq. \ref{Supp_eq12}, we obtain
\begin{subequations}\label{Supp_eq15:main}
\begin{equation}
    \centering 
    V_B = \frac{D}{A} V_M + \frac{F}{A} V_R
    \tag{\ref{Supp_eq15:main}}
\end{equation}
where $A$, $D$ and $F$ are defined in Eq. \ref{Supp_eq15:main} (a, b, c)
\begin{equation}
A = 1 - sC_{GMTx}Z_{BM} \frac{C_B}{C_{xTx}}\label{Supp_eq15:main:a} 
\end{equation}
\begin{equation}
D = 1 + s (C_{GMRx} + C_{GMTx} + C_{MG})Z_{BM}\label{Supp_eq15:main:b} 
\end{equation}
\begin{equation}
F = (sC_{GMTx}\frac{C_{xRx}}{C_{xTx}} - sC_{GMRx})Z_{BM}\label{Supp_eq15:main:c}
\end{equation}
\end{subequations}
Now, $V_{Tx}$ can be written in relation to $V_B$ and $V_T$, expressed in Eq. \ref{Supp_eq16:main}. 
\begin{subequations}\label{Supp_eq16:main}
\begin{equation}
    \centering
    V_{Tx} = V_B - V_T
    \tag{\ref{Supp_eq16:main}}
\end{equation}
Substituting $V_T$ and rearranging terms, we get 
\begin{equation}
    \centering
    V_{Tx} = UV_B + TV_R
    \label{Supp_eq16:main:a}
\end{equation}
where $U$ and $T$ are defined in Eq. \ref{Supp_eq16:main} (b, c)
\begin{equation}
U = 1 + \frac{C_B}{C_{xTx}}\label{Supp_eq16:main:b} 
\end{equation}
\begin{equation}
T = \frac{C_{xRx}}{C_{xTx}}\label{Supp_eq16:main:c}
\end{equation}
\end{subequations}
Now, $V_M$ can be expressed in terms of $V_B$ and $V_R$, presented in Eq. \ref{Supp_eq17:main} can be written as 
\begin{subequations}\label{Supp_eq17:main}
\begin{equation}
    \centering
    V_{M} = SV_B + RV_R
    \tag{\ref{Supp_eq17:main}}
\end{equation}
where $S$ and $R$ are defined in Eq. \ref{Supp_eq17:main} (a, b)
\begin{equation}
S = -\frac{C_{L(eff.)}}{    C_{GMRx}}\label{Supp_eq17:main:a} 
\end{equation}
\begin{equation}
R = +\frac{C_{xRx} + C_{L(eff.)} + C_{GMRx}}{C_{GMRx}}\label{Supp_eq17:main:b}
\end{equation}
\end{subequations}
Hence, the transfer function is formulated in Eq. \ref{Supp_eq18}
\begin{equation}
   \centering 
   T_{NTFM}(s) = \frac{V_{Rx}}{V_{Tx}} = \frac{V_B - V_R}{V_B - V_T} \approx \left(\frac{1- \frac{\left(A-DS\right)}{DR+F}}{U +T \frac{\left(A-DS\right)}{DR+F}} \right)
   \label{Supp_eq18}
\end{equation}
Similar to grounded metal, during touch-based interaction with floating metal, the transfer characteristics ($T_{TFM}(s)$) can be obtained by substituting the expression of $Z_{BM}$ from Eq. \ref{Supp_eq4}.
\subsection*{Sensitivity Analysis}
Since, the channel transfer function is defined as 
\begin{equation}
   \begin{split}
    T &= \frac{V_{\mathrm{Rx}}}{V_{\mathrm{Tx}}} 
    = \frac{V_{B}}{V_{Tx}} \times \frac{V_{Rx}}{V_{B}} \nonumber \\
    &= \frac{C_{\mathrm{retTx}}}
           {C_{\mathrm{retTx}} +C_{\mathrm{Body}}} \times \frac{C_{\mathrm{retRx}}}
           {C_{\mathrm{retRx}} + C_{\mathrm{L(eff.)}}}
    \end{split}
\end{equation}
The relative sensitivity of \(T\) to a small change in \(C_i\) is expressed in Eq. \ref{Supp_eq19}
\begin{equation}
  S_{C_i}
  = \frac{C_i}{T}\,\frac{\partial T}{\partial C_i}
  \label{Supp_eq19}
\end{equation}
\subsubsection*{Sensitivities of \(\frac{V_{B}}{V_{Tx}}\)}
\begin{subequations}\label{Supp_eq20:main:a}
    \begin{equation}
    \frac{\partial \left(\frac{V_{B}}{V_{Tx}}\right)}{\partial C_{\mathrm{retTx}}}
    = \frac{C_{\mathrm{Body}}}
            {\bigl(C_{\mathrm{retTx}} + C_{\mathrm{Body}}\bigr)^2}
            \label{Supp_eq20:main:a}
\end{equation}
\begin{equation}
  \frac{\partial \left(\frac{V_{B}}{V_{Tx}}\right)}{\partial C_{\mathrm{Body}}}
    = - \frac{C_{\mathrm{retTx}}}
               {\bigl(C_{\mathrm{retTx}} + C_{\mathrm{Body}}\bigr)^2}
               \label{Supp_eq20:main:b}
\end{equation}
Hence, the relative sensitivities are defined in Eq. \ref{Supp_eq20:main:c},  \ref{Supp_eq20:main:d}.
\begin{equation}
    S_{C_{\mathrm{retTx}}}
    = \frac{C_{\mathrm{retTx}}}{\left(\frac{V_{B}}{V_{Tx}}\right)}\,
       \frac{\partial \left(\frac{V_{B}}{V_{Tx}}\right)}{\partial C_{\mathrm{retTx}}}
    = \frac{C_{\mathrm{Body}}}
           {C_{\mathrm{retTx}} + C_{\mathrm{Body}}}
           \label{Supp_eq20:main:c}
\end{equation}
\begin{equation}
    S_{C_{\mathrm{Body}}}
    = \frac{C_{\mathrm{Body}}}{\left(\frac{V_{B}}{V_{Tx}}\right)}\,
       \frac{\partial \left(\frac{V_{B}}{V_{Tx}}\right)}{\partial C_{\mathrm{Body}}}
    = -\,\frac{C_{\mathrm{Body}}}
             {C_{\mathrm{retTx}} + C_{\mathrm{Body}}}
             \label{Supp_eq20:main:d}
\end{equation}
\end{subequations}
\subsubsection*{Sensitivities of \(\frac{V_{Rx}}{V_{B}}\)}
\begin{subequations}
    \begin{equation}
    \frac{\partial \left(\frac{V_{Rx}}{V_{B}}\right)}{\partial C_{\mathrm{retRx}}}
    = \frac{C_{\mathrm{L(eff.)}}}
            {\bigl(C_{\mathrm{retRx}} + C_{\mathrm{L(eff.)}}\bigr)^2}
            \label{Supp_eq21:main:a}
\end{equation}
\begin{equation}
    \frac{\partial \left(\frac{V_{Rx}}{V_{B}}\right)}{\partial C_{\mathrm{L(eff.)}}}
    = - \frac{C_{\mathrm{retRx}}}
               {\bigl(C_{\mathrm{retRx}} + C_{\mathrm{L(eff.)}}\bigr)^2}
               \label{Supp_eq21:main:b}
\end{equation}
\begin{equation}
  S_{C_{\mathrm{retRx}}}
    = \frac{C_{\mathrm{retRx}}}{\left(\frac{V_{Rx}}{V_{B}}\right)}\,
       \frac{\partial \left(\frac{V_{Rx}}{V_{B}}\right)}{\partial C_{\mathrm{retRx}}}
    = \frac{C_{\mathrm{L(eff.)}}}
           {C_{\mathrm{retRx}} + C_{\mathrm{L(eff.)}}}
           \label{Supp_eq21:main:c}
\end{equation}
\begin{equation}
  S_{C_{\mathrm{L(eff.)}}}
    = \frac{C_{\mathrm{L(eff.)}}}{\left(\frac{V_{Rx}}{V_{B}}\right)}\,
       \frac{\partial \left(\frac{V_{Rx}}{V_{B}}\right)}{\partial C_{\mathrm{L(eff.)}}}
    = - \frac{C_{\mathrm{L(eff.)}}}
             {C_{\mathrm{retRx}} + C_{\mathrm{L(eff.)}}}
             \label{Supp_eq21:main:d}
\end{equation}
\end{subequations}
The combined sensitivity can be obtained by putting these sensitivities expressed in Eqs. \ref{Supp_eq20:main:c}, \ref{Supp_eq20:main:d}, \ref{Supp_eq21:main:c}, \ref{Supp_eq21:main:d}, together and is expressed in Eqs. \ref{Supp_eq22:main} and \ref{Supp_eq22:main:b}.
\begin{subequations}\label{Supp_eq22:main} 
    \begin{equation}
\frac{\partial T}{\partial C_i}
= \left(\frac{V_{Rx}}{V_{B}}\right) \;\frac{\partial \left(\frac{V_{B}}{V_{Tx}}\right)}{\partial C_i},
\quad C_i\in\{C_{\mathrm{retTx}},\,C_{\mathrm{Body}}\}
\tag{\ref{Supp_eq22:main}}
\end{equation}
\\
\begin{equation}
    \frac{\partial T}{\partial C_j}
 = \left(\frac{V_{B}}{V_{Tx}}\right) \;\frac{\partial \left(\frac{V_{Rx}}{V_{B}}\right)}{\partial C_j},
\quad C_j\in\{C_{\mathrm{retRx}},\,C_{\mathrm{L(eff.)}}\}
\label{Supp_eq22:main:b}
\end{equation}
\end{subequations}

\section*{APPENDIX II}
\label{appendix:b}
\subsection*{Performance analysis of Communication Channel}
The presence of metallic objects within the leakage limit ($\sim$5 cm from the subject's body and $\sim$20 cm from communication devices, such as the transmitter (Tx) and receiver (Rx)) affects the parasitic return paths of Electro-quasistatic (EQS) Human Body Communication (HBC). This phenomenon is expected to impact the performance of wireless body-centric communication. In this analysis, we investigate the impact of nearby metallic objects on the performance of the body channel by examining variations in Signal-to-Noise Ratio (SNR), Shannon capacity, and bit error rate for different modulation schemes. These schemes include On-Off Keying (OOK), commonly referred to as Pulse Amplitude Modulation (PAM-2), Quadrature Amplitude Modulation (QAM), and Binary Phase Shift Keying (BPSK).

Let the nominal return-path capacitances and body capacitances be C$_{\mathrm{xTx}}$, C$_{\mathrm{xRx}}$, and C$_{\mathrm{B}}$, and let the metal-induced perturbations be 
C$_{\mathrm{GMTx}}$,
  C$_{\mathrm{GMRx}}$, 
 and C$_{\mathrm{Body}}$. Then the net capacitances become
C$_{\mathrm{retTx}}$,
  C$_{\mathrm{retRx}}$, 
 and C$_{\mathrm{Body}}$
The resulting EQS-HBC channel gain is presented in Eq. \ref{Supp_eq23}.
\begin{equation}
     T  = \frac{C_{\mathrm{retTx}}}
           {C_{\mathrm{retTx}} +C_{\mathrm{Body}}} \times \frac{C_{\mathrm{retRx}}}
           {C_{\mathrm{retRx}} + C_{\mathrm{L(eff.)}}}
           \label{Supp_eq23}
\end{equation}
Assume a transmit RMS voltage V$_{\mathrm{Tx}}$ and white noise power spectral density (PSD): N$_0$ = $\left( 5 \times 10 ^{-9}\right)^2 = 25 \times 10^{-18} \mathrm{V^2/ Hz}$. Over an operational channel bandwidth B, the noise variance is $\mathrm{N_0B}$. The received signal power ($P_{sig}$) is presented in Eq. \ref{Supp_eq24}
\begin{equation}
P_{\mathrm{sig}}
  = \bigl\lvert T  V_{\mathrm{Tx}}\bigr\rvert^{2} \quad
  P_{\mathrm{noise}}
  = N_{0}\,B
  \label{Supp_eq24}
\end{equation}
From the Eq. \ref{Supp_eq23} and \ref{Supp_eq24} the obtained SNR is presented in Eq. \ref{Supp_eq25}.
\begin{equation}
 SNR
  = \frac{P_{\mathrm{sig}}}{P_{\mathrm{noise}}}
  = \frac{\lvert T\rvert^{2}\,V_{\mathrm{Tx}}^{2}}
         {N_{0}\,B}
         \label{Supp_eq25}
\end{equation}
Hence, from the obtained SNR and bandwidth, the channel capacity is expressed in Eq. \ref{Supp_eq26} 
\begin{equation}
    \mathrm{Channel\, Capacity \,(bits/s)} = B \log_2 (1 + SNR)
    \label{Supp_eq26}
\end{equation}
Now, the BER expressions for various modulation schemes as a function of SNR are presented in Eqs. \ref{Supp_eq27:main:a}, \ref{Supp_eq27:main:b}, \ref{Supp_eq27:main:c}.
\begin{subequations}\label{Supp_eq27:main:a}
\begin{equation}
\mathrm{BER}_{\mathrm{OOK}} 
= Q\bigl(\sqrt{\gamma})
\label{Supp_eq27:main:a}
\end{equation}
\begin{equation}
\mathrm{BER}_{\mathrm{QPSK}}
= Q\!\bigl(\sqrt{2\,\gamma}\bigr)
\label{Supp_eq27:main:b}
\end{equation}
\begin{equation}
\mathrm{BER}_{M\text{-QAM}}
= \frac{4(\sqrt{M}-1)}{\sqrt{M}\,\log_{2}M}
  \;Q\!\Bigl(\sqrt{\frac{3\,\gamma\,\log_{2}M}{M-1}}\Bigr)
  \label{Supp_eq27:main:c}
\end{equation}
where Q is defined in Eq. \ref{Supp_eq27:main:d}
\begin{equation}
    Q(x)=\frac{1}{\sqrt{2\pi}}\int_x^\infty e^{-t^2/2}\,dt
   \label{Supp_eq27:main:d}
\end{equation}
\end{subequations}
in the above expressions $\gamma$ is related to SNR by how bit‐energy over the noise band. In particular, transmitting at bit‐rate of $R_b$ over a noise bandwidth B, the signal power is defined as $\mathrm{P_sig = E_b \, R_b}$, hence the SNR is expressed in Eq. \ref{Supp_eq28}.
\begin{subequations}
     \begin{align}
\frac{P_{sig}}{P_{noise}} = \frac{E_b \, R_b}{N_0 \, B} = \frac{E_b}{N_0} \cdot \frac{R_b}{B} \\
 \gamma = \frac{E_b}{N_0} = SNR \, \times \frac{B}{R_b}
    \end{align}
    \label{Supp_eq28} 
\end{subequations}
Assuming the bit-rate to be the same as operational bandwidth ($R_b = B$), we get $\gamma = SNR$.

\begin{figure}[ht]
\centering
\includegraphics[width=0.48\textwidth]{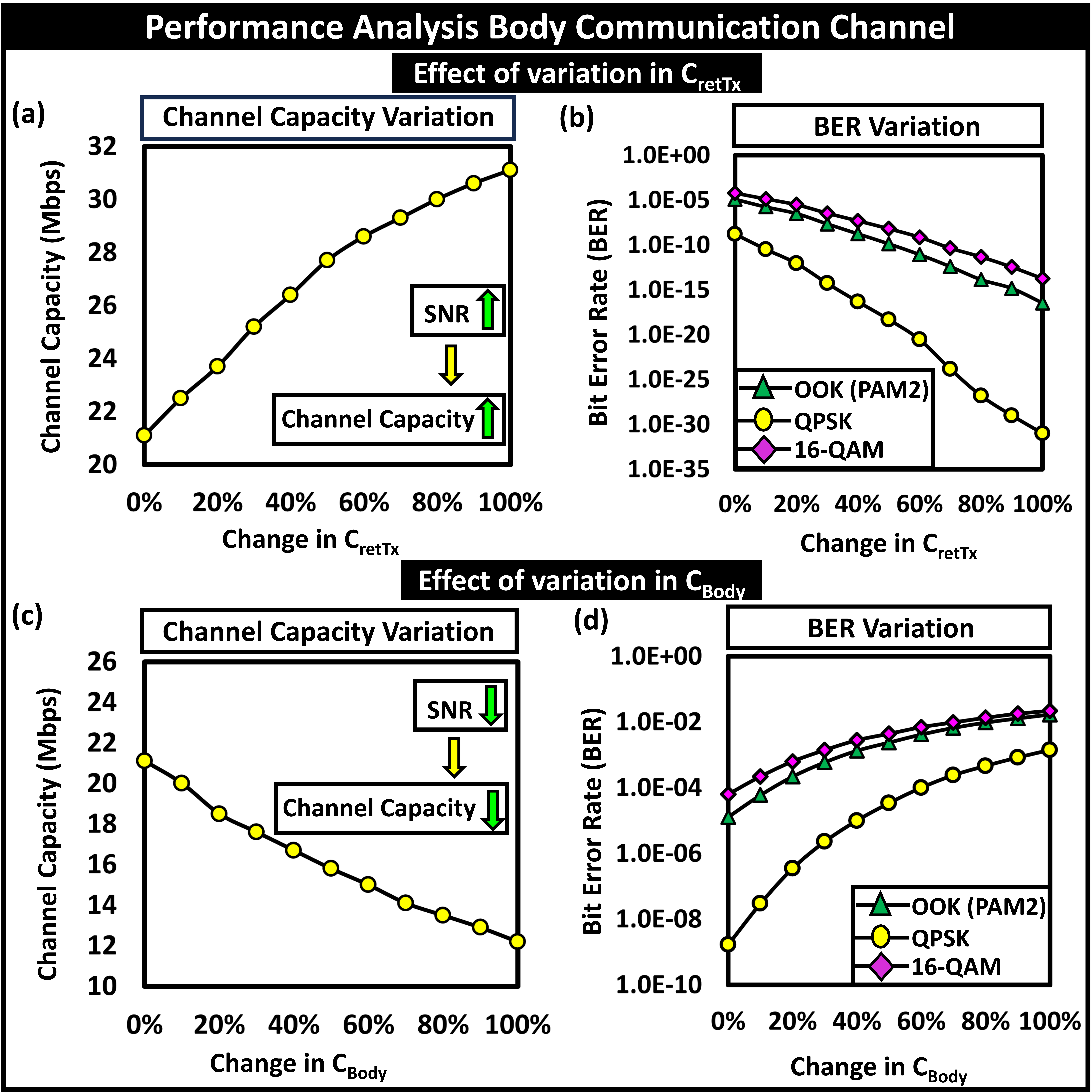}
\caption{Analyzing the effect of variation in C$_{\mathrm{retTx}}$ and C$_{\mathrm{Body}}$ on the performance metrics like channel capacity, Bit-error rate (BER) for EQS HBC: (a) Channel capacity variation with C$_{\mathrm{retTx}}$, (b) BER variation with C$_{\mathrm{retTx}}$, (c) Channel capacity variation with C$_{\mathrm{Body}}$, (d)  BER variation with C$_{\mathrm{Body}}$}
\label{fig:Performance_Analysis}
\end{figure}

Now, with operational bandwidth ($B$) of 5 MHz,  $\mathrm{V_{Tx}} = 1$V,  assuming a noise floor of -70 dBV for CMOS-based body-communication receiver, capacitances (C$_{\mathrm{xTx}}$,
  C$_{\mathrm{B}}$, C$_{\mathrm{xRx}}$, C$_{\mathrm{L(eff.)}}$) being (0.2, 150, 0.2, 5) pF in open area, if the surrounding metal induced perturbations being $\Delta$ (C$_{\mathrm{xTx}}$,
  C$_{\mathrm{B}}$, C$_{\mathrm{xRx}}$, C$_{\mathrm{L(eff.)}}$): (0.8, 0, 0.8, 0) pF, i.e., (C$_{\mathrm{retTx}}$,
  C$_{\mathrm{Body}}$, C$_{\mathrm{retRx}}$, C$_{\mathrm{L(eff.)}}$): (1, 150, 1, 5) pF,   $T \approx 0.0013$, $SNR \approx 12.49$ dB. Over a bandwidth of 5 MHz, channel capacity comes around 21.1 Mbps. The the ideal‐coherent BER for OOK comes 1 $\times$ 10$^{-5}$ that satisfies the intended uncoded BER requirements for OOK ($10^{-2}$). 
The variation in performance metrics with the change in C$_{\mathrm{retTx}}$ and C$_{\mathrm{Body}}$ is captured in  Fig. \ref{fig:Performance_Analysis}. In summary, it can be interpreted as, an increase in C$_{\mathrm{retTx}}$ leads to an enhanced channel capacity (i.e., $\sim$10 Mbps increase in channel capacity with 100$\%$ increase in C$_{\mathrm{retTx}}$) from the improved SNR, illustrated in Fig. \ref{fig:Performance_Analysis}(a).
The variation in ideal-coherent BER with C$_{\mathrm{Body}}$ for different modulation schemes like OOK, QPSK, and 16-QAM, presented in Fig. \ref{fig:Performance_Analysis}(b), shows an increase in C$_{\mathrm{Body}}$ facilitates reliable communication at reduced BER. On contrary, a rise in C$_{\mathrm{Body}}$ reduces the channel capacity owing to an attenuation in SNR, represented in Fig. \ref{fig:Performance_Analysis}(c). The BER also exhibit an increasing trend with increasing C$_{\mathrm{Body}}$, shown in Fig. \ref{fig:Performance_Analysis} (d). 

This research focuses on the scientific and system validation of the effect of nearby metals on electrostatic human body communication (EQS HBC), and data collection from humans of different generalize across body types (e.g., BMI, height, skin properties) was not needed and/or performed. A study on the variability of the communication channel for EQS HBC across different users was previously conducted. It was inferred that in EQS, with the operating wavelength ($\lambda$) being several orders of magnitude higher than the maximum body channel length (i.e., $l_{Body} << \lambda$), the electrically distributed characteristics of the body can be fairly approximated as a lumped conductor specifically as an electrical lossy wire and these subject-specific variabilities remain within the tolerance limit of the designed transceiver system. Thus, even if the channel loss varies slightly, EQS-HBC lumped model adapts seamlessly across diverse users. Calibration via a small set of measurements refines tissue resistivity and fringe-field factors, yielding a predictive, personalized HBC channel model. The variabilities can be tackled by designing adaptive transceiver system, even if the transmission loss changes with the subject body dimension.

The transmission characteristics are likely to vary with the dynamic activities of EQS HBC users, such as walking and arm swinging. These activities can alter parasitic capacitive coupling with surrounding metallic objects especially change in subject’s body posture i.e., limb angles ($\theta$) or torso posture ($h$) may alter the return path capacitances of the devices ($C_{xTx}(\theta), C_{xRx}(\theta)$), and subsequently the capacitance between the devices and user’s body i.e., $C_{GBRx} (\theta)$ and body-to-earth capacitance $C_B(h, \theta)$. A study by Yang et al. [7] examined the variability of communication channel performance for EQS HBC with subject's posture, particularly through arm movement, which can affect the extent of inter-device coupling and body shadowing, depending on the relative separation between the transmitter and receiver and their proximity to the user's torso. Our study did not account for these factors in our simulations or experiments, as the primary focus of this fundamental study was to investigate the impact of nearby metals ranging from non-grounded, grounded metallic objects to enclosed metallic surroundings like elevators and cars on transmission loss of EQS HBC. We recognize that the challenges posed by dynamic movements could be addressed by designing a posture-aware adaptive transceiver system, which can effectively handle these variabilities with an appropriate link margin. This insight motivates future work in this area.

\subsection*{Differentiating Touch and near touch}
The quantitative threshold that differentiates between two interactions can be understood in terms of the impedance \(Z_{BM}\) between a subject’s body and a metallic object. This impedance can be modeled as a resistance \(R_{BM}\) in parallel with a capacitance \(C_{BM}\), presented in Eq. \ref{Supp_eq29}.
\begin{equation}
Z_{BM} \;=\;\frac{R_{BM}}{1 + j\,\omega\,R_{BM}C_{BM}}
\label{Supp_eq29} 
\end{equation}
For simplicity, if we assume the capacitance between the subject’s body and a nearby metal, \(C_{BM}(d)\), as a function of distance \(d\)—and approximate it as a parallel-plate capacitor (neglecting fringe effects) with overlapping area \(A\)—we obtain Eq. \ref{Supp_eq30}.
\begin{equation}
C_{BM}(d)\;\approx\;\frac{\varepsilon_0\,\varepsilon_r\,A}{d}
\label{Supp_eq30} 
\end{equation}
Here \(\varepsilon_0 = 8.85\times10^{-12}\,\mathrm{F/m}\) is the permittivity of free space, and \(\varepsilon_r\) is the relative permittivity. Define \(d_c\) as the critical distance that separates “near touch” (\(d_c>0\)) from “touch” (\(d_c=0\)) interactions.  
During a “near touch” (\(d_c>0\)), the capacitive branch dominates (\(1/(j\omega C_{BM})\ll R_{BM}\)) and the impedance reduces to Eq. \ref{Supp_eq29}.
\begin{equation}
Z_{BM}\;=\;\frac{1}{j\,\omega\,C_{BM}}
\;\approx\;\frac{d}{j\,\omega\,\varepsilon_0\,\varepsilon_r\,A}
\label{Supp_eq31} 
\end{equation}
Consequently, within the electro-quasistatic regime the channel transfer function for capacitive Human Body Communication remains nearly flat.  
In contrast, during a “touch” interaction (\(d_c=0\)), the coupling impedance falls below the contact resistance \(R_{\rm con}\). We set \(R_{BM}=R_{\rm con}\), which depends on the contact area \(A_{\rm con}\):
\begin{equation}
R_{\rm con}\;\propto\;\frac{1}{A_{\rm con}}
\label{Supp_eq32} 
\end{equation}
At the boundary \(d=d_c\), \(\lvert Z_{BM}(d)\rvert \approx R_{\rm con}\), giving
\begin{equation}
d_c \;\approx\;\omega\,\varepsilon_0\,\varepsilon_r\,A\,R_{\rm con}
\label{Supp_eq33} 
\end{equation}
Thus if the subject approaches closer than \(d_c\), the capacitive impedance matches a true touch and the system can no longer distinguish “near-touch” from contact. As an example, at 
f = 5 MHz, with $A = 10\ \mathrm{cm}^2$, 
 $R_{\rm con}=1\ \mathrm{k}\Omega$.
Eq. \ref{Supp_eq33} yields
\[
d_c \approx 0.278\ \mathrm{mm}
\]
showing that the critical distance lies in the sub-millimeter to millimeter range. When the subject actually contacts a ground-connected metallic object, the presence of \(R_{\rm con}\) introduces a frequency-dependent high-pass behavior. The cutoff frequency \(f_c\) varies as
\[
f_c \;\propto\;\frac{1}{R_{\rm con}}
\]
If instead the metal is floating, one still sees high-pass characteristics; with sufficiently good contact (larger \(A_{\rm con}\), smaller \(R_{\rm con}\)), the voltage drop across \(R_{\rm con}\) becomes negligible and the transfer function again flattens.

\section*{APPENDIX III: Additional Details of Experimental Setup}
\label{appendix:c}
\subsection*{Calibration of Wearable Transmitter $\&$ Receiver} With its ground being floating, the wearable Tx requires to be calibrated against a benchtop ground-connected standard. By connecting the transmitter to a benchtop oscilloscope while varying its frequency in the EQS regime from 100 kHz to 20 MHz, the peak-peak voltage shown in the oscilloscope is recorded, and the calibration correction for the Tx is calculated. The  receiver is calibrated by connecting it to a Keysight signal generator, a benchtop standard. The output power of the benchtop signal generator varied over different power levels while varying the frequency from 100 kHz to 20 MHz, and the power difference observed between the two devices is noted to calculate the correction factor for the tinySA. The buffer's input connects to the output of a benchtop function generator, while the buffer's output is connected to a benchtop signal analyzer. The buffer's correction factor is then recorded. The buffer circuit is made by using OPA2836 from Texas Instruments which is an ultra-low power, rail-to-rail output swing, voltage-feedback operational amplifier. It operates with a supply voltage ranging in 2.5 V to 5.5 V, with its unity gain bandwidth (UGB) of 205 MHz, slew rate of 560 V/$\mu$s and its input voltage noise sensitivity of 4.6 nV/$\sqrt{Hz}$ making it suitable for handling ac signals with desired sensitivity at the receiver. For applications with battery-powered wearable devices where power is a key importance, the low-power consumption and high-frequency performance of the OPA2836 offers superior performance. 
We agree that a buffer circuit can attenuate the signal, however, with its higher UGB, this op-amp based buffer circuit designed at a unity gain configuration. 

\subsection*{Signal coupler at Tx $\&$ Rx} The signal coupler at Tx and Rx side is made of a commercially available double-sided conductive copper foil tape, measuring 3.5 cm $\times$ 4.5 cm, and of $<$ 0.1 cm thickness, affixed to the bottom surface of the cut-board, ensures contact with the subject's skin.

\subsection*{System Schematic Diagram} 
\begin{figure}[ht!]
\centering
\includegraphics[width=0.48\textwidth]{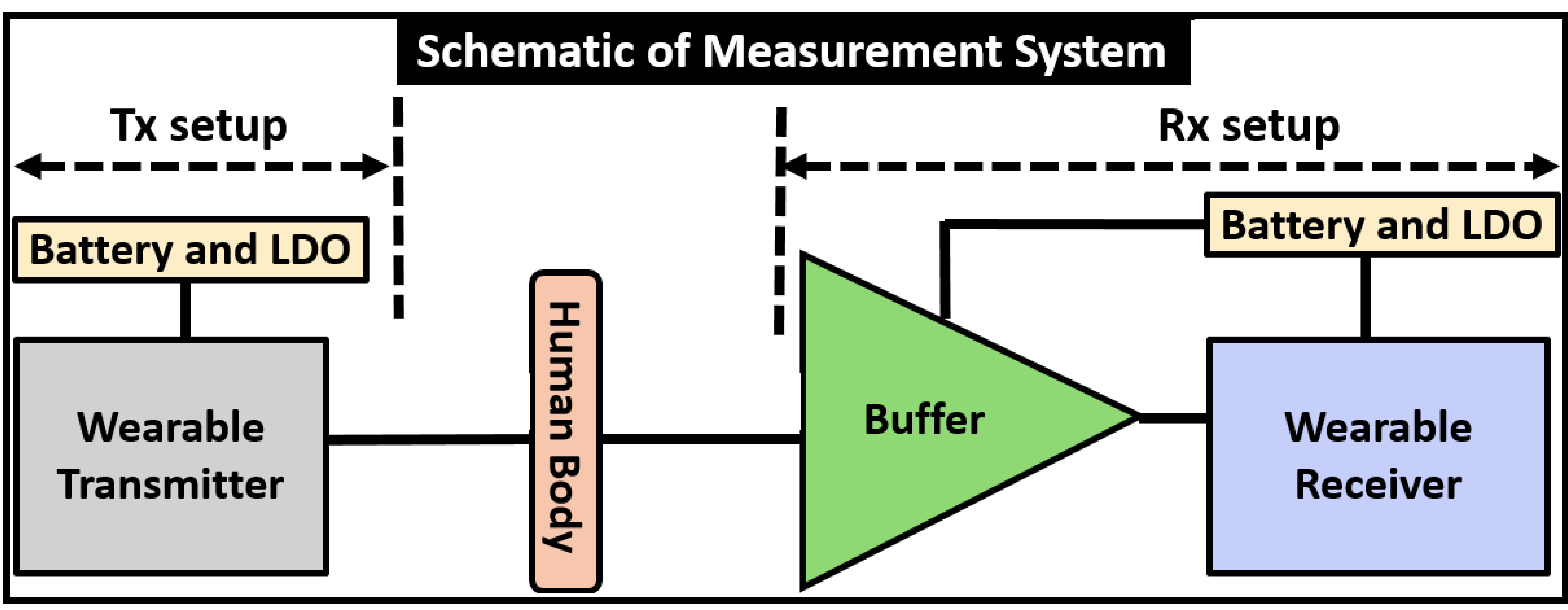}
\caption{The wearable transmitter (Tx) with its signal coupler in contact
with the user's skin couples EQS signal to the body. A capacitive receiver (Rx) at the wrist of another arm is used to measure received voltage. The receiver setup includes the Tiny SA spectrum analyzer together
with Buffer. Customized Buffer with high impedance capacitive termination is used for voltage
mode communication}
\label{fig:Measurement_Schematic}
\end{figure}

The schematic of the measurement system shown in Fig. \ref{fig:Measurement_Schematic} illustrates the signal processing pipeline. 

\subsection*{Test Conditions} Since the focus of this study is evaluating the communication channel’s performance in various metallic surroundings, experiments for grounded and non-grounded metallic objects are conducted in standard laboratory environments at an average temperature of 27$^\circ$C (80.6 $^\circ$F) and humidity 65$\%$ and the experiments in metallic enclosures are conducted in elevators and cars. We agree that performing measurements inside an anechoic chamber offers a controlled, low noise setting for acquiring precise frequency response data. Nonetheless, the chamber is surrounded by a grounded metal enclosure, which influences the results in the low-frequency EQS region. This grounded enclosure enhances the overall return path capacitance, leading to a decrease in channel loss. Consequently, the EQS region displays reduced loss, and the crossover points between the EQS and EM regions shift to higher frequencies. The anechoic chamber is capable of effectively absorbing incident electromagnetic (EM) waves above 80 MHz; thus, the findings obtained in the chamber can only be reliably compared with HFSS open-air simulations at frequencies higher than 80 MHz, as suggested by Nath et al. Moreover, skin impedance ($Z_{skin}$) is affected by humidity and sweat, which can raise the received voltage ($V_{Rx}$). However, variations in $Z_{skin}$ impact $V_{Rx}$ by less than 3 dB at EQS frequencies due to high impedance return paths of capacitive EQS HBC with wearable devices. Temperature shifts tissue permittivity and conductivity, but these effects are minor at EQS frequencies and hence these variabilities can be tackled by designing transceivers with suited link margin. Therefore, ambient Electro-magnetic Interference, humidity, and temperature do not significantly affect SNR, bandwidth, or channel capacity of EQS HBC. Signal dispersion in body channel communication can occur due to two main factors: (a) amplitude distortion, which involves frequency-dependent attenuation, and (b) phase distortion (or group-delay distortion), which refers to frequency-dependent delays. The dominance of either factor at low versus high at  EQS frequencies depends on the specific channel model being used. Considering the RC-Line model of the Body channel i.e., the body path (transmitter (Tx)-skin-tissue-skin-receiver(Rx)) as a distributed RC line, it can be shown that the amplitude attenuation grows as $\propto$ $\sqrt\omega$; phase shift also grows as $\propto$ $\sqrt\omega$ where $\omega = 2 \pi f$ , $f$ = operating frequency. In capacitive electro quasistatic human body communication (EQS HBC), the human body serves as the forward path for signal transmission, while the parasitic coupling capacitance between the device's ground and the earth's ground functions as the return path. This configuration creates a closed loop for the body channel as an electrical circuit. It is worth mentioning Maity et al. demonstrated that the transmission loss through the body in the forward path is approximately 6-10 dB, depending on whether single-ended or differential excitation and termination are used. Hence the overall transmission loss is primarily influenced by the high impedance return paths in a scenario where both the transmitter (Tx) and receiver (Rx) are wearable devices. Furthermore, it was found that using high-impedance capacitive termination at the receiver leads to flat band channel characteristics in the EQS frequency domain. Hence, this flat band channel characteristic can be leveraged to reduce the effect of signal distortion from the channel. In contrast, the galvanic mode of HBC experiences higher degree of signal distortion from the tissue properties as the principle of operation lies in dipole-dipole coupling i.e., the electric field lines of the Tx dipole get weakened by the surrounding tissues before getting picked up at the Rx dipole and the extent of this enhances at higher operating frequencies. Moreover, the nonlinearities introduced from the electrode-skin interface remain considerably lower at EQS frequencies as it was shown that the voltage drop across the band-to-skin impedance ($Z_{band-skin} < 100 \Omega$ where $Z_{band-skin} = R_{band} || C_{band}$ where $R_{band}  = 100 \Omega, C_{band}  = 200 pF$) can be ignored in comparison to the voltage across the high impedance return paths ($Z_{C_{xTx}} , Z_{C_{xRx}} >100 k\Omega$).

\begin{figure}[htp!]
\centering
\includegraphics[width=0.48\textwidth]{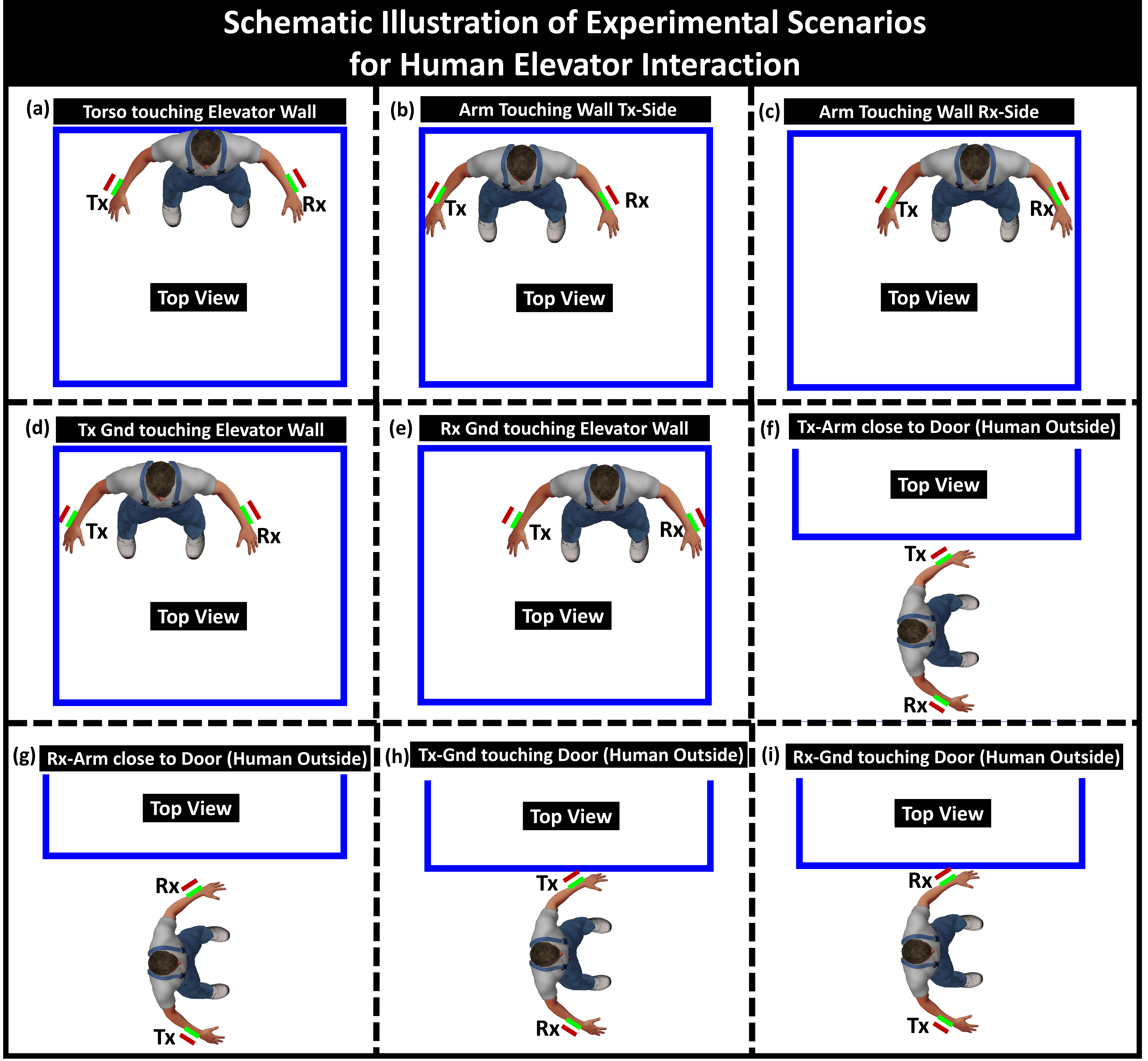}
\caption{Experimental Scenarios of interaction between EQS HBC user $\&$ Elevator: (a) Torso touching  elevator wall, (b) Arm touching wall Tx-Side, (c) Arm touching wall Rx-Side, (d) Ground of Tx touching elevator wall, (e) Ground of Rx touching elevator wall, (f) Arm with Tx close to Door (Human Outside), (g) Arm with Rx close to Door (Human Outside), (h) Ground of Tx touching Door (Human Outside), (i) Ground of Rx touching Door (Human Outside)}
\label{fig:Experimental_Scenario_Elevator}
\end{figure}
\subsection*{Schematic Illustration of the Experimental Scenarios}
This section provides a schematic representation of the experimental scenarios involving the interaction between an EQS HBC subject and the elevator walls, as illustrated in  Fig. \ref{fig:Experimental_Scenario_Elevator}. The proximity to the elevator walls affects the parasitic return path capacitances, resulting in variations in transmission loss. The scenarios depicted in Figs. \ref{fig:Experimental_Scenario_Elevator} (a, b, c) shows changes in the locations of touch-based interactions between the subject and the elevator walls. In contrast, Figs. \ref{fig:Experimental_Scenario_Elevator} (d, e) presents scenarios where the ground of the Tx and Rx components comes into contact with the elevator walls. Figs. \ref{fig:Experimental_Scenario_Elevator} (f, g) illustrates proximity-based interactions when the individual is outside. Finally, Figs. \ref{fig:Experimental_Scenario_Elevator} (h, i) depicts scenarios where the ground of the Tx and Rx components makes contact with the elevator door while the person is outside.
\begin{figure}[ht!]
\centering
\includegraphics[width=0.48\textwidth]{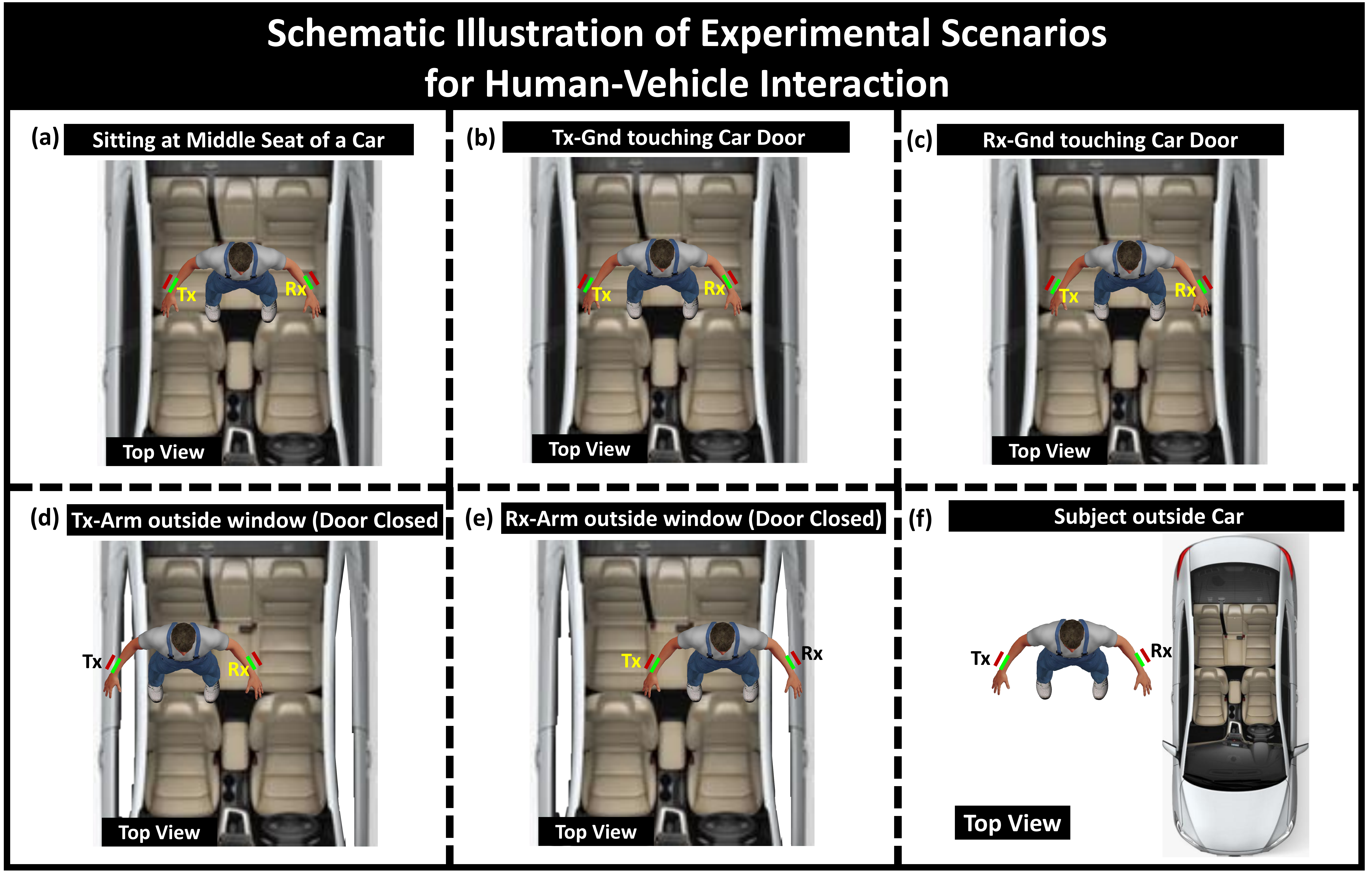}
\caption{Experimental Scenarios of interaction between EQS HBC user $\&$ Car: (a) Subject seating at middle seat of the car, (b) Ground of the Tx touching car door, (c) Ground of Rx touching car door, (d) Arm with Tx outside car window (Door Closed), (e) Arm with Rx outside car window (Door Closed), (f) Subject outside car}
\label{fig:Experimental_Scenario_Car}
\end{figure}
This section presents a diagrammatic representation of the experimental setup involving the interaction between an EQS HBC subject and a vehicle, as illustrated in Fig. \ref{fig:Experimental_Scenario_Car}. The subject's distance from the car door affects the parasitic return path capacitances, resulting in changes in transmission loss. The scenario depicted in Fig. \ref{fig:Experimental_Scenario_Car} (a) illustrates the subject seated in the middle seat of the car. In contrast, Figs. \ref{fig:Experimental_Scenario_Car} (b, c) showcase situations where the ground connections of the Tx and Rx components make contact with the car door. Figs. \ref{fig:Experimental_Scenario_Car} (d, e) demonstrates the situation when the subject extends their arm out of the car window. Lastly, Fig. \ref{fig:Experimental_Scenario_Car} (f) portrays scenarios in which the subject stands outside the vehicle.

\onecolumn

\appendices

\section*{APPENDIX IV: Comparison with Prior Studies}
\label{appendix:d}
{\scriptsize
\renewcommand{\arraystretch}{1}
\begin{longtable*}[t]{|>{\centering\arraybackslash}p{1.8cm}|
                    >{\centering\arraybackslash}p{2cm}|
                    >{\centering\arraybackslash}p{2.2cm}|
                    >{\centering\arraybackslash}p{2cm}|
                    >{\centering\arraybackslash}p{2cm}|
                    >{\centering\arraybackslash}p{2cm}|
                    >{\centering\arraybackslash}p{1.9cm}|
                    >{\centering\arraybackslash}p{1.5cm}|}
\hline
\rowcolor{headercolor}\color{white}\textbf{Authors} &
\color{white}\textbf{Frequency Range} &
\color{white}\textbf{Measurement Setup} &
\color{white}\textbf{Termination Impedance at Rx} &
\color{white}\textbf{Influence of Metal Objects} &
\color{white}\textbf{Touch-based Interaction} & 
\color{white}\textbf{Metallic Enclosure Study} & 
\color{white}\textbf{Channel Loss} \\
\hline
Lucev et al. [2012] & 100 kHz – 100 MHz & Bench-top, Ground-Connected VNA & 50 $\Omega$ Resistive & No & No & No & High-Pass, Not-Realistic \\
\hline
R. Xu et al. [2012] & 20 MHz – 100 MHz & Battery-powered Tx $\&$ Rx boards & 200 $\Omega$ Resistive & Yes & No & No & High-Pass, Not-Realistic  \\
\hline
Park et al. [2016] & 20 MHz – 150 MHz & Signal Generator \& Spectrum Analyzer,  isolating Baluns & 50 $\Omega$ Resistive \& Impedance Matching & No & No & No & Higher Loss \& Improper ground isolation\\
\hline
Maity et al. [2018] & 10 kHz – 1 MHz & Wearable Prototype & High Impedance, Capacitive & No & No & No & Realistic \\
\hline
Y. Xu et al. [2019] & 1 MHz – 50 MHz & Battery powered Tx \& Big Spectrum Analyzer (Rx) & 50 $\Omega$ Resistive & No & No & Inside Vehicle (Car) & High-Pass, Not-Realistic \\
\hline
Yang et al. [2022] & 415 kHz & Wearable Prototype & High Impedance, Capacitive & Yes & No & No & Realistic \\
\hline
Sarkar et al. [2023] & 1 MHz & Gnd-Connected Tx \& FG/Gnd-Connected Rx & High Impedance, Capacitive & Yes & No & No & Realistic with FG Rx \& Optimistic with Gnd-connected Rx\\
\hline
\textbf{This Work} [2025] & 5 MHz & Wearable Prototype & High Impedance, Capacitive & \textbf{Yes} & \textbf{Yes} & \textbf{Inside Elevator \& Car} & \textbf{Flat Band in EQS, Realistic}\\
\hline
\end{longtable*}

} 

\section*{APPENDIX V: Terminology Table}
\label{appendix:e}
\renewcommand{\arraystretch}{1}

\begin{tabularx}{\textwidth}{|>{\centering\arraybackslash}X|>{\centering\arraybackslash}X|}
\hline
\rowcolor{headerblue}
\textcolor{white}{\textbf{Terminology}} & \textcolor{white}{\textbf{Symbol}} \\
\hline
Channel Loss & $L$ \\
\hline
Self-Capacitance of electrode & $C_\text{self}$ \\
\hline
Return path Capacitance in absence of any conducting object & $C_{xTx}, C_{xRx}$  \\
\hline
Boost in Return path Capacitance from Surrounding Metals & $C_{GMTx}, C_{GMRx}$ \\
\hline
Net Return path Capacitance & $C_{\text{retTx}}, C_{\text{retRx}}$ \\
\hline
Subject’s body-to-Earth’s Ground Capacitance (no conducting object) & $C_B$ \\
\hline
Capacitance between Subject’s body and Metal object & $C_{BM}$ \\
\hline
Net Body Capacitance (with ground-connected metals) & $C_{\text{Body}}$ \\
\hline
Capacitance between device signal and ground electrodes & $C_{PP}$ \\
\hline
Fringe Capacitance & $C_F$ \\
\hline
Capacitance between device ground and subject’s body & $C_{GBTx}, C_{GBRx}$ \\
\hline
Load Capacitance at Rx & $C_L$ \\
\hline
Effective load capacitance at Rx & $C_{L(\text{eff.})} = C_L + C_{GBRx}$ \\
\hline
Transmitted Voltage & $V_{Tx}$ \\
\hline
Received Voltage & $V_{Rx}$ \\
\hline
Inter device coupling capacitance & $C_C$ \\
\hline
Body Potential & $V_B$ \\
\hline
Potential of Tx ground & $V_T$ \\
\hline
Potential of Rx ground & $V_R$ \\
\hline
Potential of Metallic Object & $V_M$ \\
\hline
Metal object to Earth’s Ground capacitance & $C_{MG}$ \\
\hline
Contact Resistance & $R_{\text{con}}$ \\
\hline
Contact Area & $A_{\text{con}}$ \\
\hline
Contact Impedance & $Z_{\text{con}}$ \\
\hline
Transfer function: \textbf{subject touching grounded metal (TGM)} & $T_{TGM}(s)$ \\
\hline
Transfer function: \textbf{subject not touching grounded metal (NTGM)} & $T_{NTGM}(s)$ \\
\hline
Transfer function: \textbf{subject touching floating metal (TFM)} & $T_{TFM}(s)$ \\
\hline
Transfer function: \textbf{subject not touching floating metal (NTFM)} & $T_{NTFM}(s)$ \\
\hline
Equivalent capacitance: \textbf{body touching metal (BTM) to earth’s ground} & $C_{BTM}$ \\
\hline
\end{tabularx}

\color{black}

\end{document}